\ifdefined\XeTeXrevision\else\pdfoutput=1\fi
\documentclass[11pt,a4paper]{article}

\usepackage[T1]{fontenc}
\usepackage{lmodern}
\usepackage[margin=2.5cm]{geometry}
\usepackage{amsmath,amssymb}
\usepackage{graphicx}
\usepackage{booktabs}
\usepackage{array}
\usepackage{xcolor}
\usepackage{cite}
\usepackage{placeins}
\usepackage{needspace}
\usepackage[colorlinks=true,linkcolor=blue!50!black,citecolor=blue!50!black,urlcolor=blue!50!black,pdftitle={Exact holographic superfluid at the Bogomolny point: transport, large charge, and R\'enyi entropies},pdfauthor={Siqing Yu}]{hyperref}

\numberwithin{equation}{section}
\newcommand{\Tc}{T_c}
\newcommand{\rhob}{\bar\rho}
\newcommand{\Bb}{\bar B}
\newcommand{\rs}{\rho_*}
\newcommand{\ee}{\mathcal{E}}
\newcommand{\Om}{\Omega}
\newcommand{\SF}{{\rm SF}}
\newcommand{\N}{{\rm N}}
\newcommand{\src}{{\rm src}}
\newcommand{\vev}{{\rm vev}}
\newcommand{\reg}{{\rm reg}}
\newcommand{\IR}{{\rm IR}}
\newcommand{\bdry}{{\rm bdry}}
\newcommand{\CA}{\mathcal{A}}
\newcommand{\CB}{\mathcal{B}}
\newcommand{\vk}{\vec k}
\newcommand{\Th}{\Theta}
\newcommand{\Rc}{\mathcal{R}}
\newcommand{\Nn}{\mathcal{N}}
\newcommand{\muc}{\mu_c}

\begin{document}

\hypersetup{pageanchor=false}
\begin{titlepage}
\begin{center}
\vspace*{1.5cm}
{\LARGE\bfseries Exact holographic superfluid at the Bogomolny point:\\ transport, large charge, and R\'enyi entropies\par}
\vspace{1.2cm}
\renewcommand{\thefootnote}{\fnsymbol{footnote}}
{\large Siqing Yu\footnote{\href{mailto:yu@siqing.ai}{\texttt{yu@siqing.ai}}}\par}
\vspace{0.3cm}
{San Diego, California, USA\par}
\end{center}
\vspace{1.0cm}
\begin{abstract}
\noindent
The probe-limit Abelian Higgs model in AdS$_5$ is a standard holographic model of a strongly coupled superfluid, usually solved numerically. With the scalar mass at the Breitenlohner--Freedman bound and the quartic coupling at the Bogomolny value, its condensed phase has a closed form on every static, uncharged AdS$_5$ background with maximally symmetric spatial sections. We prove that it is the global minimum of the free energy at fixed charge, and linearly stable, as is any nodeless condensate of the model in standard quantization with nonnegative quartic coupling and no superflow. On the planar black brane it is the superfluid at all temperatures below $T_c$: the superfluid fraction follows the Gorter--Casimir two-fluid law exactly, and the speed and attenuation of sound are in closed form. In global AdS$_5$ it is the lowest state of charge $Q$ on $\mathbb{R}\times S^3$ within the model, with its scaling dimension in closed form at all $Q$, classically in the bulk. Its excitations include towers, from an infrared fixed point of the probe sector, that the superfluid effective theory does not describe. On the hyperbolic black holes that compute charged R\'enyi entropies of a ball, it is a homogeneous hairy phase whose condensate does not fall off toward the entangling surface. With a pinning field there, the phase sets in at R\'enyi index exactly one at zero chemical potential, as found for a neutral scalar. Well below a bound on the quartic coupling, the static, homogeneous solutions computed at other couplings and masses keep an infrared fixed point with towers, but not the exact two-fluid law.
\end{abstract}
\end{titlepage}
\hypersetup{pageanchor=true}
\setcounter{footnote}{0}

\enlargethispage{\baselineskip}%
{\setlength{\parskip}{0.2em}\tableofcontents}
\newpage

\section{Introduction}
\label{sec:intro}

The application of the AdS/CFT correspondence \cite{Maldacena1997,GKP1998,Witten1998} to strongly interacting superfluids and superconductors goes back to \cite{Gubser2008,HHH2008a,HHH2008b} and is reviewed in \cite{Hartnoll2009rev,Herzog2009rev,HLS2016book}. Most of this work is numerical. Analytic approaches expand in the condensate near the critical temperature \cite{MaedaOkamura2008,HerzogPufu2009,Herzog2010}, approximate the condensate by matching or variational methods \cite{GKS2009,SiopsisTherrien2010}, or work in a limit such as large dimension \cite{EmparanTanabe2014} or a double-scaling limit \cite{ABK2010}; an analytic zero-temperature ground state is known in a related model at zero density \cite{RenXie2023}. Other exact results concern the critical point: the zero mode is known in closed form in some models \cite{BHMS2009,Herzog2010,NatsuumeOkamura2018} and can be analyzed through Heun's equation \cite{ChounCaiSin2022}, and the leading terms of the dual Ginzburg--Landau theory have been obtained around it \cite{NatsuumeOkamura2018,NatsuumeOkamura2022,Natsuume2025,Natsuume2024,Natsuume2026}. Closed-form hairy black holes are known in other models \cite{MTZ2004}, some with a superconducting interpretation \cite{KPS2009}. To our knowledge, however, the condensed phase of the standard probe-limit model has not been found in closed form away from the critical point.

\paragraph{This paper.} The probe-limit Abelian Higgs model in AdS$_5$, the version of the model of \cite{Herzog2010} with a canonical kinetic term, has a condensed phase in closed form when the scalar mass saturates the Breitenlohner--Freedman (BF) bound and the quartic coupling takes the Bogomolny value. The closed form holds on every static, uncharged AdS$_5$ background with maximally symmetric spatial sections, namely the planar black brane, global AdS$_5$ and its black holes, and the hyperbolic black holes, and also on black branes deformed by massless axions (axion branes) \cite{AndradeWithers2014}. The slicing enters only through the blackening function. Each slicing answers a different question about the dual conformal field theory (CFT): the thermal superfluid on the plane, the lowest states of large charge on the sphere and the R\'enyi entropies of a ball on the hyperbolic slicing. One fact ties all the slicings together: at every temperature the free energy of condensation has exactly the mean-field form, which fixes the thermodynamics and the scaling dimension $\Delta(Q)$ of the state of charge $Q$ and ends the curvature expansion of the pressure on the sphere at second order. This form does not fix the fluctuations. A second fact ties the plane to the sphere at large charge: apart from the phonons, the fluctuations at zero temperature are governed by one infrared fixed point. ``Exact'' in the title refers to the backgrounds; their fluctuations are computed numerically except where a closed form is stated.

\paragraph{The Bogomolny point.} The quartic coupling $\lambda=e^2/2$ (convention of this paper) is the critical or Bogomolny coupling of the Abelian Higgs model. In flat space the vector and the amplitude (Higgs) mode then have equal masses, which separates type-I from type-II superconductivity; the static energy is bounded below by the magnetic flux, and the vortices that saturate the bound obey first-order equations and exert no force on each other \cite{Bogomolny1976,deVegaSchaposnik1976,JacobsRebbi1979}. Such configurations are called BPS \cite{PrasadSommerfield1975}, and at this coupling the model admits a supersymmetric extension with a D-term potential \cite{ENS1994}.

We call this coupling, at the BF-bound mass, the Bogomolny point. In the holographic model the same coupling appears in three places, which together indicate why a closed form rational in the radial coordinate exists, at nonzero chemical potential, only at this coupling. At zero temperature the condensate interpolates between the maximum of the scalar potential at the boundary and its minimum deep in the interior, an infrared fixed point. The closed form requires the vector and the amplitude fluctuation to have equal masses at that fixed point. The solution obeys first-order equations, which reduce to a static sine-Gordon kink in the logarithm of the radial coordinate; we have not derived them from a Bogomolny bound. Near the critical temperature the condensate is a series in its amplitude whose third-order term contains a logarithmic function of the radial coordinate. Its coefficient vanishes precisely at the Bogomolny coupling, where the series is rational at every order and resums to the exact solution.

\paragraph{Large charge.} The lowest scaling dimension $\Delta(Q)$ of an operator of charge $Q$ is a basic datum of a CFT with a U(1) global symmetry. By the state-operator correspondence it is the energy, in units of $1/R$, of the lowest state of charge $Q$ on $\mathbb{R}\times S^{d-1}$ of radius $R$. In a generic CFT, theories with a moduli space being an exception, this state is expected to be a homogeneous superfluid. At large $Q$ its dimension has an expansion whose coefficients are Wilson coefficients of the effective theory of a conformal superfluid \cite{Son2002,Hellerman2015,Monin2017}; see \cite{GOR2020} for a review. In four dimensions it starts at order $Q^{4/3}$, and quantum fluctuations of the phonon add a universal term proportional to $\ln Q$ \cite{Cuomo2020}. The same effective theory fixes the scaling with $Q$ of the three-point coefficients between the lowest states of nearby charges \cite{Monin2017,Cuomo2020}, and it describes the light excitations: phonons and, at large spin, vortices \cite{CdlFMPR2017,Cuomo2019}. At weak coupling, a double-scaling limit at fixed coupling times charge connects perturbation theory to the large-charge regime \cite{BCMR2019}; at fixed coupling times charge squared the anomalous dimensions are known exactly \cite{ARR2019}.

In holographic CFTs the lowest state of large charge is a charged soliton (an AdS boson star \cite{AstefaneseiRadu2003}) or a black hole in global anti--de Sitter (AdS) space. Charged solitons have been constructed perturbatively at small charge \cite{Basu2010,BMP2010,Dias2011,IshiiNakayama2024} and numerically \cite{Dias2011,GRW2012}, also at large charge, where their masses reproduce the structure of the large-charge expansion \cite{delaFuenteZosso2020,LiuLuPang2020,GLLP2021}. As noted in \cite{LiuLuPang2020}, few exact AdS boson stars are known. Other exact charged solutions are known: the extremal Reissner--Nordstr\"om black hole gives a closed-form relation between mass and charge, with the U(1) unbroken \cite{LORS2018}; supersymmetric horizonless solutions of supergravity, such as the bubbling geometries of \cite{LLM2004,CLP2004} and the superstars of \cite{MyersTafjord2001}, carry the charge in gauge fields, with no charged scalar condensed; and an ungauged scalar with a nonpolynomial self-interaction tuned to its mass has an exact Q-ball solution in a fixed AdS background \cite{RaduSubagyo2012}. We are not aware of a closed-form soliton of a gauged charged scalar in global AdS at finite charge (a probe boson star), nor, outside supersymmetric settings, of a holographic large-charge state with a spontaneously broken U(1) and $\Delta(Q)$ in closed form at all $Q$. Here the global member of the family gives the lowest classical state of charge $Q$ within the model and its dimension in closed form at all $Q$, at leading order in $1/\Nn$ and in the probe limit, where $\Nn$ is proportional to the two-point coefficient of the current. This dimension has the form of the leading semiclassical result for the U(1) model at the Wilson--Fisher fixed point at fixed coupling times charge \cite{BCMR2019}, with the dimension of the order parameter in place of the conformal mass. The strong coupling enters through the overall coefficient and through the excitations above the state.

\paragraph{What the closed form shows.} Besides providing exact benchmarks for numerical and perturbative methods, the closed form establishes what numerical solutions can only approximate: that a relation holds exactly, that a coefficient vanishes or how two regimes connect.
\begin{itemize}
\item \emph{The two-fluid structure on the plane}. The superfluid fraction is $n_s/n=1-(T/\Tc)^4$ at every temperature below $\Tc$, the two-fluid law of Gorter and Casimir. Equivalently, the normal component carries the charge $Ts/\mu$, as the whole charge does in the normal phase. The DC conductivity of the normal component is its normal-phase value times the normal fraction. Among the couplings checked, the two relations above hold at all temperatures only at the Bogomolny coupling. The speed and the attenuation of sound and the third bulk viscosity follow in closed form, and the critical velocity analytically at low temperature. On axion branes the condensate forms even at zero chemical potential, and a normal fraction survives at zero temperature.
\item \emph{One infrared fixed point}. At zero temperature the planar solution is a domain wall between two AdS$_5$ regions. Its infrared end is a conformal fixed point at which the U(1) is broken, with a current and an order parameter of dimensions $5$ and $2+2\sqrt3$. This fixed point sets the power-law spectral weight of the optical conductivity and the continuum of the amplitude mode on the plane, and the towers of excitations above the lowest states of large charge on the sphere.
\item \emph{Large-charge data at all charges}. At leading order in $1/\Nn$, the dimension $\Delta(Q)$ connects a dilute gas of quanta with a repulsive interaction to the effective theory and gives exactly the coefficients of that theory which enter $\Delta(Q)$. The three-point coefficient of the order parameter interpolates between its value for a generalized free field and the scaling of the effective theory. Besides the phonons, the excitations include towers that the superfluid effective theory does not describe; these light states of the probe sector have gaps of order $1/R$ and do not decouple at large charge. The leading large-charge correction to the phonon energies, as a function of the chemical potential, is the same as at weak coupling \cite{BCMR2019}, and it determines a curvature coefficient of the effective theory that $\Delta(Q)$ leaves free. As in the leading weak-coupling result, the large-charge expansion converges, rather than being merely asymptotic, above a charge of order $\Nn$.
\item \emph{R\'enyi entropies}. On the hyperbolic slicing the closed form gives a homogeneous hairy phase of the black holes that compute the charged R\'enyi entropies of a ball, for the probe sector, wherever it exists. Its condensate does not vanish at the entangling surface, so the phase is not normalizable on hyperbolic space. With a pinning field on the entangling surface, its critical index at zero chemical potential is exactly one, as for a neutral scalar at the BF bound \cite{DMRS2010,BMM2013}, a consequence of the BF-bound mass that holds for any positive quartic coupling. With normalizable modes only, the normal phase is linearly stable below a critical chemical potential, computed here, which vanishes at an index of about 6.58, the onset found at zero chemical potential in \cite{BMM2013}.
\end{itemize}
Two of the main results do not rely on the closed form. First, an exact energy identity shows that any nodeless condensate of the Abelian Higgs model in the probe limit has the lowest free energy of all classical states at its charge, and is therefore linearly stable. This holds in standard quantization, for a nonnegative quartic coupling and without superflow; on $H^3$ the comparison is with states at finite energy relative to the condensate. Second, a horizon formula gives the sound attenuation for any quartic coupling. Away from the critical temperature the exact solution is an example of the first and a test of the second. Well below a bound on the quartic coupling, the static, homogeneous solutions computed at other couplings and masses have the structure of the Bogomolny point: at zero temperature the nodeless branch flows to the minimum of the potential. At these couplings and masses the spectrum on the sphere keeps its form, with towers at dimensions that depend on the couplings, and the curvature coefficient that $\Delta(Q)$ leaves free keeps its value, while the temperature independence of the charge density, the two-fluid relations and the perfect-square pressure fail where they were tested. Table~\ref{tab:summary} collects the main results.

\paragraph{Relation to earlier work.} Several of the results overlap with earlier work and are checked against it. Herzog constructed the condensed phase near the critical temperature as a series in the condensate, in a family of models that contains this one, and obtained the near-critical sound speed and, for this model at zero quartic coupling, the sound attenuation and the relaxation rate of the condensate \cite{Herzog2010}. The exact solution resums his series at the Bogomolny coupling and reproduces his published fourth-order coefficients. Natsuume and Okamura identified the dual Ginzburg--Landau theory of a Lifshitz family that contains this model at zero quartic coupling \cite{NatsuumeOkamura2018}, and added a dynamical boundary Maxwell field \cite{NatsuumeOkamura2022}. Natsuume extended it to a quartic coupling \cite{Natsuume2025} and to first order in a Gauss--Bonnet coupling \cite{Natsuume2024}, derived its critical dynamics at zero quartic coupling \cite{Natsuume2026} and computed the near-critical correlators at nonzero quartic and kinetic couplings \cite{Natsuume2026b}. Donos and collaborators derived horizon formulas for the dissipative coefficients of holographic superfluids \cite{DKP2021,DonosKailidis2022a} and the near-critical effective theory \cite{DonosKailidis2022b,DonosKailidis2025}, and Herzog, Lisker, Sur\'owka and Yarom computed the near-critical third bulk viscosity of this model at zero quartic coupling \cite{HLSY2011}. On the hyperbolic slicing, neutral hairy black holes with a scalar at the AdS$_5$ BF bound are known analytically, with backreaction, in the uncharged limit of Einstein--Maxwell--dilaton models \cite{Ren2019,Bai2022}. For a neutral scalar at the BF bound, the homogeneous hairy phase sets in exactly at index one. This was found in \cite{DMRS2010}, which gives the static mode at that index in closed form and constructs the backreacted hairy black holes perturbatively and numerically. Belin, Maloney and Matsuura computed their R\'enyi entropies and located the onset with normalizable modes only as a function of the dimension of the scalar \cite{BMM2013}. The onset of the hyperbolic superconducting instability with a chemical potential was located on charged hyperbolic black holes, numerically in AdS$_4$ \cite{BHMM2014} and analytically at zero temperature in AdS$_5$ \cite{Ren2020}, where the hairy phase was not constructed. Superconductors on axion branes were constructed numerically, with backreaction, in AdS$_4$ \cite{AndradeGentle2015,KKP2015}.

\begin{table}[t]
\centering
\footnotesize
\begin{tabular}{@{}>{\raggedright}p{0.25\textwidth} >{\raggedright}p{0.44\textwidth} >{\raggedright}p{0.1\textwidth} l@{}}
\toprule
Quantity & Result & Slicing & Section \tabularnewline
\midrule
Condensed phase & eq.~\eqref{eq:family} & all & \ref{sec:familysol} \tabularnewline
$T=0$ domain wall & first-order eqs.~\eqref{eq:firstorder}--\eqref{eq:kink}; infrared fixed point \eqref{eq:IRdims} & plane & \ref{sec:structure} \tabularnewline
First-order form & eqs.~\eqref{eq:ckappa} and \eqref{eq:firstorderT}, at every $T$ & all & \ref{sec:structure} \tabularnewline
Free energy of condensation & $\Om_\SF-\Om_\N=-\frac{L}{8e^2}(\mu^2-\muc^2)^2$ & all & \ref{sec:grandpot} \tabularnewline
Global minimum & eq.~\eqref{eq:globalmin}; lowest classical state at fixed $Q$, unique up to gauge for $\mu\neq0$; any $\lambda\ge0$, nodeless, no superflow; on $H^3$, finite relative energy & all & \ref{sec:globalmin} \tabularnewline
Linear stability & $\ee_2\ge0$, eq.~\eqref{eq:E2positive}; same conditions; on $H^3$, normalizable perturbations & all & \ref{sec:energy} \tabularnewline
Superfluid fraction, conductivity & $n_s/n=1-(T/\Tc)^4$; $\sigma_\reg=(\pi LT/e^2)(T/\Tc)^4$; ${\rm Re}\,\sigma\propto\omega^5$ at $T=0$ & plane & \ref{sec:transverse} \tabularnewline
Sound & $c_s^2=\frac13[1-(T/\Tc)^4]$, $\Gamma=5T^5/(36\pi\Tc^6)$; horizon formula, any $\lambda$ & plane & \ref{sec:sound} \tabularnewline
Axion brane & $T_c(\mu,\alpha)$, eq.~\eqref{eq:Tcaxion}; eqs.~\eqref{eq:nsaxion}--\eqref{eq:Gammaaxion} & axions & \ref{sec:axion} \tabularnewline
$\Delta(Q)$ & eq.~\eqref{eq:DeltaQ}; mean-field form; leading order in $1/\Nn$, probe limit & $S^3$ & \ref{sec:DeltaQ} \tabularnewline
Effective theory & $P=c_1(X-\frac23\Rc)^2$, $c_1=L/8e^2$; three-point coefficient, eq.~\eqref{eq:lambdaQ} & $S^3$ & \ref{sec:EFT} \tabularnewline
Excitations at large charge & phonons; towers $5+\ell+2n$, $2+2\sqrt3+\ell+2n$, \ldots & $S^3$ & \ref{sec:spectrum} \tabularnewline
Phonons at next-to-leading order & $\omega_\ell R=\sqrt{\ell(\ell+2)/3}\,[1+(\ell-1)(\ell+3)/9\mu^2R^2]$; Ricci-tensor coefficient $c_4=-L/4e^2$, any $V(\rho)$ & $S^3$ & \ref{sec:phononNLO} \tabularnewline
R\'enyi entropies of a ball & eq.~\eqref{eq:Sn}; homogeneous saddle; critical index one at $\mu=0$, eq.~\eqref{eq:Snneutral}, any $\lambda>0$ & $H^3$ & \ref{sec:renyi} \tabularnewline
\bottomrule
\end{tabular}
\caption{Summary of results, in the probe limit at $m^2L^2=-4$ and $\lambda=e^2/2$ unless marked otherwise. Here $\Nn=\pi^2L/e^2$, $X$ is the square of the chemical potential and $\Rc$ the Ricci scalar of the boundary. The sound results and the phonon correction are confirmed by the numerical computations of appendix~\ref{app:numerics}.}
\label{tab:summary}
\end{table}

Section~\ref{sec:family} presents the model and the exact family, and section~\ref{sec:every} the grand potential, the energy identity and the stability argument, which hold on every slicing. Sections~\ref{sec:plane}, \ref{sec:sphere} and~\ref{sec:renyi} treat the planar and axion branes, the sphere and the hyperbolic slicing, respectively. Section~\ref{sec:beyond} asks which results hold at other couplings and masses, on other backgrounds and in top-down models, and section~\ref{sec:discussion} discusses the towers and the probe limit and lists open questions. The appendices contain the derivations and the numerical methods.

\section{The exact family}
\label{sec:family}
\label{sec:solution}

\subsection{Setup}
\label{sec:setup}

The model has the Einstein--Maxwell--Higgs action $S=S_{\rm bulk}+S_\bdry$, with
\begin{gather}
S_{\rm bulk}=\int d^5x\,\sqrt{-g}\,\Big[\frac{1}{2\kappa_5^2}\Big(\mathcal{R}[g]+\frac{12}{L^2}\Big)-\frac{1}{4e^2}F^{ab}F_{ab}-V(|\Psi|)-|D_a\Psi|^2\Big],
\label{eq:action}\\
V(|\Psi|)=m^2|\Psi|^2+\lambda|\Psi|^4,\nonumber
\end{gather}
where $D_a=\nabla_a-iA_a$ and $\lambda\ge0$; quartic potentials and more general functions of the scalar in the action have been studied in \cite{FGR2010a,AprileRusso2010,Herzog2010}. The boundary term $S_\bdry$ is given in \eqref{eq:Sbdry}. The signature is mostly plus; $a,b,\ldots$ run over the bulk coordinates, $\mu,\nu,\ldots$ over those of the boundary, and $i,j,\ldots$ over its spatial directions. The probe limit $\kappa_5/(eL)\ll1$ \cite{HHH2008b} is taken throughout, so that the Abelian-Higgs sector does not backreact on the metric.

\paragraph{Backgrounds.} The static, uncharged solutions of Einstein's equations with a negative cosmological constant whose spatial sections are maximally symmetric are \cite{Birmingham1998,Emparan1999}
\begin{equation}
ds^2=\frac{L^2}{u^2}\Big(\frac{du^2}{f(u)}-f(u)\,dt^2+d\Sigma_\kappa^2\Big),\qquad f(u)=1+\frac{\kappa u^2}{R^2}-\Big(1+\frac{\kappa u_h^2}{R^2}\Big)\frac{u^4}{u_h^4},
\label{eq:metric}
\end{equation}
where $d\Sigma_\kappa^2$ is the metric of the sphere $S^3$ ($\kappa=1$), flat space $\mathbb{R}^3$ ($\kappa=0$) or the hyperbolic space $H^3$ ($\kappa=-1$), of curvature radius $R$. The conformal boundary is at $u\to0^+$, with the boundary metric $-dt^2+d\Sigma_\kappa^2$, and the horizon is at $u=u_h$, at the temperature
\begin{equation}
T=-\frac{f'(u_h)}{4\pi}=\frac{1}{\pi u_h}\Big(1+\frac{\kappa u_h^2}{2R^2}\Big).
\label{eq:T}
\end{equation}
For $\kappa=0$, $R$ drops out, and \eqref{eq:metric} is the AdS--Schwarzschild black brane, with $f=1-u^4/u_h^4$ and $u_h=1/\pi T$. For $\kappa=1$, the limit $u_h\to\infty$ is global AdS$_5$, with $f=1+u^2/R^2$ and a regular center at $u\to\infty$, and finite $u_h$ gives the global AdS$_5$--Schwarzschild black hole. For $\kappa=-1$ the metric is a hyperbolic black hole with its horizon at $u_h$ for $u_h\le\sqrt2R$, with $T=0$ at $u_h=\sqrt2R$; at $u_h=R$, where $T=1/2\pi R$, it is AdS$_5$ itself, in the coordinates of an AdS--Rindler wedge. Near the center of global AdS$_5$ the coordinate $r\equiv R/u$ is also used.

The scalar mass is $m^2=-4/L^2$, as in \cite{Herzog2010}, which saturates the BF bound \cite{BF1982a,BF1982b}. Near the boundary the fields behave as
\begin{equation}
\Psi=\frac{u^2}{L^{3/2}}\Big[-\Psi^{(b)*}\ln\frac{u}{u_c}+\langle O_2\rangle\Big]+\cdots,\qquad
A_\mu=A^{(b)}_\mu+\frac{e^2}{2L}\langle J_\mu\rangle u^2+\cdots,
\label{eq:expansions}
\end{equation}
where $u_c\to0$ is an ultraviolet cutoff. The coefficient $\Psi^{(b)}$ is the source of the scalar operator $O_2$ of dimension two, the order parameter, and $A^{(b)}_\mu$ is the source of the conserved U(1) current $J^\mu$ \cite{GKP1998,Witten1998}. The normalizations in \eqref{eq:expansions} follow from the variation of the renormalized on-shell action (appendix~\ref{app:asymptotics}) and are confirmed by thermodynamics in section~\ref{sec:thermo}. The chemical potential is $\mu=A^{(b)}_t$ and the charge density is $n\equiv\langle J^t\rangle=-\langle J_t\rangle$.

The boundary terms are \cite{deHaro2001,BFS2002,Skenderis2002,HKS2009,Herzog2010}
\begin{equation}
S_\bdry=\int_{u=u_c}d^4x\,\sqrt{-\gamma}\,\Big[\frac{2}{L}|\Psi|^2+\Psi\, n^a(D_a\Psi)^*+\Psi^* n^aD_a\Psi-\frac{L}{4e^2}\ln\frac{u_c}{u_0}\,F_{\mu\nu}F^{\mu\nu}\Big],
\label{eq:Sbdry}
\end{equation}
where $\gamma_{\mu\nu}$ is the induced metric at the cutoff, which also raises the indices, and $n^a$ is the outward unit normal. The scalar terms make the variational principle well posed with the coefficient of the logarithm in \eqref{eq:expansions} held fixed, and for $\Psi^{(b)}=0$ they cancel the finite surface term of the bulk scalar action (section~\ref{sec:thermo}). The last term cancels the logarithmic divergence of the Maxwell action in AdS$_5$ \cite{MarolfRoss2006}. Higher boundary terms vanish as $u_c\to0$, and changing $u_0$ changes only contact terms (appendix~\ref{app:asymptotics}).

The scalar is written as $\Psi=\rho\,e^{i\varphi}/\sqrt2$ with $\rho$ real, and the gauge-invariant vector is $B_a\equiv A_a-\partial_a\varphi$. The phase drops out of the equations of motion, which become
\begin{equation}
\nabla_a\nabla^a\rho=\big(m^2+\lambda\rho^2+B_aB^a\big)\rho,\qquad
\frac{1}{e^2}\nabla_aF^{ab}=\rho^2B^b,
\label{eq:reduced}
\end{equation}
with $F_{ab}=\partial_aB_b-\partial_bB_a$. For a static, homogeneous configuration the $u$-component of the second equation reads $\rho^2B_u=0$, so $B_u=0$ in the condensed phase. The spatial components describe a superflow, treated in section~\ref{sec:superflow}; here $B_i=0$. The equations for $\rho(u)$ and $B_t(u)$ are
\begin{equation}
\frac{u^5}{L^2}\,\partial_u\Big(\frac{f}{u^3}\,\partial_u\rho\Big)+\frac{u^2B_t^2}{L^2f}\,\rho=\big(m^2+\lambda\rho^2\big)\rho,
\qquad
u^3f\,\partial_u\Big(\frac{\partial_uB_t}{u}\Big)=e^2L^2\rho^2B_t ,
\label{eq:radial}
\end{equation}
with $B_t(u_h)=0$ and $\rho(u_h)$ finite at a horizon, and both fields regular at the center of global AdS$_5$. The slicing enters only through $f$, since $\sqrt{-g}$ has the same dependence on $u$ for every $\kappa$. The scalar source must also vanish, $\Psi^{(b)}=0$, so that the U(1) symmetry of the boundary theory is broken spontaneously.
\label{sec:reduced}

\subsection{Solution}
\label{sec:familysol}

Solutions with $\rho=0$ and $B_t=\mu(1-u^2/u_h^2)$ describe the normal phase, on every slicing. For $\rho\neq0$ the coupled equations \eqref{eq:radial} are usually solved numerically. With the additional tuning $\lambda=e^2/2$, however, they have, on every background \eqref{eq:metric}, the solution
\begin{equation}
\boxed{\;\begin{gathered}eL\rhob=\frac{p\,u^2}{1+\mu^2u^2/4},\qquad \Bb_t=\mu\,\frac{f(u)}{1+\mu^2u^2/4},\\[0.5ex]
p^2=\frac12\big(\mu^2-\muc^2\big)\Big(\mu^2+\frac{4}{u_h^2}\Big),\qquad \muc^2=4\Big(\frac{1}{u_h^2}+\frac{\kappa}{R^2}\Big),\end{gathered}\;}
\label{eq:family}
\end{equation}
for $\mu^2>\muc^2$. The overbar denotes the background. From \eqref{eq:expansions}, the condensate and the charge density of the solution are
\begin{equation}
\langle O_2\rangle=\frac{L^{1/2}}{e}\,\frac{p}{\sqrt2},\qquad n=\frac{L}{e^2}\,\frac{\mu}{2}\Big(\mu^2-\frac{4\kappa}{R^2}\Big),
\label{eq:dataK}
\end{equation}
so that at fixed $\mu$ the charge density does not depend on the temperature on any slicing. At $\mu^2=\muc^2$ the solution reduces to the normal phase.

\paragraph{Why the ansatz closes.}
\label{sec:closure} Consider $eL\rho=p\,u^2/(1+Ku^2)$ and $B_t=\mu f/(1+Ku^2)$ with constants $p$ and $K$ and with $\mu\neq0$, on the metric \eqref{eq:metric} with $m^2$ and $\lambda$ arbitrary. The Maxwell equation in \eqref{eq:radial} holds, for a given $K$, if and only if
\begin{equation}
p^2=8\Big[K^2-\frac{\kappa K}{R^2}-\frac{1}{u_h^2}\Big(\frac{1}{u_h^2}+\frac{\kappa}{R^2}\Big)\Big].
\end{equation}
The scalar equation, multiplied by its denominator, then becomes a cubic polynomial in $u^2$. For $p\neq0$ its four coefficients require $m^2L^2=-4$, $K=\mu^2/4$, $\lambda=e^2/2$ and $u_h^{-2}(u_h^{-2}+\kappa/R^2)(4K-\mu^2)=0$, which then holds automatically. Both tunings are therefore necessary for a solution of this form, and the curvature of the spatial sections plays no role in the argument: only the fact that $f$ is a quadratic polynomial in $u^2$ with $f(0)=1$ is used. This is why the closed form also holds on the axion branes below.

The denominator need not be assumed. With $x=u^2$ and $eL\rho=xG(x)$, suppose only that $B_t/f$ is proportional to $G$, with $m^2$ and $f$ arbitrary, $f(0)=1$, $G(0)=p$ and $B_t(0)=\mu$. The Maxwell equation in \eqref{eq:radial} then reads $4(fG)^{\prime\prime}=G^3$, with primes denoting $d/dx$, and at $\lambda=e^2/2$ the scalar equation reads
\begin{equation*}
4f\Big[\Big(\frac{x}{2}+\frac{\mu^2f}{p^2}\Big)G\Big]^{\prime\prime}=\Big[2(x\partial_x-1)(x\partial_x-2)(f-1)+m^2L^2+4\Big]\frac{G}{x}.
\end{equation*}
The right side vanishes for $m^2L^2=-4$ and $f$ quadratic in $u^2$, the kernel of the operator. The bracket on the left is then linear in $x$, and the Maxwell equation leaves the single pole above. This shows where each tuning enters, not why.

\begin{figure}[t]
\centering
\includegraphics[width=\textwidth]{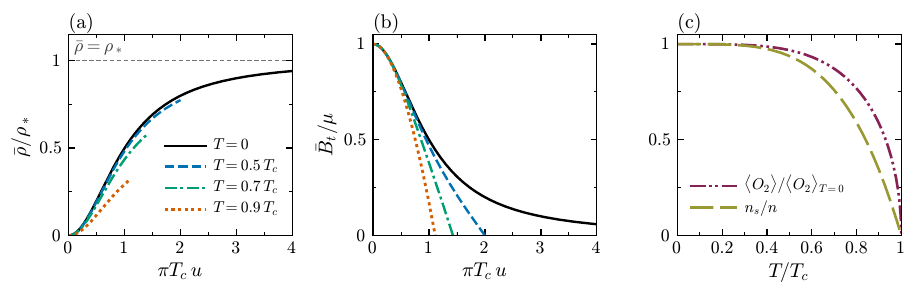}
\caption{The planar member of the family at fixed chemical potential $\mu=2\pi\Tc$. (a) The scalar, in units of the minimum $\rs$ of the potential (dashed line), and (b) the gauge potential; each finite-temperature curve ends at the horizon, $\pi\Tc u_h=\Tc/T$, where $\Bb_t$ vanishes. (c) The condensate \eqref{eq:data} and, for comparison, the superfluid fraction \eqref{eq:ns}. This and all later figures were plotted from the computed data with programs written by Claude Opus 5.5 (Anthropic).}
\label{fig:profiles}
\end{figure}

\paragraph{Planar brane.} For $\kappa=0$, $\muc=2/u_h=2\pi T$, and with $\mu=2\pi\Tc$ eq.~\eqref{eq:family} reads
\begin{equation}
\rhob=\rs\sqrt{1-\frac{T^4}{\Tc^4}}\;\frac{\pi^2\Tc^2u^2}{1+\pi^2\Tc^2u^2},\qquad
\Bb_t=\mu\,\frac{f(u)}{1+\pi^2\Tc^2u^2},\qquad \mu=2\pi\Tc,\quad \rs=\frac{2\sqrt2}{eL},
\label{eq:solution}
\end{equation}
for $T<\Tc$. The condensate and the charge density are
\begin{equation}
\langle O_2\rangle=\frac{2L^{1/2}\pi^2\Tc^2}{e}\sqrt{1-\frac{T^4}{\Tc^4}},\qquad
n=\frac{L\mu^3}{2e^2}.
\label{eq:data}
\end{equation}
The temperature $\Tc$ enters as a constant of integration: it is the temperature $\mu/2\pi$ at which the normal phase acquires a static zero mode, which does not depend on $\lambda$ \cite{Herzog2010}. The order parameter has the mean-field exponent $1/2$. The profiles and the condensate are shown in figure~\ref{fig:profiles}.

\begin{figure}[t]
\centering
\includegraphics[width=\textwidth]{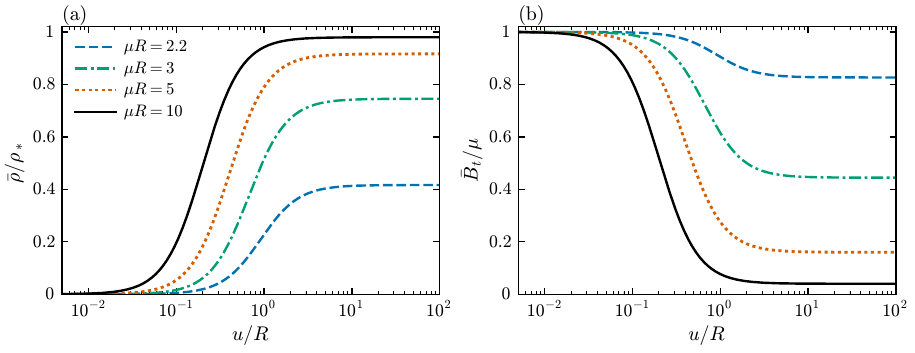}
\caption{The soliton in global AdS$_5$, eq.~\eqref{eq:family} with $\kappa=1$ and $u_h\to\infty$: (a) the scalar, in units of the minimum $\rs=2\sqrt2/eL$ of the potential, and (b) the gauge potential.}
\label{fig:profilesglobal}
\end{figure}

\paragraph{Global AdS$_5$ and its black holes.} For $\kappa=1$ and $u_h\to\infty$ the solution is a charged soliton in global AdS$_5$, which is the lowest classical state of charge $Q$ on $\mathbb{R}\times S^3$ within this model (section~\ref{sec:globalmin}). It exists for $\mu R>2$ and is regular at the center, where $eL\rhob\to2\sqrt2\,(1-4/\mu^2R^2)^{1/2}$ and $\Bb_t\to4/\mu R^2$. For $\kappa=1$ and finite $u_h$ it is the superfluid on $S^3$ at nonzero temperature, on the global black hole. The profiles of the soliton are shown in figure~\ref{fig:profilesglobal}.

\paragraph{Hyperbolic black holes.} For $\kappa=-1$ the solution is the superfluid on $H^3$. Since $\muc^2=4(1/u_h^2-1/R^2)$ is negative below $T=1/2\pi R$, where $u_h>R$, the condensate there exists for every $\mu$, including $\mu=0$. The hyperbolic black holes at $T=1/2\pi nR$, with $n$ the R\'enyi index, compute the R\'enyi entropies of a ball; this member gives their homogeneous hairy phase (section~\ref{sec:renyi}). At $\mu=0$ the scalar equation alone admits $eL\rho=p_0\,u^2$ for any $\lambda>0$ where $\muc^2<0$, with $(\lambda/e^2)\,p_0^2=-4u_h^{-2}(u_h^{-2}+\kappa/R^2)$; at $\lambda=e^2/2$ it is the $\mu=0$ member of \eqref{eq:family}.

\paragraph{Axion branes.} On the black brane of AdS$_5$ with three massless axions $\psi_I=\alpha x_I$ \cite{AndradeWithers2014}, with the action $-(1/4\kappa_5^2)\int d^5x\sqrt{-g}\,\sum_I(\partial\psi_I)^2$, the metric is \eqref{eq:metric} with flat sections and $\kappa/R^2$ replaced by $-\alpha^2/4$, so that $f=1-\alpha^2u^2/4-(1-\alpha^2u_h^2/4)\,u^4/u_h^4$. The axions do not couple to the probe fields, so eq.~\eqref{eq:family} holds there with the same replacement; its thermodynamics and transport are given in section~\ref{sec:axion}.

\subsection{Structure}
\label{sec:structure}

\paragraph{The Bogomolny coupling.} The potential $V$ has its minimum at $\rs^2=-m^2/\lambda$. Around a homogeneous configuration $\rho=\rs$ the vector and the amplitude fluctuation have the masses
\begin{equation}
m_B^2=e^2\rs^2=-\frac{e^2m^2}{\lambda},\qquad m_\rho^2=m^2+3\lambda\rs^2=-2m^2,
\label{eq:masses}
\end{equation}
which coincide precisely when $\lambda=e^2/2$, the Bogomolny coupling discussed in the introduction. At this coupling $V=(e^2/2)(|\Psi|^2+m^2/e^2)^2$ up to a constant. This is the D-term potential obtained by eliminating the auxiliary field of a U(1) vector multiplet with the Fayet--Iliopoulos term $-m^2/e^2$ \cite{FI1974}: it is quadratic in the moment map $|\Psi|^2$ of the gauged U(1), with a coefficient set by the gauge coupling. Because the gradient of the moment map has the same norm as the Killing vector that generates the U(1), the amplitude mode and the vector then have equal masses at the minimum; in flat space, where the minimum preserves supersymmetry, they form a single massive vector multiplet. With $m^2L^2=-4$ both masses equal $8/L^2$ and $\rs=2\sqrt2/(eL)$.

\paragraph{Zero temperature: a domain wall.} At $T=0$, where $f=1$, the solution \eqref{eq:solution} interpolates between the maximum of the potential, $\rho=0$, at the boundary and its minimum, $\rho=\rs$, in the deep interior $u\to\infty$. In the probe limit both ends are AdS$_5$ with the same radius, so the solution is a domain wall of the type studied with backreaction in \cite{GubserRocha2009,GubserNellore2009a,GubserNellore2009b,HorowitzRoberts2009,GPR2009}, and the infrared is an emergent conformal fixed point at which the U(1) is broken. The equal masses are what the closed form requires. In the infrared $\Bb_t\simeq4/(\mu u^2)$, and a static potential $B_t\propto u^{-2}$ of a vector of mass $m_B$ in AdS$_5$ requires $m_B^2L^2=(-2)(-2-2)=8$. The condensate, meanwhile, approaches the minimum $\rs$ of the potential, where $m_B^2=-e^2m^2/\lambda$ by \eqref{eq:masses}; with $m^2L^2=-4$ this equals $8/L^2$ only at $\lambda=e^2/2$. At the infrared fixed point the vector and the amplitude, with $m_B^2L^2=m_\rho^2L^2=8$, are dual to the current and the order parameter with the dimensions
\begin{equation}
\Delta_J^\IR=5,\qquad \Delta_O^\IR=2+2\sqrt3,\qquad (\Delta_J^\IR-1)(\Delta_J^\IR-3)=m_B^2L^2,\quad \Delta_O^\IR(\Delta_O^\IR-4)=m_\rho^2L^2 .
\label{eq:IRdims}
\end{equation}
They set the zero-temperature response on the plane (sections~\ref{sec:conductivity} and~\ref{sec:amplitude}) and the excitations above the states of large charge on the sphere (section~\ref{sec:spectrum}). The other members share this fixed point at large chemical potential: in global AdS$_5$, as $\mu R$ grows, the region $u\gtrsim2/\mu$ approaches the minimum of the potential (figure~\ref{fig:profilesglobal}), and near the boundary the soliton approaches the planar domain wall, as found numerically for charged solitons in global AdS$_4$ \cite{GRW2012}. A probe-limit zero-temperature domain wall, with a quartic potential in AdS$_4$, was studied in \cite{Yarom2009}, using a first integral of the radial equations.

\paragraph{First-order equations.} With $\hat\rho\equiv eL\rhob/\sqrt2$, which runs from $0$ at the boundary to $2$ in the infrared, the zero-temperature solution satisfies
\begin{equation}
u\,\partial_u\hat\rho=\hat\rho\,(2-\hat\rho),\qquad u\,\partial_u\Bb_t=-\hat\rho\,\Bb_t,
\label{eq:firstorder}
\end{equation}
together with the algebraic relation
\begin{equation}
(1-\hat\rho)^2+\big(u\Bb_t\big)^2=1,
\label{eq:circle}
\end{equation}
which is preserved by the flow; here $u\Bb_t/L$ is the gauge potential in an orthonormal frame. Conversely, \eqref{eq:firstorder} implies the Maxwell equation in \eqref{eq:radial} at $T=0$, and then the scalar equation is equivalent to \eqref{eq:circle} wherever $\hat\rho\neq0$. Parametrizing the circle by an angle, $1-\hat\rho=\cos\eta$ and $u\Bb_t=\sin\eta$, the system collapses to
\begin{equation}
u\,\partial_u\eta=\sin\eta\quad\Longrightarrow\quad\tan\frac{\eta}{2}=\pi\Tc u,
\label{eq:kink}
\end{equation}
the static sine-Gordon kink in $\ln u$. Dilatations, which rescale $\mu$, translate it.

\paragraph{The whole family.} At $T>0$ and on the other slicings the solution keeps this structure. At fixed $\mu$, eq.~\eqref{eq:family} is the planar domain wall at zero temperature with $\rhob$ multiplied by the constant $\sqrt2\,p/\mu^2$, which is $\sqrt{1-T^4/\Tc^4}$ on the planar brane, and $\Bb_t$ by $f(u)$: $\hat\rho=(\sqrt2\,p/\mu^2)(1-\cos\eta)$ and $u\Bb_t=f\sin\eta$, with the same kink \eqref{eq:kink}, $\tan(\eta/2)=\mu u/2$. The constant can be eliminated in favor of local fields. With $\Phi\equiv u\Bb_t/f$ and
\begin{equation}
c\equiv\frac{f\,(2-\Phi^2)-\hat\rho^2-2\kappa u^2/R^2}{2\,(2-f+\kappa u^2/R^2)},
\label{eq:ckappa}
\end{equation}
which equals $\cos\eta$ on the family, every member obeys
\begin{equation}
u\,\partial_u\hat\rho=\hat\rho\,(1+c),\qquad u\,\partial_u\Phi=\Phi\,c,\qquad c^2+\Phi^2=1,
\label{eq:firstorderT}
\end{equation}
in which the temperature and the slicing enter only through $f(u)$ and $\kappa u^2/R^2$. The constraint is preserved by the flow, $u\,\partial_u(c^2+\Phi^2-1)=4c\,(c^2+\Phi^2-1)$, and on any solution of the two first-order equations the residuals of the field equations \eqref{eq:radial} are proportional to $c^2+\Phi^2-1$. So every solution of \eqref{eq:firstorderT} solves the field equations, on every slicing and at every $T$, and on the axion branes with $\kappa/R^2\to-\alpha^2/4$; the converse fails, for instance for the nodal solutions of section~\ref{sec:nodal}. At $f=1$, \eqref{eq:firstorderT} reduces to \eqref{eq:firstorder}--\eqref{eq:circle}.

\paragraph{Superpotential.} As written, neither \eqref{eq:firstorder} nor \eqref{eq:firstorderT} is a gradient flow, one in which the momenta conjugate to $\hat\rho$ and $\Phi$ in the radial action are the gradient of a function $\mathcal{W}$ of the fields and $u$: the integrability condition fails (appendix~\ref{app:numerics}). Off the constraint, however, the flow is not unique, since terms that vanish on it can be added. At $T=0$ the gradient flow of the cubic $\mathcal{W}=(L/e^2)\,u^{-4}(\hat\rho^3/2-3\hat\rho^2/2-\hat\rho\Phi^2/2)$ preserves the circle \eqref{eq:circle} and agrees with \eqref{eq:firstorder} on it, and the radial Lagrangian equals the derivative of $\mathcal{W}$ with respect to $\ln u$ plus squares of these flow equations and a term quadratic in the constraint. The two squares have opposite signs, as they must, since the field-space metric of the radial action, which is diagonal in $\hat\rho$ and $\Phi$, is indefinite for an electric potential; the identity therefore gives no bound on the grand potential.

\paragraph{Near the critical temperature.} Near $\Tc$ the condensate is a series in its amplitude, $eL\rho=\epsilon\rho_1+\epsilon^3\rho_3+\cdots$ with $\rho_1=\xi/(1+\xi)$ and $\xi=u^2/u_h^2$ (appendix~\ref{app:nearTc}). In the model of \cite{Herzog2010}, with the kinetic function $1+\CA|\Psi|^2$ and the potential $m^2|\Psi|^2+\CB|\Psi|^4$, which, with $\Psi$ rescaled by $e$, contains the present one at $\CA=0$ and $\CB=\lambda/e^2$, the third-order term found there is
\begin{equation}
\rho_3=-(1-2\CA+\CB)\,\frac{\xi^2}{24(1+\xi)^2}+(1-8\CA-2\CB)\,\frac{\xi\ln(1+\xi)}{48(1+\xi)},
\label{eq:rho3}
\end{equation}
which we verified by substitution for general $\CA$ and $\CB$. In the present model, $\CA=0$, the logarithm drops out only at the Bogomolny coupling $\CB=\frac12$; there every order is rational in $\xi$, and the series sums to the exact solution. In the larger family the logarithm vanishes along the whole line $8\CA+2\CB=1$, but for $\mu\neq0$ the rational ansatz closes only at $\CA=0$, $\CB=\frac12$. For $\CA\neq0$ the Maxwell equation acquires a term proportional to $\CA\rho^4B_t$ that no other term cancels on the ansatz of section~\ref{sec:closure}, since it alone has a pole of order five at $u^2=-1/K$ (for $K=0$, the highest power of $u$). At $\CA=0$ that section then fixes $\CB=\frac12$.

\section{Thermodynamics and stability}
\label{sec:every}

The results of this section hold on every member of the family: the grand potential follows from the solution alone, and the energy identity and the stability argument do not use its form.

\subsection{Grand potential}
\label{sec:grandpot}
\label{sec:thermo}

The grand potential is minus the renormalized on-shell action per unit time and unit volume of $\Sigma_\kappa$. Multiplying the scalar equation by $\rho$ and integrating by parts trades the three terms of the Lagrangian quadratic in $\rho$ for a quartic term and a surface term at the cutoff. For a source-free condensate, $\rho\simeq\rho_\vev u^2$, the surface term is finite and equal to $L^3\rho_\vev^2$, and the scalar boundary terms of \eqref{eq:Sbdry} evaluate to $-L^3\rho_\vev^2$, canceling it. There are no divergences, and, for any $\lambda$ and any source-free solution at the BF bound,
\begin{equation}
\Om=-\int_0^{u_h}du\,\Big[\frac{L}{2e^2}\frac{(\partial_uB_t)^2}{u}+\frac{\lambda L^5}{4}\frac{\rho^4}{u^5}\Big],
\label{eq:Omega}
\end{equation}
with $u_h\to\infty$ for global AdS$_5$, where the surface term at the center vanishes. The curvature of the boundary does not change this result: counterterms such as $\sqrt{-\gamma}\,\mathcal{R}[\gamma]\,|\Psi|^2$, and the curvature corrections to the induced metric, are suppressed by $u^2$ and vanish at the cutoff for a source-free scalar, and the logarithmic counterterm of the Maxwell action vanishes with the boundary field strength. The gravitational contributions, including the Casimir energy on $\mathbb{R}\times S^3$, are not part of the probe sector.

On the solution \eqref{eq:family} the integrals are elementary. In the normal phase $\Om_\N=-L\mu^2/e^2u_h^2$, and in the condensed phase
\begin{equation}
\Om_\SF=\frac{L}{e^2}\Big[-\frac18\Big(\mu^2-\frac{4\kappa}{R^2}\Big)^2+\frac{2\kappa^2}{R^4}-2\Big(\frac{\kappa}{R^2}+\frac{1}{u_h^2}\Big)^2\Big],\qquad
\Om_\SF-\Om_\N=-\frac{L}{8e^2}\big(\mu^2-\muc^2\big)^2\le0 .
\label{eq:dOmega}
\end{equation}
The free energy of condensation thus has the mean-field form, the square of $\mu^2-\muc^2$, on every slicing. The condensed phase is therefore preferred whenever it exists, $\mu^2>\muc^2$, and the transition at $\mu^2=\muc^2$ is of second order. The grand potential is a sum of a part that depends only on $\mu$ and a part that depends only on $u_h$, i.e.\ on $T$ for a given branch of black holes. Hence $n=-\partial\Om/\partial\mu$ agrees with \eqref{eq:dataK}, where $n$ is read off independently from the falloff of $A_t$; this confirms the coefficient $e^2/2L$ in \eqref{eq:expansions}. The entropy density of the probe sector,
\begin{equation}
s\equiv s_\SF=-\frac{\partial\Om_\SF}{\partial T}=\frac{16\pi L}{e^2}\,\frac{1+\kappa u_h^2/R^2}{u_h^3\,(2-\kappa u_h^2/R^2)},
\label{eq:entropy}
\end{equation}
is independent of $\mu$; on $H^3$ below $T=1/2\pi R$, and on the small global black holes, it is negative, which is allowed for an $O(L/e^2)$ correction to the entropy of the black hole.

\subsection{Global minimum}
\label{sec:globalmin}

Eq.~\eqref{eq:dOmega} compares the condensed phase only with the normal phase. An exact identity compares the energy of a nodeless solution with that of every other state. Take as reference a static solution with no nodes, no source and no superflow. Its fields are $\rhob$ and $\Bb_t$, its chemical potential is $\bar\mu$, and its energy and charge are $\bar E$ and $\bar Q$. The other state is any configuration at a given time that obeys the Gauss law and has no source and no boundary field strength; it need not be static, homogeneous or nodeless. Its energy and charge are $E$ and $Q=\int d^3x\,n$. Since $\rhob$ has no nodes, one may use the complex field $w\equiv\sqrt2\,\Psi/\rhob$, a substitution familiar from the S-deformation method for black-hole stability \cite{IshibashiKodama2003,Kimura2017}. For any potential $V(\rho)$ the difference of $E-\bar\mu Q$ between the two states is
\begin{multline}
E-\bar E-\bar\mu\big(Q-\bar Q\big)=\int d^3x\,du\,\sqrt{-g}\,\Big[\frac12\rhob^2|g^{tt}|\,\big|D_tw+i\Bb_tw\big|^2+\frac12\rhob^2g^{IJ}(D_Iw)^*D_Jw\\
+\frac{1}{2e^2}|g^{tt}|g^{IJ}\big(F-\bar F\big)_{tI}\big(F-\bar F\big)_{tJ}+\frac{1}{4e^2}F_{IJ}F^{IJ}+V(\rho)-V(\rhob)-\frac{V'(\rhob)}{2\rhob}\big(\rho^2-\rhob^2\big)\Big].
\label{eq:globalmin}
\end{multline}
Here and in the rest of this section the capital indices $I,J$ run over the bulk spatial coordinates $(u,x,y,z)$. The Gauss law for both states and an integration by parts turn the cross term of the electric energy into $\bar\mu(Q-\bar Q)$ and a bulk term that completes the square in $D_t$. The substitution and the background equation give the gradient term. The surface term at the cutoff cancels, state by state, the boundary term of $E$, which comes from the scalar terms of \eqref{eq:Sbdry}. Appendix~\ref{app:energy} gives the steps. The identity holds on every slicing, by the same steps, since $\Bb_t=0$ at a horizon and the surface terms vanish at the center, and so it applies to every member of the exact family, which is nodeless. It also holds at any scalar mass in the standard quantization, with an order parameter of dimension $\Delta_O\ge2$.

The first four terms on the right side are squares. The terms in $V$ are $\frac14\lambda(\rho^2-\rhob^2)^2$ for the quartic potential, and they are nonnegative whenever $V'(\rho)/\rho$ is nondecreasing at all $\rho$. The right side is then nonnegative, and its quadratic part is the energy of the linear perturbations of section~\ref{sec:energy}. For classical states of this model in the probe limit, on a fixed background, this has five consequences.
\begin{itemize}
\item \emph{The lowest state.} The reference minimizes $E-\bar\mu Q$ over all these states. On a static solution $E-\mu Q$ is the grand potential $\Om_{\rm tot}$ of the whole of $\Sigma_\kappa$, so on a black hole $E$ is the free energy at fixed $T$. At equal charge $E\ge\bar E$: a nodeless solution without superflow is the lowest state of charge $Q$ at its temperature, and solutions with nodes and inhomogeneous, rotating and vortex states lie above it. Equality requires $D_tw+i\Bb_tw=0$, $F_{IJ}=0$, $F_{tI}=\bar F_{tI}$ and $D_Iw=0$, so $|w|$ is constant, and the Gauss law then gives $|w|=1$ for $\bar\mu\neq0$. The minimum is thus unique up to gauge, also at $\lambda=0$; at $\bar\mu=0$ it is unique unless $V$ is affine in $\rho^2$ between $\rhob^2$ and $|w|^2\rhob^2$.
\item \emph{At most one nodeless solution at each $\mu$.} Adding eq.~\eqref{eq:globalmin} with two nodeless solutions exchanged gives $(\bar\mu_1-\bar\mu_2)(\bar Q_1-\bar Q_2)\ge0$, with equality only if they coincide, apart from the flat direction at $\bar\mu=0$ noted above. Without a gauge field this is the uniqueness argument of Brezis and Oswald \cite{BrezisOswald1986}, used at $\mu=0$ in \cite{PaperDefect}. So on a given background there is at most one nodeless solution without superflow at each $\mu$, apart from that flat direction, and its charge increases with $\mu$. The normal state and, above their onsets, the solutions with nodes also solve the equations at that $\mu$, at higher $E-\mu Q$.
\item \emph{The onset.} With the normal state, of potential $B^\N_t$, as reference, the same steps and the diamagnetic inequality $|D_I\Psi|\ge|\partial_I|\Psi||$ bound the gradient and potential terms from below by the quadratic form of the scalar operator $\mathcal{D}_\N=-\nabla_a\nabla^a+m^2-|g^{tt}|(B^\N_t)^2$ on $|\Psi|$ and the term $V(\rho)-\frac12m^2\rho^2\ge0$. The normal phase thus minimizes $E-\mu Q$ exactly when $\mathcal{D}_\N\ge0$, and $\mathcal{D}_\N$ decreases with $\mu^2$, so the linear onset is the transition. By uniqueness two states cannot both be the minimum at one $(T,\mu)$, so there is no first-order transition on a fixed background; the jump at the Hawking--Page temperature (section~\ref{sec:thermoS3}) is instead a change of background.
\item \emph{Symmetry.} An isometry of the background maps a nodeless solution without superflow to another at the same $\mu$, so by uniqueness translations on the plane and SO(4) on the sphere are unbroken at zero superflow.
\item \emph{Nonlinear stability.} Without a horizon $E$ and $Q$ are conserved, so with the soliton as reference the right side of eq.~\eqref{eq:globalmin} is conserved along any source-free evolution: a state close to the soliton in this measure stays close, for as long as the solution exists.
\end{itemize}

On the plane eq.~\eqref{eq:globalmin} compares states in a periodic box or at finite energy relative to the reference, and on $H^3$ states at finite relative energy or on a compact quotient. On $H^3$ the class decides the onset: normalizable modes give $\mu_{\rm norm}$ of section~\ref{sec:renyi}, and a compact quotient gives $\muc$.

\subsection{Perturbations}
\label{sec:stability}

Linear stability and the spectra of sections~\ref{sec:transport} and~\ref{sec:sphere} rest on the perturbations $\tau$ and $b_a$ of the exact family, defined by $\rho\to\rhob+\tau$ and $B_a\to\Bb_a+b_a$, with $\Bb_a=(\Bb_t,\vec0,0)$. They are written in the coordinates of the planar brane; on the curved slicings the derivatives along the boundary become covariant derivatives on $\Sigma_\kappa$, and the perturbations are expanded in harmonics, of degree $\ell$ on $S^3$. Writing $S_{\rm bulk}=\int d^4x\,du\sqrt{-g}\,\mathcal{L}$, the part of $\mathcal{L}$ quadratic in the perturbations is
\begin{equation}
\mathcal{L}_2=-\frac{1}{4e^2}f_{ab}f^{ab}-\frac12\partial_a\tau\,\partial^a\tau-\frac12\rhob^2b_ab^a-\frac{\tau^2}{2}\big(m^2+3\lambda\rhob^2+\Bb_a\Bb^a\big)-2\rhob\,\Bb^ab_a\,\tau,
\label{eq:L2}
\end{equation}
with $f_{ab}=\partial_ab_b-\partial_bb_a$. The equations of motion are
\begin{align}
\nabla_a\nabla^a\tau&=\big(m^2+3\lambda\rhob^2+\Bb_a\Bb^a\big)\tau+2\rhob\,\Bb^ab_a,
\label{eq:tau}\\
\frac{1}{e^2}\nabla_af^{ab}&=\rhob^2b^b+2\rhob\,\tau\,\Bb^b,
\label{eq:b}
\end{align}
and the divergence of \eqref{eq:b} gives the constraint
\begin{equation}
\nabla_a\big(\rhob^2b^a+2\rhob\,\tau\,\Bb^a\big)=0.
\label{eq:constraint}
\end{equation}
Since $B_a$ is gauge invariant, all five components of $b_a$ are physical: $b_a$ is a massive vector with the position-dependent mass $e\rhob$, whose longitudinal polarization is the Goldstone mode, and $b_u$ is fixed by the $u$-component of \eqref{eq:b}, not by a choice of gauge. The spatial components split into transverse and longitudinal parts, $b^\perp_i=b_i-\partial_i\nabla^{-2}\vec\nabla\cdot\vec b$ and $b^\parallel_i=b_i-b^\perp_i$. The two transverse components decouple. The remaining fields $(\tau,b_t,b^\parallel,b_u)$ are coupled and, with the constraint \eqref{eq:constraint}, carry three degrees of freedom: the amplitude mode, the Goldstone mode and one further polarization of the massive vector.

\paragraph{Boundary conditions.}\label{sec:bcs} Let $\theta$ and $a_a$ be the perturbations of the phase $\varphi$ and of $A_a$, so that $b_a=a_a-\partial_a\theta$. A normal mode is a solution with no sources in the boundary theory and the physical behavior in the interior. The conditions are as follows; the near-boundary expansions on which the first rests are derived in appendix~\ref{app:asymptotics}.
\begin{itemize}
\item \emph{Boundary, $u\to0$.} The source of $O_2$ and the external field strength vanish,
\begin{equation}
\tau_\src=0,\qquad\theta_\src=0,\qquad f^{(0)}_{\mu\nu}=0,
\label{eq:bc}
\end{equation}
where $\tau_\src$ and $\theta_\src$ are the coefficients of $u^2\ln u$ in $\tau$ and of $\ln u$ in $\theta$, and $f^{(0)}_{\mu\nu}$ is the field strength of the boundary value of $a_\mu$. A mode then behaves as
\begin{equation}
\tau\simeq\tau_\vev u^2,\qquad b_\mu\to-\partial_\mu\theta_0,\qquad b_u=O(u),
\label{eq:bcmode}
\end{equation}
where $\tau_\vev$ and the boundary value $\theta_0$ of the phase are free, and are fixed together with the induced current by the condition in the interior. Only the transverse components of $b_\mu$ vanish at the boundary, $b^\perp_i=O(u^2)$. In harmonics on $S^3$ these conditions are \eqref{eq:bcS3}.
\item \emph{Horizon, $T>0$.} The solution is ingoing, that is, in terms of the coordinate $v=t-\int_0^udu'/f$, it is regular on the future horizon:
\begin{equation}
(\tau,\,b_t,\,b_i,\,f b_u)=\Big(1-\frac{u}{u_h}\Big)^{-i\omega/4\pi T}\times(\text{regular series in }u_h-u),\qquad f b_u+b_t\to0 .
\label{eq:ingoing}
\end{equation}
\item \emph{Regular center, global AdS$_5$, $r\to0$.} The perturbations are regular at the center.
\item \emph{Poincar\'e horizon, $T=0$, $u\to\infty$.} There $eL\rhob/\sqrt2\to2$ and $\Bb_t\simeq4/(\mu u^2)$, and the perturbations become those of a scalar and a vector of mass $8/L^2$ in AdS$_5$. For $u\gg1/\mu$ they depend on $\omega$ and $\vk$ through $\bar k=\sqrt{\vk^{\,2}-\omega^2}$, and for $|\bar k|u\gg1$
\begin{equation}
\tau\propto u^{3/2}e^{\mp\bar ku},\qquad b^\perp_i\propto u^{1/2}e^{\mp\bar ku}.
\label{eq:Poincare}
\end{equation}
The solution is $e^{-i\omega t-\bar ku}$, outgoing with $\bar k=-i\,{\rm sgn}(\omega)\sqrt{\omega^2-\vk^{\,2}}$ when $\omega^2>\vk^{\,2}$ and decaying with ${\rm Re}\,\bar k>0$ otherwise; at $\bar k=0$ the decaying powers are selected among $\tau\propto u^{2\pm2\sqrt3}$ and, for the vector, $b^\perp_i\propto u^{-2},u^4$.
\end{itemize}
Solutions that satisfy the conditions at both ends exist only for discrete frequencies: complex quasinormal frequencies with a horizon at $T>0$, computed in section~\ref{sec:transport}, and real normal-mode frequencies in global AdS$_5$, computed in section~\ref{sec:sphere}; the hyperbolic members and the global black holes are not treated.

\subsection{Linear stability}
\label{sec:energy}

The background is static, so the perturbations have an energy, which changes only through fluxes at the ends of the radial direction. The temporal component $a_t$ has no conjugate momentum and enforces the Gauss law, the $t$-component of \eqref{eq:b}. Using it, the mixed term $b_t\tau$ cancels and the sign of the $b_t^2$ term is reversed, exactly as for the temporal component of a Proca field; the surface term this produces vanishes for a normal mode (appendix~\ref{app:energy}). The result is
\begin{gather}
\ee_2=\ee_++\int d^3x\,du\,\sqrt{-g}\,\Big[\frac12g^{IJ}\partial_I\tau\,\partial_J\tau+\frac12M^2(u)\,\tau^2\Big]+\ee_\bdry,
\label{eq:E2}\\
\ee_+=\int d^3x\,du\,\sqrt{-g}\,\Big[\frac12|g^{tt}|\dot\tau^2+\frac{1}{2e^2}|g^{tt}|g^{IJ}f_{tI}f_{tJ}+\frac12\rhob^2|g^{tt}|b_t^2+\frac{1}{4e^2}f_{IJ}f^{IJ}+\frac12\rhob^2g^{IJ}b_Ib_J\Big],\nonumber
\end{gather}
where $M^2(u)\equiv m^2+3\lambda\rhob^2-|g^{tt}|\Bb_t^2$ and $\ee_\bdry$ is the contribution of \eqref{eq:Sbdry}. The part $\ee_+$ is manifestly nonnegative, but $M^2<0$ near the boundary, because $m^2<0$ and because the chemical potential lowers the effective mass, the mechanism that drives the condensation.

The remaining terms of \eqref{eq:E2} are nevertheless nonnegative, because $\ee_2$ is the quadratic part of the left side of \eqref{eq:globalmin}. On the right side, $w=(1+\tau/\rhob)e^{i\theta}$ gives at second order $|D_tw+i\Bb_tw|^2\to(\dot\tau/\rhob)^2+b_t^2$ and $(D_Iw)^*D^Iw\to\partial_I(\tau/\rhob)\partial^I(\tau/\rhob)+b_Ib^I$, and the terms in $V$ give $\lambda\rhob^2\tau^2$. The energy is therefore
\begin{equation}
\ee_2=\ee_++\int d^3x\,du\,\sqrt{-g}\,\Big[\frac12\rhob^2g^{IJ}\partial_I\Big(\frac{\tau}{\rhob}\Big)\partial_J\Big(\frac{\tau}{\rhob}\Big)+\lambda\rhob^2\tau^2\Big]\;\ge\;0 .
\label{eq:E2positive}
\end{equation}
The energy vanishes only if $b_a=0$, $\dot\tau=0$ and $\tau/\rhob$ is constant in space, with $\tau=0$ if $\lambda>0$. With $b_a=0$ the Gauss law gives $\rhob\Bb_t\tau=0$, so for $\mu\neq0$ the energy vanishes, also at $\lambda=0$, only for the global phase rotation, which does not grow in time.

The argument does not depend on the quartic form of the potential. For any potential $V(\rho)$, the term $\lambda\rhob^2\tau^2$ in \eqref{eq:E2positive} becomes $\frac12\big[V''(\rhob)-V'(\rhob)/\rhob\big]\tau^2$, so the energy is positive whenever $V'(\rho)/\rho$ is nondecreasing on the range of $\rhob$, whereas the statement of section~\ref{sec:globalmin} needs this at all $\rho$; $\lambda\ge0$ is the quartic case. The condition matters: in AdS$_4$ attractive quartic terms give unstable branches of the condensate \cite{ZZN2023}.

\paragraph{No growing modes.} Consider a mode $\propto e^{-i\omega t}$ with ${\rm Im}\,\omega>0$ and real $\vk$ that obeys the conditions of section~\ref{sec:bcs}. Its energy $\ee_2$ is finite, and no energy flows through either end of the radial direction: the sources vanish at the boundary, and the ingoing or decaying fields vanish in the interior (appendix~\ref{app:energy}). Hence $\ee_2$ is conserved. But $\ee_2$ is quadratic in the fields and grows as $e^{2\,{\rm Im}\,\omega\,t}$, so $\ee_2=0$, and by \eqref{eq:E2positive} the mode is trivial. This establishes the stability statement:
\begin{quote}
\emph{the exact solution has no exponentially growing linear mode, for any $T<\Tc$ and any real momentum, in either the transverse or the longitudinal channel.}
\end{quote}
The argument assumes the probe limit, so that the metric is not perturbed and $\ee_2$ is conserved; $\lambda\ge0$, or more generally $V'(\rho)/\rho$ nondecreasing; a nodeless condensate, so that $w$ is defined; source-free boundary conditions and ingoing or decaying behavior in the interior; and no superflow, $\Bb_i=0$. With a superflow the Gauss law does not remove the coupling $\rhob\,\tau\,\Bb^ib_i$, and a Landau-type instability is possible at $T>0$ \cite{Amado2014,GMS2023,AGMS2024}.

\paragraph{Scope of the argument.} At $T=\Tc$ the term $\lambda\rhob^2\tau^2$ disappears with the condensate, and $\tau\propto u^2/(1+\pi^2\Tc^2u^2)$ becomes a zero-energy direction, the marginal mode of the transition \cite{Herzog2010}; in the normal phase below $\Tc$ the quadratic form of the scalar is indefinite. Like eq.~\eqref{eq:globalmin}, the argument extends to any scalar mass in the standard quantization, $\Delta_O\ge2$. We do not treat the alternative quantization here. Fields outside this model are not covered: they can mix with the condensate, as through the cubic coupling on a probe D7 brane.

\paragraph{Other slicings.}
\label{sec:linstab}
Since eq.~\eqref{eq:family} is nodeless on every slicing, the argument applies without change on every member with a horizon; on $H^3$ the perturbations are taken normalizable, as the integrations by parts require. In global AdS$_5$ the surface terms at the boundary are those of the planar brane, since the curvature enters only at subleading order, and those at the regular center, $r=R/u\to0$, vanish, since $\sqrt{-g}\propto r^3$ and the fields are regular. Without a horizon $\ee_2$ is conserved for every perturbation, not only for growing ones. A normal mode with ${\rm Im}\,\omega\neq0$ would have an energy proportional to $e^{2{\rm Im}\,\omega\,t}$, which must therefore vanish:
\begin{quote}
\emph{the global soliton has no exponentially growing linear modes, and its normal-mode frequencies are real, in every channel.}
\end{quote}
For a static perturbation the field equations reduce $\ee_2$ to $\frac12\delta\mu\,\delta Q$: since $b_i$ is static at the boundary and the boundary field strength vanishes, $b_t$ tends there to a constant $\delta\mu$ on $S^3$. So $\ee_2$ vanishes unless the charge changes, which happens only at $\ell=0$. By \eqref{eq:E2positive} the static perturbation with a given $\delta\mu$ is then unique: it is $\delta\mu\,\partial/\partial\mu$ of the background, which is time independent in the variables $(\tau,b_a)$ while the boundary value of the phase grows as $-\delta\mu\,t$. It is the zero mode of the Goldstone field on $S^3$, and it has positive energy.

\paragraph{Solutions with nodes.} Whether only the solution without nodes is stable, with a mass term alone or with scalar self-interactions, was listed as an open problem by Horowitz \cite{Horowitz2010rev}. Under the assumptions above, the nodeless solution is the global minimum of the energy at fixed charge, and linearly stable, for a mass term alone and for $V^\prime(\rho)/\rho$ nondecreasing. At a solution with nodes the second variation of the energy at fixed charge is indefinite: $\rhob$ is then a zero mode with nodes of $-\nabla_a\nabla^a+V'(\rhob)/\rhob-|g^{tt}|\Bb_t^2$, which therefore has a negative eigenvalue. At $T>0$, on slices that cross the future horizon, the energy can only decrease, by flowing into the horizon, so a perturbation of negative energy cannot decay. Without a horizon, where the energy is conserved, the indefinite second variation does not by itself imply instability. Solutions with nodes have been constructed numerically in AdS$_4$, in the probe limit \cite{WHLYZ2020}, and in the same setting their onsets and their condensates near the onset have been estimated by a variational method \cite{QWOWPJ2020}; with backreaction, a time evolution passes through them as intermediate states and ends in the nodeless solution \cite{LWWZ2020}.

\paragraph{Earlier stability arguments.} For the backreacted holographic superconductor in AdS$_4$, the stability of the vector sector was proved with a positive Hamiltonian in \cite{KannoSoda2010}. The scalar sector was treated there with the same elimination of the temporal component and the same need for a nodeless condensate, and its stability was argued near the critical point and beyond it. Using the background equation for $\rhob$, eq.~\eqref{eq:E2positive} establishes it for the scalar sector at all temperatures, in the probe limit. A complementary argument, from thermodynamics in the hydrodynamic regime, is given in \cite{GouterauxMefford2024}. In the probe limit, a generalized free energy on slices that cross the future horizon decreases monotonically, and a state at its global minimum is argued to be dynamically stable \cite{TWZ2023}. Eq.~\eqref{eq:globalmin} shows that a nodeless condensate is the global minimum of the energy at fixed charge on slices of constant $t$; we have not shown the same for that functional.

\section{Transport}
\label{sec:plane}
\label{sec:transport}

On the planar brane the family is the superfluid for all $T<\Tc$. This section derives its thermodynamics, superflow, linear response and sound, and ends with the axion branes.

\subsection{Thermodynamics and superflow}
\label{sec:thermoplanar}

On the planar brane the transition is at $T=\Tc=\mu/2\pi$, and eq.~\eqref{eq:dOmega} with $\kappa=0$ and $u_h=1/\pi T$ gives the grand potential of the condensed phase,
\begin{equation}
\Om_\SF=-\frac{2\pi^4L}{e^2}\big(\Tc^4+T^4\big)=-\frac{L}{8e^2}\big(\mu^4+16\pi^4T^4\big).
\label{eq:OmegaSF}
\end{equation}
The grand potential \eqref{eq:OmegaSF}, which extends the near-critical free energy of \cite{Herzog2010} to all temperatures (appendix~\ref{app:nearTc}), is the sum of the zero-temperature superfluid, $-L\mu^4/(8e^2)$, and a thermal part, which gives the entropy density of the probe sector, $s=8\pi^4LT^3/e^2$; in the normal phase it is $s_\N=2\pi^2LT\mu^2/e^2$. The specific heat of the probe sector jumps at $\Tc$ by $16\pi^4L\Tc^3/e^2$.

\paragraph{Superflow.}\label{sec:superflow} At $T=0$ the background metric is Poincar\'e invariant, so a Lorentz boost of the zero-temperature solution is again a solution, now with a superflow. Written covariantly,
\begin{equation}
\Bb_\mu=\frac{B^{(b)}_\mu}{1+Xu^2/4},\qquad \rhob=\rs\,\frac{Xu^2/4}{1+Xu^2/4},\qquad X\equiv-\eta^{\mu\nu}B^{(b)}_\mu B^{(b)}_\nu>0,
\label{eq:boosted}
\end{equation}
where $B^{(b)}_t=\mu$ and $-B^{(b)}_i$, the boundary value of $\partial_i\varphi-A_i$, is the superflow. The boundary data must be timelike, and the boost velocity is $\beta=|\vec B^{(b)}|/\mu<1$, with $X=\mu^2(1-\beta^2)$. The condensate is $\langle O_2\rangle=L^{1/2}X/(2e)$. With the pressure $P=-\Om_\SF=L\mu^4/(8e^2)$ of the state at rest, this identifies the zero-temperature equation of state of the conformal superfluid,
\begin{equation}
P(X)=\frac{L}{8e^2}X^2,\qquad J^\mu=\frac{\partial P}{\partial B^{(b)}_\mu},
\label{eq:PX}
\end{equation}
which reproduces the charge density and the current of \eqref{eq:boosted}. It has the form $P\propto X^2$ dictated by conformal invariance for a superfluid in four dimensions \cite{Son2002,Hellerman2015,Monin2017}, with a coefficient fixed by the bulk couplings. Its sound speed is the conformal value $c_s^2=1/3$.

At fixed $\mu$ the current $\langle J^x\rangle=(L\mu^3/2e^2)\,\beta(1-\beta^2)$ is maximal at $\beta=1/\sqrt3=c_s$. The fluctuations of a boosted state are boosts of those at rest, so sound moves at ${c_\pm=(c_s\pm\beta)/(1\pm c_s\beta)}$ along and against the flow, undamped at order $k^2$ since the attenuation vanishes as $T^5$ (section~\ref{sec:sound}). The upstream speed $c_-$ vanishes at the maximum of the current. At $T>0$ the same condition marks the onset of the Landau instability \cite{GMS2023,AGMS2024}. At $T=0$ this signals no instability, since there is no normal component to take up the energy, and a Landau critical velocity arises only at $T>0$, where the horizon defines a rest frame \cite{HKS2009,BMS2009,ABK2010,Amado2014}.

The boosted states are linearly stable. A mode of a boosted state with real momentum and ${\rm Im}\,\omega>0$ corresponds, in the rest frame, to a mode with complex momentum, which the energy argument of section~\ref{sec:energy} does not cover directly. Stability in the rest frame, however, together with causality of the bulk equations, implies stability in every frame \cite{Gavassino2022,HSSW2023}.

At small $T>0$ the corrections of order $T^4$ to the boosted solutions are rational functions of $u$. With $\tilde u=\sqrt X\,u/2$, $\tilde\mu=2\mu/\sqrt X$ and $h=16\pi^4T^4/X^2$, so that $f=1-h\tilde u^4$, they are, for a flow along $x$ and with $\hat\rho=eL\rhob/\sqrt2$ as in section~\ref{sec:structure},
\begin{align}
\hat\rho&=\frac{2\tilde u^2}{1+\tilde u^2}-\frac{h\,\tilde u^2\big[\tilde\mu^2(\tilde u^4+2\tilde u^2+2)-4\tilde u^4-2\tilde u^2-2\big]}{6(1+\tilde u^2)^2},\nonumber\\
\frac{\Bb_t}{\mu}&=\frac{1}{1+\tilde u^2}-\frac{h\,\tilde u^2\big(12\tilde u^4+20\tilde u^2+12-2\tilde\mu^2\tilde u^2-3\tilde\mu^2\big)}{12(1+\tilde u^2)^2},\nonumber\\
\frac{\Bb_x}{B^{(b)}_x}&=\frac{1}{1+\tilde u^2}+\frac{h\,\tilde u^2\big(2\tilde\mu^2\tilde u^2+3\tilde\mu^2+4\tilde u^2\big)}{12(1+\tilde u^2)^2}.
\label{eq:lowTfields}
\end{align}
They are fixed by the infrared, where the regular static solutions on the black brane with $\rho=\rs$ are $B_t\propto f/u^2$ and $B_x\propto1/u^2$: the order-$h$ part of $\Bb_x$ stays bounded, and that of $\Bb_t$ grows at most as $u^2$. Their charge density and current are the derivatives of the pressure
\begin{equation}
P=\frac{L}{8e^2}\,X^2+\frac{2\pi^4L}{e^2}\,T^4\,\frac{1+\beta^2}{1-\beta^2}+O(T^8),
\label{eq:PXT}
\end{equation}
which at $\beta=0$ is $-\Om_\SF$ of \eqref{eq:OmegaSF}. Maximizing the current at fixed $\mu$ gives $\beta_c^2=\frac13-\frac94(T/\Tc)^4+\cdots$, the Landau critical velocity at low temperature.

\subsection{Superfluid density and conductivity}
\label{sec:transverse}

For perturbations $\propto e^{-i\omega t+i\vk\cdot\vec x}$ the transverse components obey, at any temperature,
\begin{equation}
\Big[u\,\partial_u\Big(\frac{f}{u}\,\partial_u\Big)+\frac{\omega^2}{f}-\vk^{\,2}\Big]b^\perp_i=\frac{e^2L^2}{u^2}\rhob^2\,b^\perp_i=\frac{8\pi^4(\Tc^4-T^4)\,u^2}{(1+\pi^2\Tc^2u^2)^2}\,b^\perp_i .
\label{eq:transverse}
\end{equation}

\paragraph{Static response and the superfluid density.} At $\omega=\vk=0$ the solution of \eqref{eq:transverse} that is regular at the horizon is
\begin{equation}
b^\perp_i=A^{(b)}_i\,Y(u),\qquad Y(u)=\frac{1+\pi^2(T^4/\Tc^2)\,u^2}{1+\pi^2\Tc^2u^2}=1-\pi^2\Tc^2\Big(1-\frac{T^4}{\Tc^4}\Big)u^2+O(u^4).
\label{eq:London}
\end{equation}
This is the linearized superflow on the background \eqref{eq:solution}, at any $T<\Tc$. Comparing with \eqref{eq:expansions} gives the London equation $\langle J_i\rangle=-(n_s/\mu)A^{(b)}_i$ with
\begin{equation}
\frac{n_s}{\mu}=\frac{2L\pi^2\Tc^2}{e^2}\Big[1-\Big(\frac{T}{\Tc}\Big)^4\Big],\qquad
\boxed{\;\frac{n_s}{n}=1-\Big(\frac{T}{\Tc}\Big)^4\;}
\label{eq:ns}
\end{equation}
where $n$ is the total charge density \eqref{eq:data}. This is the temperature dependence of the phenomenological two-fluid model of Gorter and Casimir \cite{GorterCasimir1934,Tinkham1996}, which here follows exactly from the bulk. At $T=0$ all the charge is superfluid, as it must be for a Lorentz-invariant ground state, and $n_s\propto\langle O_2\rangle^2$ holds exactly at all temperatures, not only near $\Tc$ (figure~\ref{fig:profiles}(c)). Near $\Tc$, eq.~\eqref{eq:ns} agrees with the London equation obtained perturbatively in \cite{Herzog2010}; for the holographic London equation see also \cite{HHH2008b}.

\paragraph{The normal charge.} Equivalently to \eqref{eq:ns}, the charge of the normal component satisfies $\mu\,(n-n_s)=Ts$, with $s$ the entropy density of section~\ref{sec:thermoplanar}, the relation that the whole charge obeys identically in the normal phase, where $\Om_\N\propto\mu^2T^2$. This relation holds near $\Tc$ for any $\lambda$, but at all temperatures only at $\lambda=e^2/2$ among the couplings checked, $\lambda/e^2=0$, $\frac14$, $\frac12$, $1$ and $2$ (appendix~\ref{app:numerics}): at $\lambda=0$, $\mu(n-n_s)$ exceeds $Ts$ by about $23\%$ at $0.9\,\Tc$. The relation has no simple quasiparticle reading. In a relativistic superfluid the inertia of the normal component is $Ts+\mu(n-n_s)$, and a gas of neutral excitations of speed $c$, with Landau's normal mass density $Ts/c^2$ \cite{Landau1941}, gives $\mu(n-n_s)=(1/c^2-1)\,Ts$, the form found at low temperature in holographic superfluids with a conformal infrared \cite{GouterauxMefford2020}. This vanishes for $c=1$, equals $2Ts$ for phonons and equals $Ts$ only for $c^2=\frac12$.

\paragraph{Spectrum at zero temperature.} At $T=0$, eq.~\eqref{eq:transverse} depends on $\omega$ and $\vk$ only through $\omega^2-\vk^{\,2}$, because $\rhob$ is a Lorentz scalar. Writing $b^\perp_i=\sqrt u\,Z_i$ turns it into a Schr\"odinger problem,
\begin{equation}
-Z_i''+\Big[\frac{3}{4u^2}+\frac{8\pi^4\Tc^4u^2}{(1+\pi^2\Tc^2u^2)^2}\Big]Z_i=\big(\omega^2-\vk^{\,2}\big)Z_i .
\label{eq:Schr}
\end{equation}
The potential is positive and tends to zero as $u\to\infty$. The spectrum is therefore a gapless continuum, $\omega^2-\vk^{\,2}>0$, bounded by the light cone, with no bound state: the transverse channel has no isolated mode, consistent with section~\ref{sec:energy}. Near the boundary the solutions are $uJ_1$ and $uY_1$ of argument $u\sqrt{\omega^2-\vk^{\,2}}$. In the interior the potential tends to $35/(4u^2)$, and the solutions become $uJ_3$ and $uY_3$. The change of index reflects the dimension $\Delta_J^\IR=5$ of the current at the infrared fixed point, eq.~\eqref{eq:IRdims}.

\begin{figure}[t]
\centering
\includegraphics[width=0.78\textwidth]{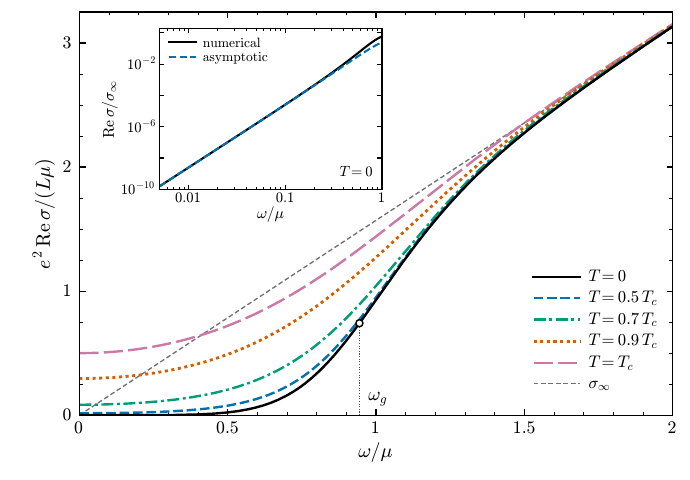}
\caption{Regular part of the optical conductivity on the exact background at fixed chemical potential $\mu=2\pi\Tc$. The $\delta(\omega)$ contribution of the superfluid, of weight $\pi n_s/\mu$, is not shown. The thin gray dashed line is the conformal asymptote $\sigma_\infty$, and $\omega_g$ marks the gap scale at $T=0$. Inset: ${\rm Re}\,\sigma/\sigma_\infty$ at $T=0$ on logarithmic axes, with the asymptotic form \eqref{eq:T0cond} (dashed).}
\label{fig:conductivity}
\end{figure}

\paragraph{Optical conductivity.}\label{sec:conductivity} The conductivity follows from \eqref{eq:transverse} at $\vk=0$. Near the boundary the solution that is ingoing at the horizon behaves as $b^\perp_x=A^{(b)}_x(1-\frac12\omega^2u^2\ln u)+b^{(2)}_xu^2+\cdots$, where $A^{(b)}_x$ is the boundary value $a^{(0)}_x$ of the perturbation. With \eqref{eq:expansions} and $E_x=i\omega A^{(b)}_x$,
\begin{equation}
\sigma(\omega)=\frac{2L}{e^2}\,\frac{b^{(2)}_x}{i\omega A^{(b)}_x},
\label{eq:sigma}
\end{equation}
up to a contact term proportional to $i\omega$ (appendix~\ref{app:asymptotics}). The real part can be written in terms of the conserved flux $\mathcal{F}={\rm Im}\big(b^{\perp*}_xf\,\partial_ub^\perp_x\big)/u$ as
\begin{equation}
{\rm Re}\,\sigma(\omega)=\frac{L}{e^2}\,\frac{\mathcal{F}}{\omega\,|A^{(b)}_x|^2}.
\label{eq:flux}
\end{equation}
In the normal phase the right-hand side of \eqref{eq:transverse} vanishes, giving the known result \cite{MST2007,HorowitzRoberts2008}
\begin{equation}
{\rm Re}\,\sigma_\N(\omega)=\frac{L}{e^2}\,\frac{\pi\omega}{2}\,\frac{\sinh(\omega/2T)}{\cosh(\omega/2T)-\cos(\omega/2T)},
\label{eq:sigmaN}
\end{equation}
which interpolates between $\pi LT/e^2$ at $\omega=0$ and the conformal value $\sigma_\infty=\pi L\omega/(2e^2)$ at $\omega\gg T$. In the condensed phase eq.~\eqref{eq:transverse} is solved numerically on the exact background; the result is shown in figure~\ref{fig:conductivity}. At $T=0$ the conductivity approaches $\sigma_\infty$ monotonically from below, with no coherence peak. Near $\Tc$ the regular part overshoots $\sigma_\infty$ slightly, as the normal-phase conductivity \eqref{eq:sigmaN} does: by $0.2\%$ at $0.9\,\Tc$ and $0.6\%$ at $\Tc$, near $\omega\approx1.8\,\mu$. At small frequency ${\rm Im}\,\sigma\to n_s/(\mu\omega)$, so that ${\rm Re}\,\sigma\supset\pi(n_s/\mu)\,\delta(\omega)$. Two limits follow in closed form.

\paragraph{DC conductivity of the normal component.} At the horizon $b^\perp_x\simeq b_h\,f^{-i\omega/4\pi T}$, which gives $\mathcal{F}=\pi T\omega|b_h|^2$. As $\omega\to0$ the solution tends to the static one \eqref{eq:London}, for which $b_h/A^{(b)}_x=Y(u_h)=(T/\Tc)^2$. Eq.~\eqref{eq:flux} then gives the exact regular part of the DC conductivity,
\begin{equation}
\sigma_\reg\equiv\sigma_\reg(\omega\to0)=\frac{\pi LT}{e^2}\,Y(u_h)^2=\frac{\pi LT}{e^2}\Big(\frac{T}{\Tc}\Big)^4 .
\label{eq:sigmaDC}
\end{equation}
It vanishes as $T^5$ at low temperature, not exponentially. On the exact solution it equals the normal-phase value $\pi LT/e^2$ times the normal fraction $(T/\Tc)^4$ of \eqref{eq:ns}. This relation is special to the Bogomolny coupling: for the other couplings checked, the ratio $Y(u_h)^2/(1-n_s/n)$ equals one only at leading order near $\Tc$, and at $\lambda=0$ it falls to about $0.6$ at $T\approx0.8\,\Tc$.

\paragraph{Low frequency at zero temperature.} At $T=0$ and $\omega\ll\mu$ the conductivity follows from matching a static solution near the boundary to the outgoing infrared wave $b^\perp_x\propto uH^{(1)}_3(\omega u)$ (appendix~\ref{app:T0cond}), which gives
\begin{equation}
{\rm Re}\,\sigma(\omega)=\frac{\pi L}{8e^2}\,\frac{\omega^5}{\mu^4}\Big[1+\frac43\,\frac{\omega^2}{\mu^2}\ln\frac{\mu}{\omega}+O\Big(\frac{\omega^2}{\mu^2}\Big)\Big]\qquad(T=0,\ \omega\ll\mu),
\label{eq:T0cond}
\end{equation}
or ${\rm Re}\,\sigma/\sigma_\infty\simeq\frac14(\omega/\mu)^4$. The exponent is $2\Delta_J^\IR-5$ with $\Delta_J^\IR=5$, as the infrared scaling predicts \cite{GubserRocha2009,HorowitzRoberts2009}; the same $\omega^5$ law holds for the zero-temperature domain wall of a type IIB truncation \cite{GPR2009}, whose infrared fixed point has the same spectrum (section~\ref{sec:topdown}). The logarithm comes from the leading correction to the infrared potential, of relative order $1/(\mu u)^2$: together with the frequency, it turns the behavior $u^{-2}$ of the infrared solution at $1/\mu\ll u\ll1/\omega$ into $u^{-2+2\omega^2/3\mu^2+\cdots}$, and the leading logarithms exponentiate (appendix~\ref{app:T0cond}). There is no hard gap, because the condensate does not end the geometry: the zero-temperature solution flows to the infrared fixed point, where the current is irrelevant. Defining a gap scale by ${\rm Re}\,\sigma(\omega_g)=\frac12\sigma_\infty(\omega_g)$ gives $\omega_g=0.94404\,\mu=5.9316\,\Tc$; it should not be compared with $\omega_g\approx8\Tc$ of the standard holographic superconductors \cite{HHH2008a,HorowitzRoberts2008}, defined in the latter by the minimum of ${\rm Im}\,\sigma$, a scheme-dependent quantity in AdS$_5$. At finite momentum the transverse spectral function follows from $\omega\,{\rm Re}\,\sigma$ at $\vk=0$ with $\omega\to\sqrt{\omega^2-\vk^{\,2}}$; it vanishes outside the light cone and, by \eqref{eq:T0cond}, as $(\omega^2-\vk^{\,2})^{\Delta_J^\IR-2}$ at its edge.

\subsection{Sound and the amplitude mode}
\label{sec:sound}
\label{sec:longitudinal}

\paragraph{Hydrodynamic prediction.} The gapless mode of the longitudinal channel is the Goldstone boson. In the probe limit the metric, and with it the temperature and the normal-fluid velocity, does not fluctuate, so the hydrodynamic mode is the one in which the superfluid component oscillates against a normal component at rest. This is fourth sound \cite{Yarom2009,HerzogYarom2009}, the mode called second sound in the probe-limit literature \cite{HKS2009,AKL2009,Herzog2010}. Its dispersion relation at leading order, $\omega=\pm c_sk$, follows from charge conservation, $\partial_t\delta n=-(n_s/\mu)\nabla^2\theta$, and from the Josephson relation, $\partial_t\theta=-\delta\mu$, which give
\begin{equation}
c_s^2=\frac{n_s/\mu}{\chi},\qquad\chi\equiv\Big(\frac{\partial n}{\partial\mu}\Big)_T .
\label{eq:cshydro}
\end{equation}
For the exact solution $n=L\mu^3/(2e^2)$ is independent of $T$, so $\chi=3n/\mu$, and with \eqref{eq:ns}
\begin{equation}
c_s^2=\frac13\Big[1-\Big(\frac{T}{\Tc}\Big)^4\Big].
\label{eq:cs}
\end{equation}
At $T=0$ this is the conformal value $1/3$, as required by \eqref{eq:PX} and shown for holographic superfluids in any dimension in \cite{EGP2016} (in AdS$_4$ the corresponding value $\frac12$ was found at low temperature with a quartic potential in \cite{Yarom2009}); it decreases monotonically and vanishes at $\Tc$. The next term at $T=0$, $\omega=(k/\sqrt3)(1+k^2/9\mu^2+\cdots)$, determines the phonon gaps on the sphere at order $(\mu R)^{-2}$ (section~\ref{sec:phononNLO}). Near the transition $c_s^2\simeq\frac43(1-T/\Tc)$, which coincides with the near-critical result of \cite{Herzog2010} at $\CA=0$, $\CB=1/2$. At the next order, probe-limit superfluid hydrodynamics adds to the conservation law a normal current $-\sigma_\reg\vec\nabla\delta\mu$, where $\sigma_\reg$ is the DC conductivity of the normal component, and to the Josephson relation a third bulk viscosity $\zeta_3$ \cite{HLSY2011}. With $k=|\vk|$, the sound mode then has the dispersion relation
\begin{equation}
\omega=\pm c_sk-i\Gamma k^2+O(k^3),\qquad\Gamma=\frac12\Big(\frac{\sigma_\reg}{\chi}+\zeta_3\,\frac{n_s}{\mu}\Big).
\label{eq:Gammahydro}
\end{equation}

\paragraph{A horizon formula for the attenuation.} The attenuation can be computed from horizon data, by energy balance (appendix~\ref{app:attenuation}). At leading order in $\omega\sim k$ the mode is a superposition of two static solutions, the response of the background to a change of the chemical potential at fixed temperature, with $\delta\mu=i\omega\theta_0$ by the Josephson relation, and the London profile \eqref{eq:London} for $b_x$. Its energy \eqref{eq:E2positive} reduces to boundary terms, $\langle\ee_2\rangle=\frac12(n_s/\mu)k^2|\theta_0|^2$ averaged over a period, and is lost only through the horizon, where the ingoing flux is a sum of squares of the horizon values of $\partial_t\tau$, $f_{tx}$ and $\rhob\,b_t$. The horizon value of $b_t$, needed at second order, follows by integrating the constraint \eqref{eq:constraint} from the boundary. Energy balance, $2\Gamma k^2\langle\ee_2\rangle=\langle\text{flux}\rangle_{\rm horizon}$, gives
\begin{equation}
\Gamma=\frac{\sigma_\reg}{2\chi}+\frac{n_s/\mu}{2\chi^2}\Big[\frac{L^3}{u_h^3}\,(\partial\rhob_h/\partial\mu)^2+\frac{u_h}{e^4L}\,\frac{(\partial E_h/\partial\mu)^2}{\rhob_h^{\,2}}\Big],
\label{eq:horizonGamma}
\end{equation}
where $\rhob_h$ and $E_h=\partial_u\Bb_t|_{u_h}$ are the condensate and the radial electric field at the horizon, and the derivatives are taken at fixed $T$. Eq.~\eqref{eq:horizonGamma} does not use the exact solution and holds for any $\lambda$. It is the probe-limit, AdS$_5$ form of the horizon formula derived by Donos and Kailidis for backreacted superfluids at finite density in AdS$_4$ \cite{DonosKailidis2022a}, which generalizes the zero-density result of \cite{DKP2021}. The derivation here, from energy balance, is independent of theirs; the relation is spelled out in appendix~\ref{app:attenuation}.

On the exact solution all the ingredients of \eqref{eq:horizonGamma} are elementary (appendix~\ref{app:attenuation}). The normal current gives $3/5$ of the attenuation and the horizon term $2/5$, at every temperature, and
\begin{equation}
\Gamma=\frac{5}{36\pi T}\Big(\frac{T}{\Tc}\Big)^6.
\label{eq:Gamma}
\end{equation}
Backreaction limits the validity of \eqref{eq:cs} at low $T$, and of \eqref{eq:Gamma} more strongly (section~\ref{sec:discussion}).

\paragraph{Third bulk viscosity.} Comparison of \eqref{eq:horizonGamma} with \eqref{eq:Gammahydro} gives
\begin{equation}
\zeta_3=\frac{1}{\chi^2}\Big[\frac{L^3}{u_h^3}\,(\partial\rhob_h/\partial\mu)^2+\frac{u_h}{e^4L}\,\frac{(\partial E_h/\partial\mu)^2}{\rhob_h^{\,2}}\Big]\ \ge\ 0,
\label{eq:zeta3}
\end{equation}
the sum of the dissipation of the amplitude of the condensate and of the charge carried into the horizon by the Goldstone mode. On the exact solution
\begin{equation}
\zeta_3\,\frac{n_s}{\mu}=\frac23\,\frac{\sigma_\reg}{\chi},\qquad\zeta_3=\frac{e^2}{18\pi^3L}\,\frac{T^5}{\Tc^4(\Tc^4-T^4)},
\label{eq:zeta3exact}
\end{equation}
which diverges as $(\Tc-T)^{-1}$ at the transition, while $\zeta_3n_s/\mu$ stays finite. The divergence reflects the slow relaxation of the amplitude of the order parameter, whose rate $\gamma$ (below) vanishes at $\Tc$: eq.~\eqref{eq:Gammahydro} holds for $c_sk\ll\gamma$, and closer to $\Tc$ the amplitude must be kept as an additional slow variable of the effective theory \cite{DonosKailidis2022b}.

\paragraph{Comparison with the near-critical regime.} Near $\Tc$, eq.~\eqref{eq:horizonGamma} can be evaluated for any $\lambda$ on the perturbative background of appendix~\ref{app:nearTc}, which gives
\begin{equation}
\lim_{T\to\Tc}\pi\Tc\,\Gamma=\frac{33+48\lambda^2/e^4}{4\,(7+4\lambda/e^2)^2}.
\label{eq:GammaTc}
\end{equation}
At $\lambda=0$ this is $33/196$, the result obtained in \cite{Herzog2010} from the poles of the retarded correlators, and at $\lambda=e^2/2$ it is the limit $5/36$ of \eqref{eq:Gamma}. Quasinormal modes on numerically constructed backgrounds, extrapolated to $T\to\Tc$, agree with \eqref{eq:GammaTc} for $\lambda/e^2=0$, $\frac14$, $\frac12$, $1$ and $2$ (appendix~\ref{app:numerics}). At $\lambda=0$ the horizon formula \eqref{eq:zeta3} also reproduces the near-critical third bulk viscosity of \cite{HLSY2011} (appendix~\ref{app:attenuation}), an independent check. The near-critical effective theory of this AdS$_5$ model at zero quartic coupling has also been studied in \cite{BGGL2024,BuYang2026,Natsuume2026}. The model F equations of \cite{Natsuume2026} give $33/196$. Their quartic coefficient, $7/96$ in the normalization of appendix~\ref{app:nearTc}, is the one at fixed charge density. By Landau theory it is the one at fixed $\mu$, $(1+4\lambda/e^2)/96$, times the ratio $\chi/\chi_\N=(7+4\lambda/e^2)/(1+4\lambda/e^2)$ of the susceptibilities just below $\Tc$ and in the normal phase (appendix~\ref{app:attenuation}). With $(7+4\lambda/e^2)/96$ in place of $7/96$, they reproduce \eqref{eq:GammaTc}, and the relaxation rate of the amplitude mode below, for any value of $\lambda$. Both also follow from the poles of the near-critical correlators of \cite{Natsuume2026b}, which gives this coefficient at nonzero quartic coupling.

\paragraph{Quasinormal modes.} The predictions for sound are tested against the longitudinal quasinormal modes of \eqref{eq:tau}--\eqref{eq:constraint} on the exact background, computed by Chebyshev collocation (appendix~\ref{app:numerics}). At small momentum there is a pair of modes with the dispersion relation \eqref{eq:Gammahydro}. The sound speed and the attenuation agree with \eqref{eq:cs} and \eqref{eq:Gamma} at all temperatures computed (appendix~\ref{app:numerics}). This confirms the hydrodynamic argument, the exact superfluid density \eqref{eq:ns} and the horizon formula \eqref{eq:horizonGamma}.

\begin{figure}[t]
\centering
\includegraphics[width=\textwidth]{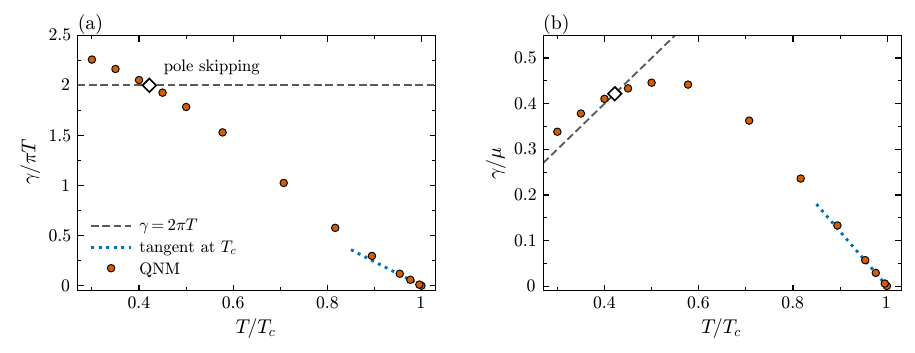}
\caption{Relaxation rate of the purely damped amplitude mode at $\vk=0$, from the quasinormal modes (QNM) on the exact background: (a) in units of $\pi T$; (b) in units of $\mu$, at fixed $\mu=2\pi\Tc$. The dashed line is $\gamma=2\pi T$, and the diamond marks the exact pole-skipping point $(\Tc/T)^2=3+4\sqrt{21}/7$, where the rate crosses it. The dotted lines are the near-critical tangents $\gamma/\pi T=\frac{12}{5}(1-T/\Tc)$ and $\gamma/\mu=\frac65(1-T/\Tc)$.}
\label{fig:qnm}
\end{figure}

\paragraph{The amplitude mode.}\label{sec:amplitude} The closest mode to the origin after the sound pair is purely damped, $\omega=-i\gamma(T)+O(k^2)$, and describes the relaxation of the amplitude of the order parameter (figure~\ref{fig:qnm}). The relaxation rate $\gamma$ vanishes at $\Tc$, and $\gamma/\pi T$ grows as the temperature is lowered: it is $0.2979$, $0.5781$, $1.0264$, $1.5307$ and $1.7842$ at $(\Tc/T)^2=\frac54$, $\frac32$, $2$, $3$ and $4$, respectively. In units of $\mu$ the rate is largest near $\Tc/2$, about $0.45\,\mu$, and falls to $0.34\,\mu$ at $0.3\,\Tc$ (figure~\ref{fig:qnm}(b)). Near $\Tc$ the amplitude relaxes at fixed charge density, and the expansion of appendix~\ref{app:nearTc} gives $\gamma\simeq\frac{4\pi}{5}(\chi/\chi_\N)(\Tc-T)$ for any $\lambda$. At $\lambda=0$ this is the result of \cite{Herzog2010}, and at $\lambda=e^2/2$ it is the tangent $\gamma/\pi T\simeq\frac{12}{5}(1-T/\Tc)$ in figure~\ref{fig:qnm}. Away from $\Tc$ we have not found a closed form. One value is exact: the near-horizon analysis of pole skipping \cite{NatsuumeOkamura2026}, applied to this background, gives $\gamma=2\pi T$ at $(\Tc/T)^2=3+4\sqrt{21}/7$, that is, at $T\approx0.4219\,\Tc$ (appendix~\ref{app:attenuation}). At $T=0$ the amplitude fluctuation is dual to the operator of dimension $\Delta_O^\IR=2+2\sqrt3$ of the infrared fixed point, eq.~\eqref{eq:IRdims}. Like the transverse current it therefore has a gapless continuous spectrum; numerically, its spectral function has no sharp Higgs pole, only a broad maximum, of $0.85$ times the conformal value at $\omega\approx0.96\,\mu$ (appendix~\ref{app:numerics}). Each infrared operator that the amplitude overlaps contributes $\omega^{2\Delta-4}$, with a coefficient that follows in closed form from the static solutions (appendix~\ref{app:T0cond}). At $\vk=0$,
\begin{equation}
\frac{{\rm Im}\,G^R_{O_2O_2}(\omega)}{{\rm Im}\,G^R_\infty}=\frac{1}{36}\Big(\frac{\omega}{\mu}\Big)^6+\frac14\Big[\frac{(2+\sqrt3)\,\Gamma(\sqrt3)}{\Gamma(2\sqrt3)}\Big]^4\Big(\frac{\omega}{\mu}\Big)^{4\sqrt3}+O\Big(\Big(\frac{\omega}{\mu}\Big)^{4+2\sqrt3}\Big),
\label{eq:T0amp}
\end{equation}
where ${\rm Im}\,G^R_\infty$ is the conformal value, reached for $\omega\gg\mu$. The first term comes from mixing with the longitudinal vector through $\Bb_t$, and the second, whose coefficient is $0.326$, from the infrared scalar; the two terms are equal at $\omega\approx0.070\,\mu$. The amplitude mode in holographic superfluids has been studied in \cite{AKL2009,Bhaseen2013,ABGL2021,DonosPantelidou2022,FGMT2024}.

\subsection{Axion branes}
\label{sec:axion}
\label{sec:transverseaxion}
\label{sec:soundaxion}

In the probe limit the axions do not couple to the charged sector, whose current already relaxes into the neutral plasma; they change only the blackening function of the neutral background (section~\ref{sec:familysol}). On the axion brane the temperature is $T=(1-\alpha^2u_h^2/8)/\pi u_h$, and the condensed phase \eqref{eq:family}, with $\kappa/R^2$ replaced by $-\alpha^2/4$, exists for $\mu^2>\muc^2=4/u_h^2-\alpha^2$, that is below
\begin{equation}
T_c(\mu,\alpha)=\frac{2\mu^2+\alpha^2}{4\pi\sqrt{\mu^2+\alpha^2}} .
\label{eq:Tcaxion}
\end{equation}
This grows monotonically with $\alpha$, from $\mu/2\pi$ at $\alpha=0$, and approaches $\alpha/4\pi$ at large $\alpha$; below $T=\alpha/4\pi$ the condensate exists even at $\mu=0$. At $T=0$ the horizon is extremal, AdS$_2\times\mathbb{R}^3$ with AdS$_2$ radius $L_2=L/2$, and the scalar has $m^2L_2^2=-1$ at $\mu=0$, below the AdS$_2$ bound $-1/4$: the infrared instability of \cite{DenefHartnoll2009}. With backreaction in AdS$_4$, a phase induced by the axions at $\mu=0$ was found in \cite{KKP2015}; here the corresponding phases are given by \eqref{eq:family}. The transport of sections~\ref{sec:transverse}--\ref{sec:sound} carries over in closed form (appendix~\ref{app:attenuation}). At $T=0$, where the extremal horizon has a finite area, a normal fraction $1-n_s/n=\alpha^2/2(\mu^2+\alpha^2)$ survives; $c_s^2=(2\mu^2+\alpha^2)/2(3\mu^2+\alpha^2)$ lies between the conformal value $\frac13$ and $\frac12$; and $\sigma_\reg/\sigma_\N(0)$ and $\Gamma$ vanish as $T^2$, not $T^4$ and $T^5$.

\section{Large charge}
\label{sec:sphere}
\label{sec:largecharge}
\label{sec:conformal}

In global AdS$_5$ the zero-temperature member of the family describes a conformal superfluid on $\mathbb{R}\times S^3$, in the lowest classical state of charge $Q$ of this model. After the phase diagram, this section derives from this solution the dimension of the lowest operator of charge $Q$, matches it to the effective theory of the conformal superfluid and computes the excitations above it.

\subsection{Phase diagram}
\label{sec:thermoS3}

On $S^3$ at $T=0$, in global AdS$_5$, the condensate forms at $\mu R=2$, where the chemical potential reaches the dimension of the lightest charged operator, $O_2$, and the corresponding normal mode $u^2/(1+u^2/R^2)$ of the scalar becomes static in the rotating frame \cite{BMP2010}. The pressure is then
\begin{equation}
P=-\Om_\SF=\frac{L}{8e^2}\Big(\mu^2-\frac{4}{R^2}\Big)^2 .
\label{eq:PS3}
\end{equation}
At $T>0$ on $S^3$ the relevant background is set by the gravitational sector, through the Hawking--Page transition at $T_{\rm HP}=3/2\pi R$ \cite{HawkingPage1983,Witten1998thermal}. Below it the background is thermal AdS$_5$, on which the probe sector is classically independent of $T$ and condenses for $\mu R>2$; above it the background is the large black hole, $u_h\le R$, on which $\muc^2R^2=4(1+R^2/u_h^2)$. With \eqref{eq:T} the phase boundary on $S^3$ is
\begin{equation}
\muc R=2\quad(T<T_{\rm HP}),\qquad \muc^2R^2=4+\Big(\pi TR+\sqrt{\pi^2T^2R^2-2}\Big)^2\quad(T>T_{\rm HP}).
\label{eq:mucS3}
\end{equation}
It jumps from $2$ to $2\sqrt2$ at $T_{\rm HP}$ and approaches the planar value $2\pi TR$ at high temperature. For $2<\mu R<2\sqrt2$ the condensate therefore disappears at $T_{\rm HP}$, in a first-order transition of the probe sector driven by the gravitational one. For $\mu R>2\sqrt2$ it persists, and $\langle O_2\rangle^2$ drops there by $8L/e^2R^4$, independently of $\mu$, while the charge density \eqref{eq:dataK}, independent of $T$, does not jump. The small black holes, $u_h>\sqrt2R$, are thermodynamically unstable; at $u_h=\sqrt2R$, the minimum of $T$, the entropy \eqref{eq:entropy} of the probe sector diverges. The stability argument and eq.~\eqref{eq:globalmin} concern only the fields of the probe sector. In AdS$_4$, without a quartic coupling, the probe superfluid on the sphere has been constructed numerically on large and small black holes, together with its quasinormal modes \cite{GNTZ2024}.

\subsection{Charge and dimension}
\label{sec:DeltaQ}

\begin{figure}[t]
\centering
\includegraphics[width=\textwidth]{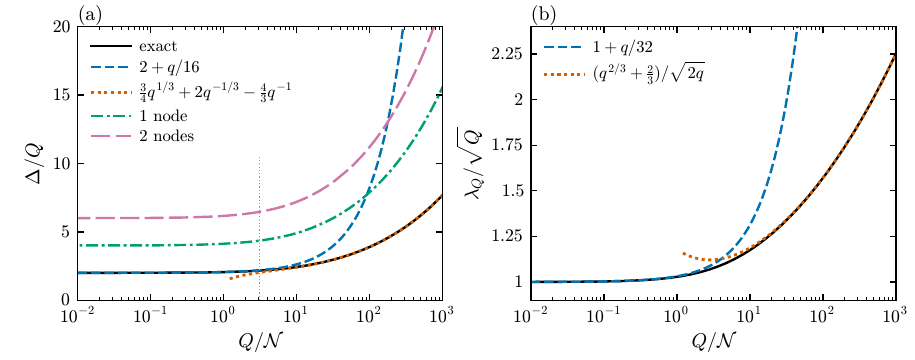}
\caption{(a) The dimension per unit charge of the lowest state of charge $Q$, eq.~\eqref{eq:DeltaQ}, with $q=Q/\Nn$, its small-charge expansion \eqref{eq:smallQ} to second order and its large-charge expansion \eqref{eq:largeQ} to third order. The dotted vertical line marks $q_*=16/(3\sqrt3)$, where the radii of convergence of the two expansions meet. The dash-dotted and long-dashed curves are the solutions with one and two nodes, computed numerically. (b) The three-point coefficient \eqref{eq:lambdaQ} divided by $\sqrt Q$, its value for a generalized free field, with its small- and large-charge expansions.}
\label{fig:DeltaQ}
\end{figure}

The scaling dimension of an operator is the energy of the corresponding state on $\mathbb{R}\times S^3$, $\Delta=ER$, with the energy of the vacuum subtracted. In the probe sector the vacuum is $\rho=0$, $A_t=\mu$, whose energy and charge vanish; the Casimir energy of the gravitational sector is common to all states and drops out. At zero temperature the soliton of section~\ref{sec:familysol} carries the charge $Q=2\pi^2R^3n$ and the energy $E=\Om_{\rm tot}+\mu Q=2\pi^2R^3(\mu n-P)$, where $\Om_{\rm tot}=-2\pi^2R^3P$ is its grand potential and $P$ is given by \eqref{eq:PS3}. With $\hat\mu\equiv\mu R$ and $\Nn\equiv\pi^2L/e^2$,
\begin{equation}
\boxed{\;Q=\Nn\,\hat\mu\,(\hat\mu^2-4),\qquad \Delta=\Nn\Big(\frac34\hat\mu^4-2\hat\mu^2-4\Big)=\frac{\Nn}{4}\,(\hat\mu^2-4)(3\hat\mu^2+4),\qquad \hat\mu\ge2.\;}
\label{eq:DeltaQ}
\end{equation}
Eq.~\eqref{eq:DeltaQ} satisfies the first law, $d\Delta/dQ=\mu R$, which is equivalent to $n=-\partial\Om/\partial\mu$. In terms of the charge, $\Delta/Q=(3\hat\mu^2+4)/4\hat\mu$, with $\hat\mu\ge2$ the root of $\hat\mu^3-4\hat\mu=Q/\Nn$ (figure~\ref{fig:DeltaQ}).

Since the normal-phase conductivity at $\omega\gg T$ is $\sigma_\infty=\Nn\,\omega/2\pi$, eq.~\eqref{eq:sigmaN}, $\Nn$ is proportional to the coefficient $C_J$ of the two-point function of the current. We identify $\Delta$ with the dimension of the lowest operator of charge $Q$ within this model, at leading order in $1/\Nn$ at fixed $q\equiv Q/\Nn$ and in the probe limit: by eq.~\eqref{eq:globalmin} the soliton is the lowest classical state of that charge. A bottom-up model cannot exclude other charged operators of the dual CFT with lower dimensions.

Black holes do not compete in the probe limit. A global black hole that carries the charge $Q$, with the probe fields in the normal phase, has $\Delta\ge Q\,(3e^2L^2/2\kappa_5^2)^{1/2}\gg Q$: with the normalizations of \eqref{eq:action}, this is $L$ times the mass of an extremal Reissner--Nordstr\"om black hole of charge $Q$ in five flat dimensions, and the Reissner--Nordstr\"om--AdS$_5$ black holes \cite{CEJM1999} of that charge approach it only when small and extremal. At large charge the extremal planar Reissner--Nordstr\"om brane has the pressure $(4\kappa_5^2/3e^2L^2)\,L\mu^4/8e^2$, below that of the soliton at the same $\mu$, so that at the same $Q$ its dimension is larger by the factor $(3e^2L^2/4\kappa_5^2)^{1/3}$. Hairy black holes do not compete either. At fixed $Q$ and fixed total entropy, which is the area of the backreacted horizon and includes the entropy of the probe sector, the energy changes at first order by the probe grand potential at fixed $T$ and $\mu$ plus $\mu Q$. By \eqref{eq:dOmega} the probe sector on a global black hole therefore contributes less to $\Delta$ than the soliton by $4\Nn(2R^2/u_h^2+R^4/u_h^4)$, while the black hole itself adds $(3\pi^2L^3/\kappa_5^2)(R^2/u_h^2+R^4/u_h^4)$, which is larger by a factor of order $e^2L^2/\kappa_5^2$.

\paragraph{Nodal branches.}\label{sec:nodal} By eq.~\eqref{eq:globalmin} the static, spherically symmetric solutions with nodes lie above the soliton at equal charge. The branches with $j=1$, $2$, $3$ nodes, constructed numerically (appendix~\ref{app:numerics}), start at $\mu R=2+2j$. The identity also places above the soliton the inhomogeneous solutions, which branch off at $\mu R=2+\ell$ from the modes of $O_2$ with angular momentum $\ell$ and start with $\Delta\simeq(2+\ell)Q$, and the solutions with angular momentum, such as vortices; we have not constructed these.

\paragraph{Small charge.} Near the onset, where $\hat\mu\to2$,
\begin{equation}
\frac{\Delta}{\Nn}=2q+\frac{q^2}{16}-\frac{q^3}{256}+\frac{q^4}{2048}-\frac{21\,q^5}{262144}+O(q^6).
\label{eq:smallQ}
\end{equation}
The leading term is the dimension of $Q$ quanta of $O_2$, each in the ground state on $S^3$, and the second is their interaction, $U\,Q(Q-1)/2\simeq\frac12UQ^2$ with $U=1/8\Nn$ the tree-level energy of each pair, which is repulsive. At leading order in $1/\Nn$ the same $U$ is the anomalous dimension of the lowest operator of charge two, $O_2^2$, whose dimension is $4+U$. The coefficient $1/16$ also follows from a direct expansion of the field equations around the vacuum, which does not use the exact solution and holds for any $\lambda$. The first-order solution is the normal mode at the onset (section~\ref{sec:thermoS3}), and at third order the condition that the correction to $\rho$ be source free fixes the shift of $\mu$, which gives
\begin{equation}
\frac{\Delta}{\Nn}=2q+\frac{1}{24}\Big(1+\frac{\lambda}{e^2}\Big)q^2+O(q^3),
\label{eq:smallQlambda}
\end{equation}
in which the exchange of the gauge field contributes $1/24$ and the contact interaction $\lambda/24e^2$. For any $\lambda$, $U=(1+\lambda/e^2)/12\Nn$ is the tree-level anomalous dimension of the double-trace operator of a charged scalar of dimension two in AdS$_5$ \cite{FitzpatrickShih2011}; see also \cite{AharonyPalti2021} and, for $Q$ quanta, \cite{FFL2024}. In the probe limit at $\lambda=0$, eq.~\eqref{eq:smallQlambda} also follows from the energy of the perturbative soliton of Ishii and Nakayama \cite{IshiiNakayama2024}.

\paragraph{Large charge.} For $q\gg1$, $\hat\mu=q^{1/3}\big(1+\frac43q^{-2/3}-\frac{64}{81}q^{-2}+\cdots\big)$ and
\begin{equation}
\frac{\Delta}{\Nn}=\frac34\,q^{4/3}+2\,q^{2/3}-\frac43+\frac{32}{27}\,q^{-2/3}-\frac{64}{81}\,q^{-4/3}+0\cdot q^{-2}+\frac{2048}{2187}\,q^{-8/3}+O\big(q^{-10/3}\big).
\label{eq:largeQ}
\end{equation}
Since $\hat\mu(q)$ is a root of $\hat\mu^3-4\hat\mu-q=0$, whose discriminant $256-27q^2$ vanishes at $q=\pm q_*$ with $q_*=16/(3\sqrt3)\approx3.08$, the physical root is singular at $q=-q_*$ and, as a function of $q^{-2/3}$, at complex values of modulus $q_*^{-2/3}$; at $q=q_*$, where $\mu R=4/\sqrt3$, it is regular. So the small-charge expansion \eqref{eq:smallQ} converges for $Q<q_*\Nn$ and the large-charge expansion \eqref{eq:largeQ}, a series in $q^{-2/3}$, for $Q>q_*\Nn$. The large-charge expansion is thus convergent, not merely asymptotic, at this order in $1/\Nn$, as is that of the leading weak-coupling result, whose cubic is the same up to rescaling (section~\ref{sec:EFT}). Its first three terms reproduce $\Delta$ to $1\%$ for $Q>9.5\,\Nn$, and the first two terms of \eqref{eq:smallQ} do so for $Q<2.6\,\Nn$.

\paragraph{Convexity.} By the first law, $d^2\Delta/dQ^2=d\hat\mu/dQ=1/[\Nn(3\hat\mu^2-4)]>0$. The dimension is therefore a convex function of the charge, with $\Delta(0)=0$, so that $\Delta(Q_1+Q_2)\ge\Delta(Q_1)+\Delta(Q_2)$: the lowest state of charge $Q_1+Q_2$ is not bound relative to the lowest states of charges $Q_1$ and $Q_2$. At leading order in $1/\Nn$ convexity is equivalent to a positive charge susceptibility, $dQ/d\mu>0$, which holds for every nodeless solution (section~\ref{sec:globalmin}). At charges of order one, convexity is the sign of the pair energy: since $U>0$, the dimension $4+U$ of $O_2^2$ lies above twice that of $O_2$, so the convexity conjectured in \cite{AharonyPalti2021} (see also \cite{Antipin2021conv}, and \cite{IshiiNakayama2024} for a holographic study) holds here already at unit charge, at leading order in $1/\Nn$. Superadditivity fails in a weakly coupled model \cite{SharonWatanabe2023} and can fail through quantum corrections at large charge \cite{CFGMR2025}.

\subsection{Effective theory}
\label{sec:EFT}

At large charge the lowest state is described by the effective theory of a conformal superfluid, in which a Goldstone field $\chi$, the phase of the order parameter, has the Lagrangian $P(X)$ with $X=-\partial_\mu\chi\partial^\mu\chi$, completed by curvature terms required by Weyl invariance \cite{Son2002,Hellerman2015,Monin2017}. The ground state on $\mathbb{R}\times S^3$ is $\chi=\mu t$, on which $X=\mu^2$, the Ricci scalar of the boundary is $\Rc=6/R^2$, and the Weyl tensor vanishes. Through second order in derivatives the pressure of this state is
\begin{equation}
P=c_1X^2+c_2\,\Rc X+c_3\,\Rc^2+O\big(\Rc^3/X\big),
\label{eq:PEFT}
\end{equation}
where the coefficients $c_i$ are combinations of Wilson coefficients; terms involving derivatives of $X$ do not contribute to the homogeneous state. Comparing with the exact pressure \eqref{eq:PS3} gives
\begin{equation}
c_1=\frac{L}{8e^2},\qquad c_2=-\frac{L}{6e^2},\qquad c_3=\frac{L}{18e^2},\qquad
P=c_1\Big(X-\frac23\,\Rc\Big)^2,
\label{eq:EFTcoeffs}
\end{equation}
and at leading order in $1/\Nn$ all higher terms in the curvature vanish. The coefficient $c_1$ is the zero-temperature equation of state of the planar superfluid, eq.~\eqref{eq:PX}. The pressure is a perfect square, $c_3=c_2^2/4c_1$, and vanishes quadratically at the onset $X=\frac23\Rc$, i.e.\ at $\mu R=\Delta_O$, with $\Delta_O=2$ the dimension of $O_2$; equivalently $c_2=-\Delta_O^2c_1/3$. Thus the effective theory, truncated to three terms, reproduces $\Delta(Q)$ at leading order in $1/\Nn$ for all $Q$, including the regime $Q\ll\Nn$ where the derivative expansion has no reason to apply, a consequence of the mean-field form. At $m^2L^2=-4$ and other $\lambda$ this form holds only at first order in $\lambda-e^2/2$ (section~\ref{sec:othercouplings}).

The coefficient $c_2$ can also be obtained without the global solution. Since the renormalized on-shell action is stationary under variations of the fields, the change of the pressure to first order in $1/R^2$, at fixed $\mu$, is the explicit change of the Lagrangian evaluated on the planar solution \eqref{eq:solution} at $T=0$. Writing global AdS$_5$ in Fefferman--Graham coordinates, $ds^2=L^2z^{-2}[dz^2-(1+z^2/4R^2)^2dt^2+R^2(1-z^2/4R^2)^2d\Omega_3^2]$, the derivative of the static Lagrangian density with respect to $1/R^2$ integrates on the planar zero-temperature solution to $\partial P/\partial(R^{-2})=-L\mu^2/e^2$, i.e.\ $6c_2=-L/e^2$, in agreement with \eqref{eq:EFTcoeffs}.

For a general pressure \eqref{eq:PEFT}, the large-charge expansion of the dimension is, for $R=1$,
\begin{equation}
\Delta=\frac{3}{8\pi^{2/3}c_1^{1/3}}\,Q^{4/3}-\frac{3\pi^{2/3}c_2}{c_1^{2/3}}\,Q^{2/3}+12\pi^2\Big(\frac{c_2^2}{c_1}-6c_3\Big)+O\big(Q^{-2/3}\big),
\label{eq:alphabetagamma}
\end{equation}
and with \eqref{eq:EFTcoeffs} the three coefficients are $\frac34\Nn^{-1/3}$, $2\Nn^{1/3}$ and $-\frac43\Nn$, as in \eqref{eq:largeQ}. In four dimensions the term of order $Q^0$ is not universal, since it depends on $c_3$; the universal term of order $Q^0\ln Q$ \cite{Cuomo2020} arises from the one-loop fluctuations of the phonon and is of relative order $1/\Nn$.

\paragraph{Comparison with weak coupling.} Eq.~\eqref{eq:DeltaQ} has the same form as the leading semiclassical result for the lowest operator of charge $Q$ in the U(1) model at the Wilson--Fisher fixed point in $d=4-\epsilon$, at fixed $\lambda_*Q$ \cite{BCMR2019} (for the limit at fixed $\lambda_*Q^2$ see \cite{ARR2019}). There the homogeneous state on $\mathbb{R}\times S^3$ is a solution of a Ginzburg--Landau theory with the conformal mass $1/R$, whose pressure is $P\propto(\mu^2-1/R^2)^2$, and the charge and dimension obey $\mu R(\mu^2R^2-1)=\lambda_*Q/8\pi^2$ and $\Delta/Q=(3\mu^2R^2+1)/4\mu R$. Eq.~\eqref{eq:DeltaQ} is the same with $1/R$ replaced by $2/R$, the dimension of $O_2$, and with $\lambda_*/8\pi^2$ replaced by $1/\Nn$. At leading order, the large-charge sector of the holographic superfluid at the Bogomolny point is thus that of a mean-field theory of an order parameter of dimension two. The termination of the curvature expansion in \eqref{eq:EFTcoeffs} is a property of the mean-field form, which the leading order of \cite{BCMR2019} shares; in both cases the fluctuations at the next order add terms that are not polynomial. What the holographic model adds at this order is the realization of the mean-field form in a strongly coupled CFT, with its coefficient $c_1$, and the spectrum of section~\ref{sec:spectrum}, with states beyond the effective theory.

\paragraph{Three-point coefficient.}\label{sec:OPE} At leading order in $1/\Nn$ the condensate of the global soliton gives the matrix element of the order parameter between the lowest states of charges $Q$ and $Q+1$, i.e.\ a three-point coefficient of the CFT; by \eqref{eq:dataK}, $R^2\langle O_2\rangle=(L^{1/2}/e)\,\hat\mu\sqrt{\hat\mu^2-4}/2$. Its normalization requires the two-point function of $O_2$: with the regular solution $u^2K_0(ku)$ in Euclidean Poincar\'e AdS$_5$, the on-shell action with the boundary terms \eqref{eq:Sbdry} gives $\langle O_2O_2^\dagger\rangle(k)=-\ln k$ up to terms analytic in $k^2$, the Fourier transform of $C_O/|x|^4$ with $C_O=1/2\pi^2$, independent of $L$ and $e$. For the unit-normalized operator, the three-point coefficient is therefore
\begin{equation}
\lambda_Q=\frac{R^2\langle O_2\rangle}{\sqrt{C_O}}=\sqrt{\frac{\Nn}{2}}\;\hat\mu\sqrt{\hat\mu^2-4}
=\begin{cases}\sqrt Q\,\Big(1+\dfrac{Q}{32\Nn}+\cdots\Big), & Q\ll\Nn,\\[2ex]
\sqrt{\dfrac{\Nn}{2}}\,\Big[\Big(\dfrac{Q}{\Nn}\Big)^{2/3}+\dfrac23-\dfrac29\Big(\dfrac{Q}{\Nn}\Big)^{-2/3}+\cdots\Big], & Q\gg\Nn.\end{cases}
\label{eq:lambdaQ}
\end{equation}
At small charge (figure~\ref{fig:DeltaQ}(b)) this is the matrix element $\sqrt Q$ of a generalized free field between states of $Q-1$ and $Q$ quanta, which fixes the normalization independently. At large charge it scales as $Q^{2/3}=Q^{\Delta_O/3}$, as the effective theory predicts for an operator of dimension $\Delta_O$ \cite{Monin2017,Cuomo2020}. As $R\to\infty$ at fixed $\mu$, the condensate tends to $\langle O_2\rangle=L^{1/2}\mu^2/2e$, its value in the zero-temperature planar superfluid, section~\ref{sec:superflow}.

\subsection{Excitations}
\label{sec:spectrum}

The normal modes of the global soliton give the low-lying states above the lowest state of charge $Q$. A mode with $\omega\neq0$ carries no charge, since the charge is conserved and its perturbation oscillates, so its frequency is the gap $\omega R=\Delta_{\rm exc}-\Delta_Q$ at fixed $Q$. The perturbations $\tau$ and $b_a$ are expanded in harmonics on $S^3$ (appendix~\ref{app:perturbations}); the transverse vector harmonics of $b_i$ decouple.

As $Q\to0$ the condensate disappears and the spectrum becomes that of the free fields in global AdS$_5$, in the frame rotating with $\mu R=2$. As $Q\to\infty$ the interior approaches the infrared fixed point (figure~\ref{fig:profilesglobal}), and the frequencies of modes localized there approach the normal modes in global AdS$_5$ of fields with the dimensions \eqref{eq:IRdims}, which do not depend on $Q$: besides the phonons, the spectrum at large charge contains towers of states built on the infrared fixed point.

\paragraph{Vector modes.}\label{sec:transverseS3} The transverse components obey the Sturm--Liouville problem \eqref{eq:transverseS3}, whose mass term, in the global angle $\zeta$ defined by $u=R\cot\zeta$, is $W=e^2L^2\rhob^2\sec^2\zeta$ and depends on position. Two properties of $W$ bound the spectrum. First, $0\le W<8\sec^2\zeta$, since $eL\rhob<2\sqrt2$. Second, at fixed $\zeta$, $W$ increases with $\mu$. The lower bound is the Maxwell field, and the upper bound is a Proca field of mass squared $8$; their spectra are $3+\ell+2n$ and $5+\ell+2n$, $\ell\ge1$, $n\ge0$. The weight $\tan\zeta$ does not depend on $\mu$, and the form domain of the Proca field lies inside the common domain of the Maxwell and soliton problems. By the min-max principle, the frequency of the $n$-th mode at angular momentum $\ell$ therefore increases monotonically with the charge, between these limits:
\begin{equation}
3+\ell+2n<\omega_{\ell n}R<5+\ell+2n .
\label{eq:transversebounds}
\end{equation}
As $Q\to0$ the lower limit is approached. These states are the descendants of the current $J_\mu$, of dimension three, added to the dilute gas. As $Q\to\infty$ the upper limit is approached: the current has dimension $\Delta_J^\IR=5$ at the infrared fixed point. Both properties of $W$ use the exact solution. At the other couplings computed at $m^2L^2=-4$, the bounds and the monotonic rise persist, with $5$ replaced by $\Delta_J^\IR$, only for $\lambda<e^2/2$ (table~\ref{tab:generic}).

The approach to the upper limit follows from first-order perturbation theory. At large $K=\mu^2R^2/4$ and at a fixed angle $\zeta$, $W=8\sec^2\zeta\,[1-(1+2\tan^2\zeta)/K+O(K^{-2})]$. The expansion fails near the boundary, $\tan^2\zeta\gtrsim K$, but the Proca modes vanish there as $\cos^4\zeta$, and that region contributes at order $K^{-3}$. The first-order shift, with the Proca modes of appendix~\ref{app:perturbations} and the weight $\tan\zeta$, gives
\begin{equation}
\omega_{\ell n}R=5+\ell+2n-\frac{2}{3}\,\frac{(5+\ell+2n)^2-(\ell+1)^2}{\mu^2R^2}+O\big((\mu R)^{-4}\big),
\label{eq:transverseshift}
\end{equation}
for all $\ell$ and $n$, with the matrix elements evaluated in closed form there. Since $\mu R\simeq(Q/\Nn)^{1/3}$, the gaps approach their infrared values as $Q^{-2/3}$. Near the onset, first-order perturbation theory around the Maxwell modes gives the slope
\begin{equation}
\frac{d\,\omega_{\ell n}R}{d\,\mu R}\Big|_{\mu R=2}=\frac{8(n+1)(n+\ell+2)}{(3+\ell+2n)(2n+\ell+2)(2n+\ell+4)} .
\label{eq:onsetslope}
\end{equation}

\begin{figure}[t]
\centering
\includegraphics[width=\textwidth]{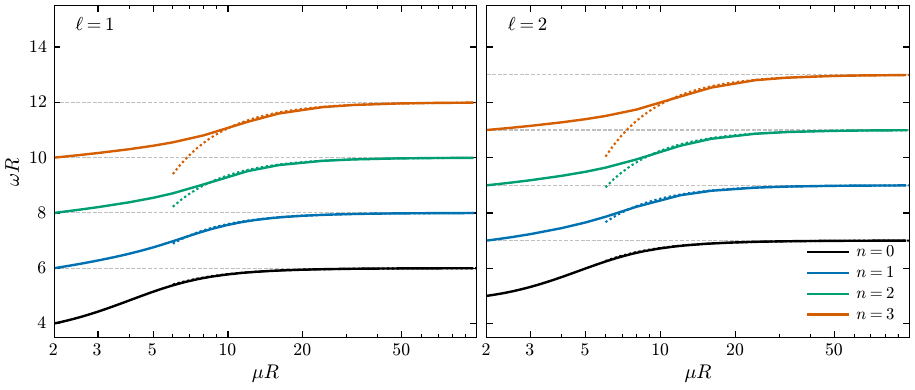}
\caption{Gaps of the transverse vector modes above the lowest state of charge $Q$, for $\ell=1$ and $2$ and the four lowest modes, $n=0$ to $3$, as functions of $\mu R$. The gaps increase from $3+\ell+2n$, the descendants of the current at $Q=0$, to $5+\ell+2n$ (dashed), the normal modes of the current at the infrared fixed point. The dotted curves are the large-charge expansion \eqref{eq:transverseshift}.}
\label{fig:transverse}
\end{figure}

A spectral method and a shooting method (appendix~\ref{app:numerics}) give the spectrum of figure~\ref{fig:transverse}. It lies within \eqref{eq:transversebounds} and approaches \eqref{eq:transverseshift}, whose coefficient the numerics reproduce. Modes with different $\ell$ do not repel and can cross. For instance, $(\ell,n)=(3,0)$ and $(1,1)$, which are degenerate at both ends, cross at $\mu R=3.4226$.

\begin{figure}[t]
\centering
\includegraphics[width=\textwidth]{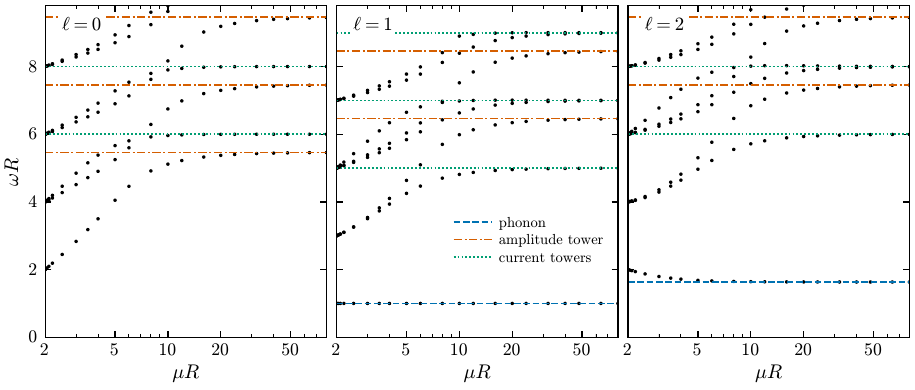}
\caption{Frequencies of the scalar-channel modes above the lowest state of charge $Q$, for $\ell=0$, $1$, $2$, as functions of $\mu R$. At each $\mu R$ the eight lowest modes are shown (seven at $\ell=0$, where the zero mode is omitted), so for $\mu R\lesssim12$ some modes near the top of the range are missing. The horizontal lines are their large-charge limits: the phonon $\sqrt{\ell(\ell+2)/3}$ (dashed), the amplitude tower $2+2\sqrt3+\ell+2n$ (dash-dotted), and the towers $4+\ell+2n$ and $6+\ell+2n$ of the infrared current (dotted). The mode at $\omega R=1$ for $\ell=1$ is exact.}
\label{fig:scalar}
\end{figure}

\paragraph{Scalar modes.}\label{sec:scalarS3} In the scalar channel $\tau$, $b_t$, the radial component of $b_a$ and the longitudinal part of $b_i$ are coupled, with the conditions \eqref{eq:bcS3}. The spectrum, computed by collocation for $\ell\le4$ and $2<\mu R\le80$, and by shooting up to $\mu R=40$ (appendix~\ref{app:numerics}), is shown in figure~\ref{fig:scalar}, and the two methods agree. All frequencies are real, as section~\ref{sec:linstab} requires, and $\omega=0$ occurs only at $\ell=0$, for the static change of the charge.

\paragraph{The descendant.} For $\ell=1$ the lowest mode has $\omega R=1$ exactly, for every $Q$. It is generated by an isometry of AdS$_5$ acting on the soliton, a boost that moves it away from the center, and it is the descendant $P_\mu$ of the lowest state of charge $Q$. Its eigenfunction is \eqref{eq:descendant}. At $\ell=0$ there is no mode at $\omega R=2$: the descendant $P^2$ of the lowest state is a state of two quanta of the $\ell=1$ mode, which does not appear in the linear spectrum.

\paragraph{Small charge.} As $Q\to0$ the spectrum tends to $\omega R=\ell+2n$, $4+\ell+2n$ and, for $\ell\ge1$, $2+\ell+2n$. In the dilute gas, the first is the replacement of one quantum of $O_2$ by a descendant $\partial^\ell(\partial^2)^nO_2$, which costs $\ell+2n$. The second is the addition of a pair of quanta of $O_2^\dagger$ and $O_2$, at fixed charge, with dimension $2+(2+\ell+2n)$. The third is the current $J_\mu$, of dimension three, or one of its descendants, in the scalar harmonics. The lowest mode at $\ell=0$ starts at $\omega R=2$, the replacement of one quantum by $\partial^2O_2$, and rises monotonically with the charge to $2+2\sqrt3$.

\paragraph{Large charge.} As $Q\to\infty$ the modes with $\omega R<12$ and $\ell\le4$ approach the following limits.
\begin{itemize}
\item The phonons, the lowest modes for $\ell\ge1$, with $\omega_\ell R\to\sqrt{\ell(\ell+2)/3}$, the spectrum of the conformal superfluid on $S^3$ with $c_s^2=\frac13$ \cite{Monin2017,Cuomo2020}. The approach is as $(\mu R)^{-2}\propto Q^{-2/3}$, with the coefficient derived below, eq.~\eqref{eq:phononS3}.
\item The amplitude tower $\omega R\to2+2\sqrt3+\ell+2n$, the normal modes of the order parameter at the infrared fixed point, approached from below as $(\mu R)^{-2}$; for the lowest mode, $\ell=n=0$, perturbation theory in $1/K$ gives the coefficient $-\frac{24}{11}(8+5\sqrt3)\approx-36.35$ (appendix~\ref{app:perturbations}), which the numerical spectrum reproduces.
\item Two towers of the massive vector, $\omega R\to4+\ell+2n$ ($\ell\ge1$) and $6+\ell+2n$ ($\ell\ge0$), which are the states of a vector primary of dimension $5$ in the scalar harmonics, as in the transverse channel. For $\ell=1$, $n=0$ the gap tends to $5=\Delta_J^\IR$: this is the infrared current itself.
\end{itemize}
Together with the vector modes, for $\omega R<12$ and $\ell\le4$ the spectrum at large charge thus consists of the phonons and of the single-particle states of the two fields at the infrared fixed point, with gaps that approach constants as $Q\to\infty$. Some gaps in the scalar channel overshoot their limits by up to $1\%$ and approach them from above. The phonons are the counterpart on the sphere of zero-temperature sound on the plane, eq.~\eqref{eq:cs}, and the amplitude tower that of the continuum of the amplitude mode, which has the same infrared dimension (section~\ref{sec:amplitude}).

\paragraph{Phonons at next-to-leading order.}\label{sec:phononNLO} The phonon gaps approach their limits as
\begin{equation}
\omega_\ell R=\sqrt{\frac{\ell(\ell+2)}{3}}\,\Big[1+\frac{(\ell-1)(\ell+3)}{9\mu^2R^2}+O\big((\mu R)^{-4}\big)\Big],
\label{eq:phononS3}
\end{equation}
which is derived in appendix~\ref{app:phonon} in two steps.

The first step holds for any conformal superfluid in four dimensions: symmetry, the flat-space limit and the descendant at $\ell=1$, with $\omega R=1$ at every charge, fix the relative correction to $\omega_\ell$ as $\Xi(\ell-1)(\ell+3)/\mu^2R^2$, as in the dispersion relation of the effective action \cite{Cuomo2020}, with $\Xi$ the coefficient of zero-temperature sound, $\omega=(k/\sqrt3)(1+\Xi k^2/\mu^2+O(k^4))$.

The second step computes $\Xi$ on the zero-temperature domain wall of section~\ref{sec:structure}. Expanded in $\omega\sim k$, the longitudinal perturbations reduce to static problems solved in closed form, and the conservation of the charge gives $c_s^2=\frac13$ at second order and $\Xi=1/9$ at fourth, eq.~\eqref{eq:soundk3}. The infrared fixed point can enter only from relative order $(k/\mu)^6$.

In the effective action \eqref{eq:Lnlo}, $\Xi=(c_4-2c_2)/6c_1$, where $c_4$ multiplies a Ricci-tensor term that $\Delta(Q)$ leaves free; with \eqref{eq:EFTcoeffs}, $\Xi=1/9$ gives $c_4=-2c_1=-L/4e^2=-\Nn/4\pi^2$, at leading order in $1/\Nn$ and in the probe limit. This value does not come from the Bogomolny coupling: for a canonical Maxwell term and any potential $V(\rho)$, the Maxwell equation on the zero-temperature flow fixes it, given the derivative expansion, eq.~\eqref{eq:c4general}. The phonons on $S^3$ confirm the resulting $\Xi$ at every coupling and mass computed (appendix~\ref{app:numerics}). The two-point coefficient of the current, proportional to $\Nn$, does not fix $c_4$ alone, however: with a kinetic function $Z(\rho)$ of the gauge field, $c_4$ is an average of $Z$ over the wall, eq.~\eqref{eq:c4Z}. What depends on the potential is $\Xi$, through $c_1/\Nn$ and $c_2/\Nn$, so the relation $c_4=-2c_1$ holds only where $c_1=L/8e^2$.

The same value $\Xi=1/9$, and with it eq.~\eqref{eq:phononS3}, follows from the leading-order Goldstone dispersion of the U(1) model at the Wilson--Fisher fixed point at fixed $\lambda_*Q$ \cite{BCMR2019}, expanded at large $\mu R$; there the pressure has $c_2=-c_1/3$, so $c_4=0$. The two dispersions differ at the next order: the term of order $k^4/\mu^4$ in $\sqrt3\,\omega/k$ has the coefficient $-5/162$ there and $19/162$ here, eq.~\eqref{eq:soundk5}. Since $\Xi$ is the same in both, $c_4=2c_2+\frac23c_1$ differs only through $c_2=-\Delta_O^2c_1/3$, with $\Delta_O$ replaced by $1$ there.

Fits of the phonon gaps on $S^3$ reproduce $(\ell-1)(\ell+3)/9$ (appendix~\ref{app:numerics}).

\subsection{Validity}
\label{sec:validity}

Eq.~\eqref{eq:DeltaQ} is the leading term of $\Delta=\Nn h_0(q)+h_1(q)+O(1/\Nn)$ at fixed $q$, the classical limit of the bulk theory, and it is obtained in the probe limit. The fluctuations of the bulk fields contribute $h_1(q)$ (section~\ref{sec:discussion}). The matter fields of the solution are bounded in an orthonormal frame uniformly in $\mu$: $e\rhob\le2\sqrt2/L$, and the field strength in that frame, $u^2|\partial_u\Bb_t|/L^2$, is bounded by $9/(4\sqrt3L^2)$, so that the stress tensor of the matter fields is bounded by a number of order $1/e^2L^4$. The metric is therefore corrected at relative order $\kappa_5^2/e^2L^2$ everywhere, for all $Q$, and we expect the corrections to $\Delta/\Nn$ to be of the same relative order uniformly in $q$; they are not computed here. Since $\kappa_5^2/e^2L^2=\Nn\kappa_5^2/\pi^2L^3$, and $L^3/\kappa_5^2$ is proportional to the central charge $C_T$, the backreaction is of relative order $\Nn/C_T$, and the probe limit requires $\Nn\ll C_T$, as for a flavor sector. The backreaction correction to $\Delta$, of order $\Nn^2/C_T$, exceeds the one-loop term $h_1$ once $\Nn\gtrsim C_T^{1/2}$. At large charge the backreaction modifies the coefficients $c_i$ of the effective theory and, through the change of the radius of the infrared region, the dimensions of section~\ref{sec:spectrum}.

\section{R\'enyi entropies}
\label{sec:renyi}

The R\'enyi entropies $S_n=(1-n)^{-1}\ln{\rm Tr}\,\rho_A^n$ of a ball of radius $R$ in the vacuum of a CFT on $\mathbb{R}^{1,3}$ are thermal quantities on the hyperbolic cylinder $\mathbb{R}\times H^3$ of curvature radius $R$ \cite{CHM2011,HMSY2011}: the reduced density matrix is conjugate to a thermal state at $T_0=1/2\pi R$, and
\begin{equation}
\frac{S_n}{V_{H^3}}=\frac{n}{1-n}\,\frac{1}{T_0}\Big[\Om(T_0)-\Om(T_0/n)\Big]=\frac{n}{n-1}\,\frac{1}{T_0}\int_{T_0/n}^{T_0}dT\,s(T),
\label{eq:HMSY}
\end{equation}
where $\Om$ is the free energy and $s$ the entropy, both per unit volume of $H^3$. With a chemical potential $\mu$ for the U(1), held fixed while $T$ varies, the same formulas, with $\Om$ the grand potential, give the charged R\'enyi entropies of \cite{BHMMMS2013}. In a holographic CFT the relevant bulk solutions are the hyperbolic black holes \eqref{eq:metric} with $\kappa=-1$ and $T=T_0/n$, i.e.\ with $u_h(n)=4nR/(1+\sqrt{1+8n^2})$, and a hairy phase of these black holes changes the R\'enyi entropies \cite{BMM2013,BHMM2014}.

\paragraph{The homogeneous saddle.} The member $\kappa=-1$ of the exact family is the homogeneous hairy phase of these black holes. Its condensate does not fall off toward the boundary of $H^3$, which is the entangling surface, so it is not a normalizable deformation of the normal phase. As noted for the constant mode in \cite{BMM2013}, it computes the R\'enyi entropies of a ball when operators inserted at the entangling surface source it. At $\mu=0$ and $n<n_*$ it does so only with such sources, since the normal phase is the only normalizable static saddle there; $n_*$ is the onset of the normalizable modes, discussed below. For $1<n<n_*$ at $\mu=0$, these operators are a pinning field for a relevant charged operator on the entangling surface, and the homogeneous saddle is the infrared fixed point of the flow it starts: the field sets the distance from the entangling surface at which the flow reaches the saddle, not the amplitude of its condensate \cite{PaperDefect}. With these sources, or on a compact quotient $H^3/\Gamma$, where it is normalizable and gives thermal quantities rather than entanglement entropies \cite{DMRS2010,BelinMaloney2014}, it has the lower grand potential and, being nodeless, it minimizes $E-\mu Q$ among states at finite relative energy (section~\ref{sec:globalmin}).

\begin{figure}[t]
\centering
\includegraphics[width=\textwidth]{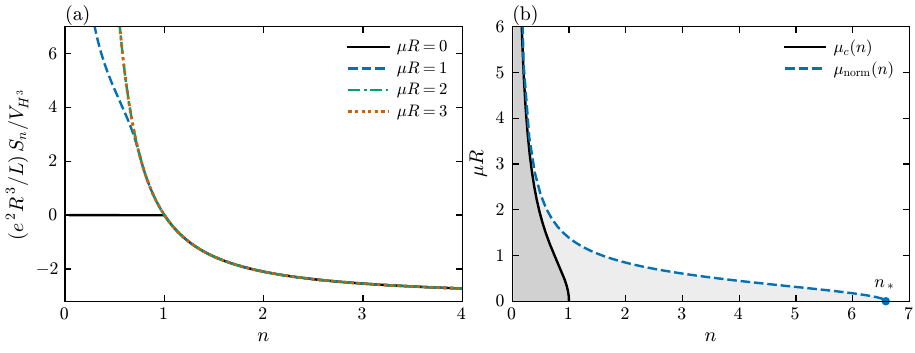}
\caption{(a) The contribution of the probe sector to the charged R\'enyi entropies of a ball, eq.~\eqref{eq:Sn}, per unit volume of $H^3$. In the hairy phase it does not depend on $\mu$, and the curves for different $\mu$ coincide; for $n<1$ the curve at a given $\mu$ leaves the common one at the index where $\muc(n)=\mu$, below which the black hole is in the normal phase. At $\mu=0$ the entropy vanishes for $n\le1$. (b) The phase boundary $\mu=\muc(n)$ of the homogeneous saddle, eq.~\eqref{eq:mucn}, and the onset $\mu_{\rm norm}(n)$ of the static mode with $-\nabla^2=1/R^2$, which reaches $\mu=0$ at $n_*\approx6.58$. Below $\mu_{\rm norm}$ (light and dark shading) the normal phase is linearly stable against normalizable perturbations; below $\muc$ (dark shading) it is also the only homogeneous saddle.}
\label{fig:renyi}
\end{figure}

The probe sector contributes to the grand potential at order $L/e^2$, while the gravitational sector contributes at order $L^3/\kappa_5^2$; the former is computed here. From \eqref{eq:dOmega}, its grand potential per unit volume of $H^3$ is $\Om=-(L/e^2)[\mu^2/u_h^2+\frac18(\mu^2-\muc^2)_+^2]$, where $(x)_+\equiv\max(x,0)$, with the critical chemical potential
\begin{equation}
\muc(n)^2R^2=4\Big[\frac{R^2}{u_h(n)^2}-1\Big]=\frac{1+\sqrt{1+8n^2}-4n^2}{2n^2}.
\label{eq:mucn}
\end{equation}
It vanishes at $n=1$, is positive for $n<1$ and negative for $n>1$. The hairy phase therefore exists for every $\mu$ when $n>1$, including $\mu=0$, and for $\mu>\muc(n)$ when $n<1$; the phase boundary in the $(n,\mu)$ plane is $\mu=\muc(n)$, which ends at $n=1$, $\mu=0$. Inserting $\Om$ in \eqref{eq:HMSY}, the contribution of the homogeneous saddle of the probe sector to the charged R\'enyi entropy per unit volume of $H^3$ is
\begin{equation}
\boxed{\;\frac{S_n(\mu)}{V_{H^3}}=\frac{L}{e^2R^3}\,\frac{\pi n}{4(1-n)}\Big[\muc^4-\big(\muc^2-\mu^2\big)_+^2\Big]R^4,\;}
\label{eq:Sn}
\end{equation}
with $\muc=\muc(n)$ from \eqref{eq:mucn}. Its main features, shown in figure~\ref{fig:renyi}, are the following.

\begin{itemize}
\item \emph{Independence of $\mu$ in the hairy phase.} For $n\ge1$, and for $n<1$ when $\mu>\muc(n)$, $S_n$ does not depend on $\mu$. This follows from the separation of $\Om_\SF$ noted in section~\ref{sec:grandpot}: its $\mu$-dependent part cancels in \eqref{eq:HMSY} at fixed $\mu$. For $n<1$ and $\mu<\muc(n)$, $S_n$ grows with $\mu$ and joins the hairy value with a continuous first derivative at $\mu=\muc(n)$: the transition is of second order.
\item \emph{Zero chemical potential.} $S_n=0$ for $n\le1$, and for $n>1$
\begin{equation}
\frac{S_n}{V_{H^3}}=-\frac{L}{e^2R^3}\,\frac{\pi n}{4(n-1)}\,\muc(n)^4R^4=-\frac{L}{e^2R^3}\Big[\frac{16\pi}{9}(n-1)-\frac{224\pi}{81}(n-1)^2+\cdots\Big],
\label{eq:Snneutral}
\end{equation}
which decreases monotonically to $-\pi L/e^2R^3$ as $n\to\infty$. The hairy phase lowers the free energy, so its contribution is negative. The R\'enyi entropy has a kink at $n=1$ itself: the critical index of the homogeneous neutral hairy phase is exactly one, as found, with backreaction, for a neutral scalar at the BF bound in \cite{DMRS2010,BMM2013}. As noted there and in \cite{BelinMaloney2014}, at the BF bound the black hole at $T_0$, which is AdS$_5$ itself, is at the threshold of the instability. Since the neutral solution exists for any $\lambda>0$ (section~\ref{sec:familysol}), with $\Om_\SF-\Om_\N=-(L/16\lambda)\muc^4$, eq.~\eqref{eq:Snneutral} holds for any $\lambda>0$ with $L/e^2$ replaced by $L/2\lambda$: the onset of the homogeneous saddle at $n=1$ is a consequence of the BF-bound mass, not of the Bogomolny coupling, and the kink comes from the second-order transition of the probe sector. The transition is also at $n=1$, but of third order and with $\partial_nS_n$ continuous, in the closed-form backreacted solutions of \cite{Ren2019,Bai2022}, and an instability of the R\'enyi entropies at exactly $n=1$ appears in the large-$N$ O($N$) vector model in three dimensions, there with normalizable modes \cite{HLS2018}. The slope at $n=1^+$, of order $L/e^2$, does not contradict the relation of $\partial_nS_n$ at $n=1$ to $C_T$ \cite{Perlmutter2013}, to which the probe sector does not contribute at this order: that relation assumes analyticity at $n=1$, which the transition there breaks, and with normalizable modes only the contribution vanishes for $n<n_*$ and is analytic at $n=1$.
\item \emph{Entanglement entropy.} As $n\to1$ at fixed $\mu>0$, the system is in the hairy phase on both sides and $S_n\propto\muc^4/(1-n)\to0$. In the homogeneous saddle, the probe sector therefore contributes nothing to the charged entanglement entropy at order $L/e^2$, for any $\mu$, because the thermal entropy \eqref{eq:entropy} is proportional to $\muc^2$, which vanishes at $T_0$. The normal phase, which is disfavored at $n=1$, would give $(L/e^2R^3)\,\frac43\pi\mu^2R^2$ per unit volume; with normalizable modes only, it is linearly stable for $\mu<\mu_{\rm norm}(1)\approx1.39/R$ (below).
\end{itemize}
The volume of $H^3$ requires a cutoff at a distance $\delta$ from the entangling surface. Its universal part is $V_{H^3}\supset-2\pi R^3\ln(2R/\delta)$, so the probe sector contributes $-2\pi(L/e^2)\hat s_n\ln(2R/\delta)$ to the universal term of $S_n$, where $\hat s_n\equiv(e^2R^3/L)\,S_n/V_{H^3}$ is the function plotted in figure~\ref{fig:renyi}(a); for $n>1$ at $\mu=0$ this contribution is positive.

\paragraph{Normalizable modes.} Without operators inserted at the entangling surface, the bulk fields must be normalizable on $H^3$, and normalizable perturbations have $-\nabla^2\ge1/R^2$. With normalizable modes only, the normal phase at $\mu=0$ is linearly stable up to $n_*\approx6.58$, as found in \cite{BMM2013} for $\Delta_O=2$: the static scalar mode with $-\nabla^2=1/R^2$ appears only at $T_*\approx0.152\,T_0$, well below $T_0$, where the homogeneous mode sets in (appendix~\ref{app:numerics}). In the probe sector the normal phase is then the only normalizable static saddle: at $\mu=0$ the Maxwell equation forces $B_t=0$, and multiplying the scalar equation by $\rho$ and integrating, as for \eqref{eq:Omega}, equates the quadratic form of the normal phase, positive for $n<n_*$, to minus $\lambda$ times a positive quartic integral, so $\rho=0$. The contribution of the probe sector to $S_n$ at $\mu=0$ therefore vanishes identically for every index below the onset, $n<n_*$. At $\mu\neq0$ the mode with $-\nabla^2=1/R^2$ sets in at $\mu_{\rm norm}(n)$, figure~\ref{fig:renyi}(b), and higher modes set in later, since $-\nabla^2$ enters the radial equation as a positive mass. For $n<1$ this curve lies above $\muc(n)$, at $n=1$ it has $\mu_{\rm norm}R\approx1.39$, and it vanishes at $n_*$. Beyond $n_*$ the normal phase is unstable at every $\mu$, since $\mu$ only lowers the effective mass of the scalar. The nonlinear phase beyond $n_*$, left open in \cite{BMM2013,BHMM2014}, is constructed at $\mu=0$ in the probe limit in \cite{PaperDefect}. By section~\ref{sec:globalmin} the probe sector has no nodeless normalizable hairy saddle without superflow at any $\mu<\mu_{\rm norm}(n)$; with backreaction, a first-order transition there is not excluded. As $n\to\infty$ the instability matches the near-horizon criterion of \cite{BMM2013,BHMM2014,Ren2020}, by which the mode with $-\nabla^2=1/R^2$ violates the AdS$_2$ BF bound at every $\mu$ in the probe limit. In AdS$_4$ with backreaction the corresponding onset was found numerically in \cite{BHMM2014}, where at large scalar charge $\mu$ is likewise destabilizing.

\section{Beyond the exact family}
\label{sec:beyond}

This section asks how much of the preceding results depends on the Bogomolny point and on the family of section~\ref{sec:family}: which results hold at other couplings and masses of the model (section~\ref{sec:othercouplings}), whether the closed form extends to other backgrounds and dimensions (section~\ref{sec:beyondfamily}), and what carries over to top-down models (section~\ref{sec:topdown}).

\subsection{Other couplings and masses}
\label{sec:othercouplings}

Table~\ref{tab:generic} lists which results hold away from $\lambda=e^2/2$ and $m^2L^2=-4$. Each pair $(\Delta_O,\lambda/e^2)$, with $m^2L^2=\Delta_O(\Delta_O-4)$, is a different bulk theory, so the table compares models, not states of one model. Eq.~\eqref{eq:globalmin} holds at every pair, so the tuning is not what makes a nodeless solution without superflow the lowest state of its charge, or the linear onset the transition; it gives the closed form and the relations that fail in the table. At general $\Delta_O$ the small-charge expansion \eqref{eq:smallQlambda} becomes $\Delta/\Nn=\Delta_Oq+(\Delta_O-1)(\Delta_O-1+\lambda/e^2)\,q^2/[8(2\Delta_O-1)]+O(q^3)$, the two-particle result of \cite{FitzpatrickShih2011}. Below the infrared bound $\lambda<|m^2|L^2e^2/3$, which is $\lambda<4e^2/3$ at $m^2L^2=-4$, the static, homogeneous solutions found at every pair computed still have the structure of the Bogomolny point: the condensate stays nodeless from $\Tc$ down to $10^{-3}\Tc$, and at $T=0$ the flow ends at the minimum of the potential.

\begin{table}[!tp]
\centering
\footnotesize
\begin{tabular}{@{}>{\raggedright}p{0.36\textwidth} >{\raggedright}p{0.48\textwidth} >{\raggedright}p{0.10\textwidth}@{}}
\toprule
Result & Away from the Bogomolny point & Basis \tabularnewline
\midrule
Global minimum at fixed charge; nodeless soliton lowest & any $\Delta_O\ge2$ and $V'(\rho)/\rho$ nondecreasing, nodeless, no superflow, over all classical states & proof \tabularnewline
Linear stability, real normal modes on the sphere & same conditions & proof \tabularnewline
Horizon formulas \eqref{eq:horizonGamma} and \eqref{eq:zeta3} & any $\lambda$ & derivation \tabularnewline
Second-order transition at $\Tc$; one nodeless branch & every $\lambda\ge0$; for $V'(\rho)/\rho$ nondecreasing, at most one nodeless solution without superflow at each $\mu$ (at $\mu=0$, unless $V$ is affine), and no first-order transition & proof \tabularnewline
Small charge, eq.~\eqref{eq:smallQlambda} & any $\lambda$; at general $\Delta_O$, with another coefficient & derivation \tabularnewline
$c_4=-L/4e^2$ & any $V(\rho)$, given the derivative expansion; with $Z(\rho)F^2$, the average \eqref{eq:c4Z} & derivation \tabularnewline
Onset of the homogeneous saddle at $n=1$ and $\mu=0$, eq.~\eqref{eq:Snneutral} & any $\lambda>0$ at $m^2L^2=-4$, with $L/e^2\to L/2\lambda$ & derivation \tabularnewline
\midrule
Flow to the minimum of $V$; $\Delta_J^\IR$, $\Delta_O^\IR$; ${\rm Re}\,\sigma\propto\omega^{2\Delta_J^\IR-5}$ & below the bound; above it, $\Delta_J^\IR=4$ and a preferred frame & numerics \tabularnewline
Effective-theory form of $\Delta(Q)$ through order $Q^0$; phonon limit and factor $(\ell-1)(\ell+3)$ of eq.~\eqref{eq:phononalpha} & below the bound, with coefficients that depend on $\lambda$ and $\Delta_O$; above it, with a term in $\ln\mu R$, eq.~\eqref{eq:lnmuR}; the phonon limit at every pair & numerics \tabularnewline
Towers at $\Delta_O^\IR+\ell+2n$, $\Delta_J^\IR\pm1+\ell+2n$, $\Delta_J^\IR+\ell+2n$ & below the bound; above it, gaps of order $1/R$ not set by one dimension & numerics \tabularnewline
Spectral bounds \eqref{eq:transversebounds}, monotonic gaps, $\rhob<\rs$ & with $5\to\Delta_J^\IR$, for $\lambda\le e^2/2$ at $m^2L^2=-4$; for $\lambda>e^2/2$, only the lower bound & numerics \tabularnewline
\midrule
$n$ independent of $T$; $n_s\propto\langle O_2\rangle^2$; $\mu(n-n_s)=Ts$ and $\sigma_\reg/\sigma_\N(0)$ equal to the normal fraction away from $\Tc$; the $3/5$ to $2/5$ split of $\Gamma$ & fails at every pair computed & numerics \tabularnewline
Rational closed form; first-order equations and kink & rational closed form only at the Bogomolny point for $\mu\neq0$, in the ansatz of section~\ref{sec:closure}; the first-order equations and the kink also at $\lambda=8e^2/3$, $T=0$, on the plane & derivation \tabularnewline
Perfect-square pressure, $c_2=-\Delta_O^2c_1/3$, $c_3=c_2^2/4c_1$, end of the curvature expansion & at first order in $\lambda-e^2/2$ only; at $\Delta_O\neq2$, at no coupling computed & numerics \tabularnewline
$\Xi=1/9$ & at no other pair computed & numerics \tabularnewline
\bottomrule
\end{tabular}
\caption{Results away from the Bogomolny point, in the probe limit, with a canonical Maxwell term unless stated. The bound is the infrared bound $\lambda/e^2<|m^2|L^2/3$. Rows run from results that hold at other pairs, through results that keep their form with changed values, to results that fail. Numerical entries hold at the pairs $(\Delta_O,\lambda/e^2)$ computed (appendix~\ref{app:numerics}), for static, homogeneous solutions; a sharp change of structure between the pairs is not excluded, and very near the bound some entries are not resolved.}
\label{tab:generic}
\end{table}

At the minimum of the potential, $m_B^2L^2=|m^2|L^2e^2/\lambda$, so $\Delta_J^\IR=2+(1+m_B^2L^2)^{1/2}$ and ${\rm Re}\,\sigma\propto\omega^{2\Delta_J^\IR-5}$. The exponent $5$ thus requires $\lambda/e^2=|m^2|L^2/8$, which at $m^2L^2=-4$ is $\lambda=e^2/2$, while $\Delta_O^\IR=2+(4+2|m^2|L^2)^{1/2}$ does not depend on $\lambda$. The infrared bound is necessary, since near the minimum $u\Bb_t\propto u^{2\pm(1+m_B^2L^2)^{1/2}}$ must vanish in the infrared; that the flow does end there is a numerical result, obtained at $m^2L^2=-4$ up to $\lambda/e^2=1.33$ with the infrared condition imposed on the stable manifold of the minimum (appendix~\ref{app:numerics}). The bound is the probe limit of the condition that the current be irrelevant in the infrared, $\Delta_J^\IR>4$, given with backreaction in \cite{ALY2014}; in AdS$_4$ the condition reads $\Delta_J^\IR>3$, as stated in \cite{GubserNellore2009b}. At $\lambda=0$, the case without a quartic term, as in \cite{HHH2008a,HHH2008b}, and that of real $\vec z$ on the D7 branes of section~\ref{sec:topdown}, the potential has no minimum, so we do not expect an infrared fixed point. At $m^2L^2=-4$ the flows at $T=0$, followed down to $\lambda/e^2=1/256$, have $c_1$ and $|c_2|$ still growing; at $\lambda=0$ on the black brane, $\rhob_h\propto(\mu/T)^2$ down to $0.06\,\Tc$, and at the BF-bound mass the condensate grows at low $T$ as a power of $\ln(\Tc/T)$ \cite{STM2012}.

Above the bound the flow ends instead at the other constant solution of \eqref{eq:radial}, $e^2L^2\rho^2=3$, $u\Bb_t=(3\lambda/e^2-|m^2|L^2)^{1/2}$. It has a preferred frame and $\Delta_J^\IR=4$, and is a probe counterpart of the Lifshitz-like infrared of \cite{GubserNellore2009b,ALY2014}. On the sphere the flow stays near this solution for $1/\mu\ll u\ll R$, and with $b=u\Bb_t$ at this solution the pressure gains the term
\begin{equation}
\frac{L}{e^2}\,\frac{3b^2\,(2+|m^2|L^2+3b^2)}{4\,(4+2|m^2|L^2+5b^2)}\,\frac{\ln\mu R}{R^4}.
\label{eq:lnmuR}
\end{equation}
Its coefficient is exact, from the series about this solution in global AdS$_5$, and the charge at large $\mu R$ reproduces it without fits (appendix~\ref{app:numerics}); the matching that makes it a term in the pressure is argued. On the plane at $T=0$ and $m^2L^2=-4$, one flow to it is exact: at $\lambda=8e^2/3$, $eL\rho=\sqrt3\,Ku^2/(1+Ku^2)$ and $\Bb_t=\mu(1+Ku^2)^{-1/2}$ with $K=\mu^2/4$, so that $e^2L^2\rho^2\to3$ and $u\Bb_t\to2$. It obeys first-order equations of the form \eqref{eq:firstorder}, $u\,\partial_u\hat\rho=\hat\rho\,(2-2\sqrt{2/3}\,\hat\rho)$ and $u\,\partial_u\Bb_t=-\sqrt{2/3}\,\hat\rho\,\Bb_t$, with the same kink $\tan(\eta/2)=\mu u/2$. On this flow $c_1=L/16e^2$, $c_2=-5L/48e^2$ and $\Xi=-\frac19$, a second value of $\Xi$ in closed form from \eqref{eq:c4general}; the phonons on $S^3$ agree. At $m^2L^2=-4$, $\Xi$ runs from $0.276$ at $\lambda/e^2=\frac18$ to $-0.193$ at $4$. We have found no extension to $T>0$, where \eqref{eq:rho3} keeps its logarithm. At $T=0$ the two flows are the members of one class: for $\Bb_t=\mu(1+Ku^2)^{-\nu}$ the Maxwell equation fixes $eL\rho=2\sqrt{\nu(\nu+1)}\,Ku^2/(1+Ku^2)$, and of this class only $\nu=1$, at $\lambda=e^2/2$, and $\nu=\frac12$ also solve the scalar equation.

The closed form also gives the pressure near the Bogomolny point, to first order in $\lambda-e^2/2$. Since the on-shell action is stationary under variations of the fields, $\partial\Om/\partial\lambda=(L^5/4)\int_0^{u_h}du\,\rhob^4/u^5$ at fixed $\mu$ and $T$, which at $\lambda=e^2/2$ equals $(4\pi^4LT^4/3e^4)(\Th-1)^2(\Th+3)/(\Th+1)$, with $\Th=(\Tc/T)^2$; at $T=0$ the pressure is $P=(L\mu^4/8e^2)[1-\frac23(\lambda/e^2-\frac12)+O((\lambda/e^2-\frac12)^2)]$. On the sphere the same step gives $P=(L/8e^2)(\mu^2-4/R^2)^2$ times the same bracket, so the perfect square survives at this order, and $c_2=-\Delta_O^2c_1/3$ and $c_3=c_2^2/4c_1$ fail only at the next.

\subsection{Other backgrounds and dimensions}

\paragraph{Other backgrounds.}
\label{sec:beyondfamily}
The closure of section~\ref{sec:familysol} fails on two other backgrounds. On an AdS$_5$--Reissner--Nordstr\"om background, charged under a second U(1) so that the probe U(1) remains a probe, $f$ acquires a term $q_{\rm RN}^2u^6$; with the same ansatz, the Maxwell equation of \eqref{eq:radial}, cleared of denominators, has a highest-power coefficient proportional to $\mu\,q_{\rm RN}^4$, so $q_{\rm RN}=0$ for $\mu\neq0$. Nor does the AdS$_5$ soliton \cite{HorowitzMyers1998}, on which the holographic insulator-to-superconductor transition takes place \cite{NRT2009}, admit the form that the solution takes on the black holes, $eL\rho=p\,u^2/(1+Ku^2)$, $B_t=\mu(1+d_1u^2+d_2u^4)/(1+Ku^2)$: the coefficient equations have no real solution with $p\neq0$.

\paragraph{Other dimensions.} On the AdS$_{d+1}$ black brane with the scalar mass at the BF bound and $\lambda=e^2/2$, the vector and amplitude masses at the minimum of the potential still coincide, $m_B^2=m_\rho^2=d^2/(2L^2)$, and the stability argument and eq.~\eqref{eq:globalmin} still apply, with $2/L\to d/(2L)$ in the scalar boundary terms of \eqref{eq:Sbdry}. We have found no closed form for $d\neq4$, and three checks tie this form to $d=4$ (appendix~\ref{app:numerics}). At $T=0$ a profile $\rho\propto\hat u^{d/2}/(1+\hat u^{d/2})$, with $\hat u\propto u$, forces $B_t\propto u^{d/4-1}/(1+\hat u^{d/2})$, which has a finite, nonzero chemical potential only for $d=4$. The zero mode at $\Tc$ is rational only if its local exponents $\pm iB_t(u_*)/f'(u_*)$ at the complex zeros $u_*$ of $f$ are integers, which happens only for $d=4$, where they are $\pm\mu_cu_h/2=\pm1$ at $u_*=\pm iu_h$. And $n/\mu^{d-1}$, constant in the condensed phase for $d=4$, varies with $T$ for $d=3$, $5$, $6$. These checks do not exclude a closed form of another type. The closed form does extend to the Lifshitz black branes with $3z$ spatial dimensions and a scalar of dimension $2z$ at the Lifshitz BF bound. Their static, homogeneous equations reduce to those of the planar brane with $(u/u_h)^2$ replaced by $(u/u_h)^{2z}$, as Natsuume and Okamura noted in a model without a quartic coupling, whose condensed phase they constructed near $\Tc$ \cite{NatsuumeOkamura2018}. With a quartic coupling the reduction divides $m^2$, $\lambda$ and $e^2$ by $z^2$, so the closed form holds there at $\lambda=e^2/2$.

\subsection{Top-down models}
\label{sec:topdown}

The type IIB truncation of \cite{GHPT2009} has $m^2L^2=-3$ and a nonpolynomial potential and does not reduce to the present one, but its zero-temperature domain wall \cite{GPR2009} ends at an infrared fixed point with the same spectrum: at the infrared extremum of its potential $m_\rho^2L_\IR^2=m_B^2L_\IR^2=8$, with $L_\IR$ the infrared radius, which gives the infrared dimensions $5$ and $2+2\sqrt3$. The two masses are equal because of the D-term structure of section~\ref{sec:structure}: there too the potential is quadratic in the moment map of the U(1). We are not aware of a consistent truncation of supergravity that gives the present model, and its quartic coupling is tuned, so we do not expect the closed form to carry over to dimension-two operators of top-down models, such as the $\mathbf{20}'$ of $\mathcal{N}=4$ super Yang--Mills theory. In an $\mathcal{N}=1$ superconformal theory the unitarity bounds \cite{DobrevPetkova1985,Minwalla1998} allow a scalar superconformal primary of dimension two only at zero R-charge, where it saturates both bounds and is the bottom of a current multiplet \cite{Ceresole2000}, or at R-charge $\pm\frac43$, where it is chiral or antichiral. The double trace of a chiral operator is chiral, so its dimension is fixed by its R-charge, and the energy $U$ of a pair of quanta, eq.~\eqref{eq:smallQlambda}, vanishes. In the bulk the field dual to the chiral operator has the D-term quartic $\lambda=e^2/2$, and the neutral partner sourced by the cubic coupling cancels the photon and the quartic: $\Nn U=\frac1{12}+\frac1{24}-\frac18=0$ (appendix~\ref{app:numerics}). Without that partner, as in the present model, we know of nothing that protects $\lambda=e^2/2$. In the D3/D7 system with an isospin chemical potential \cite{CPVZ2013}, whose scalar of dimension two condenses along a unit vector $\vec z$ of the SU(2)$_R$ triplet, real $\vec z$ gives the present model at $\lambda=0$, but section~\ref{sec:energy} does not cover the other fields on the brane. There the scalar is the bottom of a current multiplet, and at tree level a condensate along $\vec z$ has $U=|\vec z\cdot\vec z|^2/12\Nn\le1/12\Nn$, below its Bogomolny value $1/8\Nn$. As a quartic form in $\vec z$ and its conjugate, invariant under SU(2)$_R$ and the phase of $\vec z$, $U$ is linear in $|\vec z\cdot\vec z|^2$, so its values at real and null $\vec z$, $1/12\Nn$ and $0$, fix it. For real $\vec z$, $U$ agrees with the tree-level gluon correlators of \cite{ABFZ2021,ABZ2022}.

\section{Discussion}
\label{sec:discussion}

\paragraph{Beyond the superfluid effective theory.} The towers of section~\ref{sec:spectrum} are the normal modes of the infrared fixed point. At leading order in $1/\Nn$ every heavy state of a holographic CFT carries towers of multi-trace states built from the light single-trace operators. Those built from the stress tensor, with the integer gaps $4,5,6,\ldots$ of its conformal family, do not see the soliton in the probe limit. Those found here, built from the current and the order parameter, do: their gaps are set by the condensate and, at $\lambda=e^2/2$, move with the charge from the values of the free fields to those of the infrared fixed point. They are the counterpart on the sphere of the infrared conformal sector of zero-temperature holographic superfluids \cite{GubserNellore2009a,HorowitzRoberts2009}. For backreacted charged solitons in global AdS$_4$, whose planar limit also ends in an infrared AdS$_4$ region, the lowest radial normal mode was found numerically to keep a finite frequency as the charge grows \cite{AriasMasSerantes2016}.

The superfluid effective theory is valid below the scale $\mu\propto Q^{1/3}$, so these states lie within its regime of validity but are not described by it: the large-charge sector here is a conformal superfluid coupled to a second, infrared CFT. Heavy states of CFTs with spontaneously broken boosts require low-lying primaries other than descendants \cite{KMPR2021}, which in a superfluid are the phonons; the bootstrap at large charge also admits solutions with several Regge trajectories \cite{JMZ2017}, further constrained by probes of the U(1) current and of the stress tensor \cite{KiaeeMonin2025,KiaeeMonin2026}. The towers are thus an explicit example of light states beyond the Goldstone mode, with gaps that depend on $Q$ only through corrections of order $(\mu R)^{-2}\propto Q^{-2/3}$; whether they persist at finite $\Nn$ for $Q\gg\Nn$ is beyond the classical computation of this paper.

\paragraph{Probe limit.} Backreaction matters most at low temperature: the brane holds the normal component at rest only while the inertia $\mu n_s$ of the superfluid component is small compared with the enthalpy density $2\pi^4L^3T^4/\kappa_5^2$ of the brane. Their ratio is $4\kappa_5^2(\Tc^4-T^4)/(e^2L^2T^4)$, so the sound speed \eqref{eq:cs} requires $T^4\gg4\kappa_5^2\Tc^4/(e^2L^2)$; below this temperature it no longer applies, although $c_s^2=1/3$ at $T=0$ is fixed by conformal invariance. The attenuation \eqref{eq:Gamma} requires more. In the sound mode the momentum density vanishes, so the normal component moves against the superfluid component, at a velocity smaller by this ratio, and the shear viscosity of the brane damps it. This adds $(2\kappa_5^2/3e^2L^2)(\Tc^4-T^4)/(\pi T^5)$ to $\Gamma$, the shear-viscosity term of the second-sound attenuation in \cite{HLSY2011} when the brane dominates the enthalpy. At low temperature this term is small compared with \eqref{eq:Gamma} only if $(\kappa_5^2/e^2L^2)(\Tc/T)^{10}\ll5/24$, that is $T\gg(\kappa_5^2/e^2L^2)^{1/10}\,\Tc$; for $\kappa_5^2/e^2L^2=10^{-2}$ it is larger below $T\approx0.72\,\Tc$. At $T=0$ the backreaction gives the two AdS$_5$ regions of the domain wall different radii, which changes the infrared dimensions and with them the exponent of the conductivity; the closed form is then presumably lost. Near $\Tc$ the leading backreaction is known analytically at zero quartic coupling \cite{HLSY2011,BBM2011}, and the near-critical dynamics is known beyond the probe limit \cite{DonosKailidis2025,DonosKailidis2026}; the full correction at first order in $\kappa_5^2/e^2L^2$ away from $\Tc$ is not computed here. Eq.~\eqref{eq:globalmin} holds on a fixed background, and we know of no counterpart with backreaction.

\needspace{4\baselineskip}
\paragraph{Open questions.}
\begin{itemize}
\item On every slicing the free energy of condensation follows from the charge density \eqref{eq:dataK} alone, which at fixed $\mu$ does not depend on $T$ and equals that of the normal phase at $\muc$: integrating the difference from $\muc$ gives the mean-field form \eqref{eq:dOmega}. The question is therefore why the charge density does not depend on the temperature; on the solution, at fixed $\mu$, neither does the kink \eqref{eq:kink}. On the plane just below $\Tc$, $(T\partial_T\ln n)_\mu=4(2\lambda-e^2)/(4\lambda+e^2)$, which vanishes only at the Bogomolny coupling. Conversely, at $m^2L^2=-4$, if $uB_t/f$ at fixed $\mu$ does not depend on the temperature, then $\lambda=e^2/2$ and the expansion of $B_t/f$ in $u^2$ is that of the closed form, through the order $u^{10}$ computed. Explaining the temperature independence therefore amounts to explaining the closed form.
\item The origin of the first-order equations \eqref{eq:firstorder} and of the sine-Gordon structure \eqref{eq:kink} is not understood beyond the equal-mass condition of section~\ref{sec:structure}, which does not hold at $\lambda=8e^2/3$, where at $T=0$ the same kink appears (section~\ref{sec:othercouplings}). At $T=0$ the superpotential of the flow gives no bound, and we have neither derived the first-order equations from a Bogomolny bound nor found a supersymmetric embedding that explains them, although first-order flows with a charged scalar at finite density do arise in supersymmetric settings \cite{BKPW2011}. Eq.~\eqref{eq:globalmin} is a sum of squares, but it needs a solution as its reference and holds at every $\lambda\ge0$, so it is not such a bound. The analogy with the flat-space Bogomolny vortices is also limited: the exact solution is a homogeneous radial profile rather than a vortex (for vortices in holographic superfluids see, e.g., \cite{DMPSS2010,DHIS2014}), and the boundary U(1) is a global symmetry.
\item Away from $\Tc$ we have the relaxation rate of the amplitude mode only numerically, apart from one value (section~\ref{sec:amplitude}); the exact background may make a closed form accessible, as it did for the attenuation \eqref{eq:Gamma}. The zero-temperature conductivity may be expressible through confluent Heun functions (appendix~\ref{app:T0cond}; see also \cite{ArnaudoWithers2025}). At $T>0$ the near-horizon pole-skipping points of \cite{NatsuumeOkamura2026} depend only on $\rhob_h$ and $E_h$, which \eqref{eq:horizondata} gives in closed form; whether the nonuniversal points found near $\Tc$ in \cite{Natsuume2026b} persist away from $\Tc$ is open.
\item At $T>0$ superflow is treated here only in linear response, eq.~\eqref{eq:London}, and at low temperature. The nonlinear flowing states \cite{HKS2009,BMS2009,ABK2010} and their quasinormal modes, which locate the Landau instability \cite{Amado2014,GMS2023,AGMS2024}, are not computed here; sections~\ref{sec:globalmin} and~\ref{sec:energy} assume a background without superflow.
\item The corrections of relative order $1/\Nn$, including the Casimir energy of the phonon and the universal logarithm of \cite{Cuomo2020}, follow from the fluctuation spectrum of section~\ref{sec:spectrum}. The renormalized sum of the zero-point energies of the normal modes of the soliton, to which the towers contribute alongside the phonon, gives $h_1(q)$ at all $q$; it is not computed here, nor is the effect of the towers on the universal logarithm.
\item The charged R\'enyi entropies \eqref{eq:Sn} also determine the symmetry-resolved entanglement, through their continuation to imaginary chemical potential \cite{BHMMMS2013,GoldsteinSela2018}, as in the holographic computation of \cite{ZNM2021} in AdS$_3$; here this requires continuing them across the phase boundary, which we have not done. At large charge, the symmetry-resolved R\'enyi entropy of a disk has been obtained from the superfluid effective theory \cite{Watanabe2025}.
\item With normalizable modes only, the R\'enyi entropies above the onset $\mu_{\rm norm}(n)$, or above $\mu_{\rm norm}(1)$ for $n<1$ since \eqref{eq:HMSY} also involves the black hole at $T_0$, require a hairy phase.
\item On the line $8\CA+2\CB=1$, where the third-order logarithm also vanishes (section~\ref{sec:structure}), further exact solutions, if they exist, would need a form other than the rational ansatz.
\item States with angular momentum, in particular vortices \cite{CdlFMPR2017,Cuomo2019}, lie above the soliton at fixed charge (section~\ref{sec:globalmin}), but the lowest state at fixed angular momentum is open: the isometries of AdS$_5$ map the soliton to its descendants, which oscillate about the center, but not to rotating solutions.
\end{itemize}

\section*{Acknowledgments}
The author is grateful to Gregory Gabadadze for guidance and many discussions. Part of this work was done at the Center for Cosmology and Particle Physics, New York University. The exact solution on the planar black brane was found by the author in 2017, in unpublished work. Claude Opus 5.5 (Anthropic) assisted with the scientific reasoning, the derivations, the literature search and the drafting of the text. The author directed it by choosing the questions and deciding what to pursue and include. The author rederived the analytic results, checked the arguments, reviewed the numerical methods and their controls, and is solely responsible for the content, including any errors. The programs used in this work are available from the author upon reasonable request.

\appendix
\FloatBarrier
\section{Boundary terms and energy}
\label{app:asymptotics}

This appendix supplies the variation behind the normalizations in \eqref{eq:expansions}, the near-boundary expansions behind \eqref{eq:bc} and \eqref{eq:bcmode}, the proof of eq.~\eqref{eq:globalmin}, and the Hamiltonian and energy flux of the perturbations.

\paragraph{Variational principle.} The on-shell variation of the gauge-field action is the surface term $e^{-2}\int d^4x\sqrt{-g}\,F^{u\mu}\delta A_\mu$ at $u=u_c$, which gives the normalization of $\langle J_\mu\rangle$ in \eqref{eq:expansions}. For the scalar, write $\Psi=u^2[\Psi_\src\ln(u/u_c)+\Psi_\vev]$; the bulk action and the scalar terms of \eqref{eq:Sbdry} then vary as $-L^3\int d^4x\,(\Psi_\vev\,\delta\Psi_\src^*+{\rm c.c.})$, so the coefficient of the logarithm is the source \cite{Papadimitriou2007}. With $\Psi_\src=-\Psi^{(b)*}/L^{3/2}$, whose sign makes $\langle O_2\rangle$ positive for real positive $\Psi$, $\langle O_2\rangle=L^{3/2}\Psi_\vev$. Because the logarithm in \eqref{eq:expansions} is normalized at the cutoff, the on-shell action is finite as a function of $\Psi^{(b)}$ and $\langle O_2\rangle$; with a source, $\langle O_2\rangle$ so defined contains a term proportional to $\Psi^{(b)*}\ln u_c$, the familiar logarithm at the BF bound \cite{BFS2002,Papadimitriou2007}, which is local in the source. Higher boundary terms, such as $|\Psi|^4$ or $|D_\mu\Psi|^2$, vanish as $u_c\to0$. The length $u_0$ is arbitrary; changing it, or normalizing the logarithm in \eqref{eq:expansions} at a fixed scale, adds local counterterms, which change only contact terms: a term proportional to $i\omega$ in the conductivity and terms polynomial in the momenta in the correlators of $O_2$. The results of this paper, obtained from ${\rm Re}\,\sigma$, spectral functions and poles of correlators, do not depend on these choices.

\paragraph{Near the boundary.} In the gauge $a_u=0$ the perturbations $\tau$, $\theta$, $a_a$ of section~\ref{sec:stability} behave, at any temperature, as $\tau=u^2[\tau_\src\ln u+\tau_\vev]$, $\theta=\theta_\src\ln u+\theta_0$ and $a_\mu=a^{(0)}_\mu+u^2[a^{(2)}_\mu-\frac12\partial^\nu f^{(0)}_{\nu\mu}\ln u]$, up to terms smaller by $u^2$ (and logarithms), since near the boundary $\rhob\simeq\rho_\vev u^2$, $\Bb_t\simeq\mu$ and $f=1+O(u^4)$ suppress the background terms of \eqref{eq:tau}--\eqref{eq:constraint}; $f^{(0)}_{\mu\nu}$ is the field strength of $a^{(0)}_\mu$. With $\delta\Psi=(\tau+i\rhob\,\theta)/\sqrt2$ and \eqref{eq:expansions}, $\tau_\src$ and $\theta_\src$ vary the source of $O_2$, $\tau_\vev$ and $\theta_0$ the condensate, $a^{(0)}_\mu$ is an external field and $\delta\langle J_\mu\rangle=(2L/e^2)\,a^{(2)}_\mu$ is the induced current. The normal-mode conditions \eqref{eq:bc} set the sources to zero, so $a^{(0)}_\mu$ is pure gauge and is removed, and $b_\mu\to-\partial_\mu\theta_0$. The $u$-component of \eqref{eq:b}, the linearized Ward identity, then gives $\partial^\mu a^{(2)}_\mu=0$, and the constraint \eqref{eq:constraint} gives $b_u=O(u)$, hence \eqref{eq:bcmode}; the interior fixes $\tau_\vev$, $\theta_0$ and $a^{(2)}_\mu$ up to normalization.

\paragraph{The energy identity.}\label{app:energy} Let $\Pi^I\equiv e^{-2}\sqrt{-g}\,|g^{tt}|g^{IJ}F_{tJ}$, with $\Pi^u\to n$ at the boundary and $I,J$ over $(u,x,y,z)$ as in section~\ref{sec:globalmin}. The Gauss law and the energy are
\begin{gather*}
\partial_I\Pi^I=2\sqrt{-g}\,|g^{tt}|\,{\rm Im}(\Psi^*D_t\Psi),\\
E=\int d^3x\,du\,\sqrt{-g}\,\Big[\frac{1}{2e^2}|g^{tt}|g^{IJ}F_{tI}F_{tJ}+\frac{1}{4e^2}F_{IJ}F^{IJ}+|g^{tt}||D_t\Psi|^2+g^{IJ}(D_I\Psi)^*D_J\Psi+V\Big]\\
+\int d^3x\,\frac1L\sqrt{-\gamma}\,\rho^2\Big|_{u_c},
\end{gather*}
and on a static solution $E-\mu Q=\Om_{\rm tot}$. The reference has $\bar F_{tI}=-\partial_I\Bb_t$. The electric energy differs from that of the reference by the square in \eqref{eq:globalmin} and the cross term $-\partial_I\Bb_t\,(\Pi-\bar\Pi)^I$. Integrating by parts, the surface term is $\bar\mu(Q-\bar Q)$ at the boundary, where $\Bb_t=\bar\mu$ is constant because the reference has no superflow and the boundary field strength vanishes; it vanishes at a horizon, where $\Bb_t=0$, and at a center. By the Gauss law for both states the bulk term is $\sqrt{-g}\,|g^{tt}|\Bb_t\big[2\,{\rm Im}(\Psi^*D_t\Psi)+\rhob^2\Bb_t\big]$. With $|D_t\Psi+i\Bb_t\Psi|^2=|D_t\Psi|^2+\Bb_t^2|\Psi|^2+2\Bb_t\,{\rm Im}(\Psi^*D_t\Psi)$ and $\Psi=\rhob w/\sqrt2$, the kinetic terms of both states and this bulk term sum to the first term of \eqref{eq:globalmin} and the remainder $-\frac12|g^{tt}|\Bb_t^2(\rho^2-\rhob^2)$. Altogether,
\begin{multline*}
\int d^3x\,du\,\Big[\frac12F_{tI}\Pi^I+\sqrt{-g}\,|g^{tt}|\,|D_t\Psi|^2\Big]-\int d^3x\,du\,\Big[\frac12\bar F_{tI}\bar\Pi^I+\frac12\sqrt{-g}\,|g^{tt}|\,\Bb_t^2\rhob^2\Big]=\bar\mu\big(Q-\bar Q\big)\\
+\int d^3x\,du\,\sqrt{-g}\,\Big[\frac{1}{2e^2}|g^{tt}|g^{IJ}\big(F-\bar F\big)_{tI}\big(F-\bar F\big)_{tJ}\\
+\frac12\rhob^2|g^{tt}|\,\big|D_tw+i\Bb_tw\big|^2-\frac12|g^{tt}|\Bb_t^2\big(\rho^2-\rhob^2\big)\Big].
\end{multline*}

For the gradient, $(D_I\Psi)^*D^I\Psi=\frac12\rhob^2(D_Iw)^*D^Iw+\frac12\partial_I\rhob\,\partial^I(\rhob|w|^2)$, with $\rhob|w|^2=\rho^2/\rhob$. An integration by parts, with the background equation $\big(-\nabla_a\nabla^a+V'(\rhob)/\rhob-|g^{tt}|\Bb_t^2\big)\rhob=0$ in the bulk term, gives
\begin{multline*}
\int d^3x\,du\,\sqrt{-g}\,g^{IJ}(D_I\Psi)^*D_J\Psi=\int d^3x\,du\,\sqrt{-g}\,\Big[\frac12\rhob^2g^{IJ}(D_Iw)^*D_Jw-\Big(\frac{V'(\rhob)}{2\rhob}-\frac12|g^{tt}|\Bb_t^2\Big)\rho^2\Big]\\
-\int d^3x\,\frac12\sqrt{-g}\,g^{uu}\rho^2\,\partial_u\ln\rhob\,\Big|_{u_c},
\end{multline*}
with the second term of \eqref{eq:globalmin} on the right; the surface terms at a horizon and at a center vanish and are left out. After subtracting the same expression for the reference, the $\Bb_t^2$ terms cancel those of the previous display. At the cutoff the surface term tends to $-L^3\rho_\vev^2$ for $\rho\simeq\rho_\vev u^2$, for any source-free state, which the boundary term of $E$ cancels state by state: for such a state the scalar terms of \eqref{eq:Sbdry} reduce to $-(2/L)\sqrt{-\gamma}\,|\Psi|^2=-(1/L)\sqrt{-\gamma}\,\rho^2$. At $\Delta_O>2$ both vanish.

For a perturbation, the part of the surface term quadratic in $\tau$ is $\frac12\sqrt{-g}\,g^{uu}\tau^2\partial_u\ln\rhob$. It vanishes at a $T>0$ horizon, where $g^{uu}\propto f$ and $\partial_u\ln\rhob$ is finite, if $f\tau^2\to0$, and as $u\to\infty$ at $T=0$, where on the exact family $\sqrt{-g}\,g^{uu}\partial_u\ln\rhob\propto u^{-6}$, if $\tau$ grows more slowly than $u^3$. At the cutoff it tends to $-L^3\tau_\vev^2$, which is canceled by $\ee_\bdry=+L^3\tau_\vev^2$, since the scalar terms of \eqref{eq:Sbdry} have the quadratic part $-(1/L)\sqrt{-\gamma}\,\tau^2$. In \eqref{eq:E2} the two scalar bulk terms behave as $\pm2L^3\tau_\vev^2/u$, and only their sum is integrable, a consequence of the BF bound; in \eqref{eq:E2positive} each term is integrable, since $\tau/\rhob\to\tau_\vev/\rho_\vev$.

\paragraph{Hamiltonian.} The canonical variables are $\tau$, $\theta$ and $a_I$, and $\mathcal{L}_2$ depends on $\theta$ and $a_a$ only through $b_a=a_a-\partial_a\theta$. The momenta are $P_\tau=\sqrt{-g}\,|g^{tt}|\dot\tau$, $P^I=e^{-2}\sqrt{-g}\,|g^{tt}|g^{IJ}f_{tJ}$ and $P_\theta=-\sqrt{-g}\,|g^{tt}|\big(\rhob^2b_t+2\rhob\Bb_t\tau\big)$, and $a_t$ enforces the Gauss law $\partial_IP^I=P_\theta$, the $t$-component of \eqref{eq:b}. With $\dot a_I=f_{tI}+\partial_Ia_t$ and the Gauss law, $P^I\dot a_I+P_\theta\dot\theta=P^If_{tI}-P_\theta\,b_t+\partial_I(P^Ia_t)$. The term $-P_\theta b_t$ cancels the mixed term $b_t\tau$ of $-\sqrt{-g}\,\mathcal{L}_2$ and reverses the sign of its $b_t^2$ term, as for a Proca field. The total derivative leaves $P^ua_t$ at the ends of the radial direction. At the boundary $P^u\to\delta n$, so this term is $-\delta n\,a^{(0)}_t$, which vanishes for a normal mode; it involves $a_t$, not $b_t\to-\partial_t\theta_0\neq0$. This gives \eqref{eq:E2}.

\paragraph{Energy flux.} The flux through a surface of constant $u$, toward increasing $u$, is
\begin{equation}
S^u=-\sqrt{-g}\,g^{uu}\Big(\partial_u\tau\,\partial_t\tau+\frac{1}{e^2}g^{\mu\nu}f_{u\mu}f_{t\nu}+\rhob^2b_ub_t\Big),
\label{eq:Su}
\end{equation}
and $\dot\ee_2$ is the flux at the cutoff minus the flux into the interior, plus $\dot\ee_\bdry$. \emph{At the boundary}, with $\tau_\src=\theta_\src=0$, the expansions above give $S^u\to-\partial_t\big(L^3\tau_\vev^2\big)-\delta\langle J^\mu\rangle f^{(0)}_{t\mu}$, up to terms local in $a^{(0)}_\mu$: the first term is compensated by $\ee_\bdry$, and the second, the work of the external field, vanishes for a normal mode. This holds term by term although $b_t(0)\neq0$, since $b_\mu(0)$ is a gradient, so that $f_{t\mu}=O(u^2)$, and $b_u=O(u)$. The integrand of \eqref{eq:E2positive} is integrable at $u=0$, where each of its terms vanishes at least as $u$. \emph{At the horizon}, the ingoing behavior \eqref{eq:ingoing} gives $S^u\ge0$ for real fields, and for a mode $\propto e^{-i\omega t}$
\begin{equation}
\text{energy density}\ \propto\Big(1-\frac{u}{u_h}\Big)^{-1+{\rm Im}\,\omega/2\pi T},\qquad
S^u,\ P^ua_t\ \propto\Big(1-\frac{u}{u_h}\Big)^{{\rm Im}\,\omega/2\pi T}.
\end{equation}
For ${\rm Im}\,\omega>0$ the energy is finite, and the flux, $P^ua_t$ and $f\tau^2$ vanish at the horizon; for ${\rm Im}\,\omega\le0$ the energy diverges there, as for any quasinormal mode on a constant-$t$ slice, and the argument of section~\ref{sec:energy} does not concern such modes. At the extremal horizons of $H^3$ and of the axion brane, where $f$ has a double zero, the ingoing factor $e^{i\omega\int du/f}$ vanishes faster than any power for ${\rm Im}\,\omega>0$, and the same conclusions hold. \emph{At $T=0$}, ${\rm Im}\,\omega>0$ and real $\vk$ give ${\rm Re}\,\bar k>0$, so by \eqref{eq:Poincare} all fields decay as $e^{-u\,{\rm Re}\,\bar k}$. For the full fields the same steps apply: with no source and no boundary field strength the flux at the cutoff is $-\partial_t(L^3\rho_\vev^2)$, which the boundary term of $E$ compensates, and the radial current $\sqrt{-g}\,g^{uu}\rho^2B_u$ vanishes there, so with a regular center in place of a horizon $E$ and $Q$ are conserved.

\section{Near-critical expansion}
\label{app:nearTc}

This appendix gives the expansion quoted in section~\ref{sec:structure} and used in sections~\ref{sec:thermoplanar} and~\ref{sec:sound} and appendix~\ref{app:attenuation}, and compares it with the exact solution.

\paragraph{Expansion.} Near $\Tc$ the fields are expanded in the amplitude of the condensate, for general $\lambda$ and in units $e=L=u_h=1$, so that $\xi=u^2$ and $f=1-\xi^2$. The static, homogeneous equations take the form
\begin{equation}
\frac{-4\rho+\lambda\rho^3}{\xi}=4\xi^2\partial_\xi\Big(\frac{f}{\xi}\,\partial_\xi\rho\Big)+\frac{B_t^2}{f}\,\rho,\qquad
\frac{\rho^2B_t}{\xi}=4\xi f\,\partial_\xi^2B_t,
\label{eq:static}
\end{equation}
with $B_t(1)=0$, $\rho(1)$ finite and $\Psi^{(b)}=0$. About the normal phase $B_t=\mu(1-\xi)$,
\begin{equation}
B_t=\mu_c(1-\xi)+\epsilon^2B_2+O(\epsilon^4),\qquad\rho=\epsilon\rho_1+\epsilon^3\rho_3+O(\epsilon^5).
\end{equation}
Regular, source-free solutions $\rho_1$ are known in closed form at $\mu_c=2$, $6$, $10$, \ldots\ \cite{Herzog2010}; by the Sturm oscillation theorem only the one at $\mu_c=2$ is nodeless. To order $\epsilon^2$
\begin{equation}
\rho_1=\frac{\xi}{1+\xi}\,\Sigma,\qquad B_2=\delta\mu_2(1-\xi)-\frac{\xi(1-\xi)}{8(1+\xi)}\,\Sigma^2,
\label{eq:rho1}
\end{equation}
with $\Sigma$ and $\delta\mu_2$ free; normalizing $\epsilon$ as the exact coefficient of $u^2$ in $\rho$ for $\Sigma=1$, the higher orders start at $O(\xi^2)$. At order $\epsilon^3$,
\begin{equation}
D_\xi\rho_3=\frac{\xi\,(\nu_1\xi^2+\nu_2\xi+\nu_3)}{2(1+\xi)^3},\qquad
D_\xi\equiv-4\Big[\frac1\xi+\frac{1-\xi}{1+\xi}-(1+\xi^2)\partial_\xi+\xi(1-\xi^2)\partial_\xi^2\Big],
\end{equation}
where $D_\xi\rho_1=0$, $\nu_1=\Sigma^3-8\delta\mu_2\Sigma$, $\nu_2=-(1+2\lambda)\Sigma^3$ and $\nu_3=8\delta\mu_2\Sigma$. Since $\xi^{-2}D_\xi$ is symmetric with respect to $d\xi$, a regular, source-free $\rho_3$ exists only if $\nu_1+2\nu_2+7\nu_3=0$, that is
\begin{equation}
\delta\mu_2\,\Sigma-\frac{1+4\lambda}{48}\,\Sigma^3=0 .
\label{eq:solvability}
\end{equation}
Here $\mu-\mu_c=\mu(1-T/\Tc)=\epsilon^2\delta\mu_2+O(\epsilon^4)$, and a homogeneous condensate has $\Sigma=1$ and $\delta\mu_2=\mu_2^{\rm hom}\equiv(1+4\lambda)/48$. A slow dependence on the boundary coordinates adds $\frac12\nabla^2\Sigma$ to \eqref{eq:solvability}, which for $\psi\equiv|\langle O_2\rangle|=\epsilon\Sigma/\sqrt2+O(\epsilon^3)$ becomes the static Ginzburg--Landau equation
\begin{equation}
\frac12\nabla^2\psi+\epsilon^2\delta\mu_2\,\psi-\frac{1+4\lambda}{24}\,\psi^3=O(\epsilon^5),
\end{equation}
with coherence length $(2\epsilon^2\delta\mu_2)^{-1/2}\propto(1-T/\Tc)^{-1/2}$, in agreement with the dual Ginzburg--Landau theory of \cite{NatsuumeOkamura2018,NatsuumeOkamura2022,Natsuume2025}; the last gives its quartic coefficient for the model of \cite{Herzog2010} (section~\ref{sec:structure}) as $(1+4\CA+4\CB)/96$ in the same normalization of the order parameter.

\paragraph{Amplitude relaxation.} At $\vk=0$ the expansion extends to a slow time dependence through $\epsilon^2v$, with $v$ the ingoing coordinate of \eqref{eq:ingoing}; a dot denotes $d/d(\epsilon^2v)$. In these coordinates the metric is $u^{-2}(-f\,dv^2-2\,dv\,du+d\vec x^{\,2})$ and the ingoing condition is regularity at the horizon; the scalar equation in \eqref{eq:reduced} acquires $-2u^2\partial_v\partial_u\rho+3u\,\partial_v\rho$, and the radial component of the Maxwell equation reads $u^2\partial_vF_{uv}=\rho^2(fB_u-B_v)$. At order $\epsilon^2$, $B_v$ is $B_2$ of \eqref{eq:rho1}, with $\Sigma$ and $\delta\mu_2$ now time dependent, plus $h\,\xi$, where $h$ is its value at the horizon, zero for a static state; the charge density is $n=2\mu_c+\epsilon^2n_2$ with $n_2=2\delta\mu_2-2h+\frac14\Sigma^2$. At the horizon the radial equation gives $\partial_vE_h=-\rho_h^{\,2}B_v|_{u_h}$, with $E_h=F_{uv}|_{u_h}$. No charge enters at $\vk=0$, so with the Gauss law $E_h=-n+\frac38\epsilon^2\Sigma^2$ this fixes $h=-3\dot\Sigma/\Sigma$. At order $\epsilon^3$ the time derivatives add a total derivative to the solvability integral, evaluated at the horizon, and \eqref{eq:solvability} becomes $\frac12\dot\Sigma=(\delta\mu_2+\frac12h)\Sigma-\frac{1+4\lambda}{48}\Sigma^3$. At fixed $n$ this is
\begin{equation*}
5\,\dot\Sigma=\frac{n_2}{2}\,\Sigma-\frac{7+4\lambda}{48}\,\Sigma^3,
\end{equation*}
whose linearization about $\Sigma=1$ gives $\gamma=(7+4\lambda)\,\epsilon^2/120$, that is, with units restored, $\gamma=\frac25(\chi/\chi_\N)(\mu-\mu_c)$. This is the amplitude equation of the model F of \cite{Natsuume2026} at fixed density, with the relaxation coefficient found there, which does not depend on $\lambda$, and the fixed-density quartic coefficient $(7+4\lambda)/96$ of section~\ref{sec:sound}, which the bulk derivation thus confirms. The same rate is the pole at $\vk=0$ of the near-critical correlators of \cite{Natsuume2026b} at $\CA=0$, $\CB=\lambda$; here it follows directly from the solvability condition at fixed density.

\paragraph{Relation to the exact solution.} In the model of \cite{Herzog2010}, the homogeneous condensate has $\mu_2^{\rm hom}=(1+4\CA+4\CB)/48$ and the third-order term \eqref{eq:rho3}. At $\lambda=\frac12$, with $\epsilon=2\sqrt2\,[(\Tc/T)^4-1]^{1/2}$ the exact coefficient of $u^2$ in \eqref{eq:solution}, the solution has $\mu=2(1+\epsilon^2/8)^{1/4}$ and $\rho_3=-\xi^2/[16(1+\xi)^2]$, in agreement with \eqref{eq:rho3}. At fourth order it reproduces the shift $-3\epsilon^4/1024$ of the chemical potential and the free energy of condensation $-(1+4\CA+4\CB)\,\epsilon^4/384$ of \cite{Herzog2010}, which at $\CA=0$, $\CB=\frac12$ is $\Om_\SF-\Om_\N=-\epsilon^4/128$. Eq.~\eqref{eq:dOmega} is the exact homogeneous free energy whose quartic expansion in $\psi$ is the Ginzburg--Landau potential above.

\section{Low-frequency response at zero temperature}
\label{app:T0cond}

At $T=0$ and $\vk=0$, with $\hat\xi=\mu^2u^2/4$ and $\hat\omega=\omega/\mu$, eq.~\eqref{eq:transverse} for $b\equiv b^\perp_x$ becomes
\begin{equation}
\hat\xi\,\partial_{\hat\xi}^2b+\hat\omega^2\,b-\frac{2\hat\xi}{(1+\hat\xi)^2}\,b=0,
\label{eq:Heun}
\end{equation}
an equation of confluent Heun type, with regular singular points at $\hat\xi=0,-1$ and an irregular one at infinity. For $\hat\omega\ll1$ an inner solution ($u\ll1/\omega$) is matched to an outer one ($u\gg1/\mu$).

\paragraph{Inner region.} At $\hat\omega=0$ the solutions are $y_1=1/(1+\hat\xi)$ and $y_2=(1+\hat\xi)^2$, with Wronskian $3$. To first order, $b=C_1\big(y_1+\hat\omega^2y_1^{(1)}\big)+C_2y_2$, and variation of parameters gives, up to terms that only redefine $C_1$ and $C_2$,
\begin{equation}
y_1^{(1)}=\frac13\Big[\frac{\ln\hat\xi+\hat\xi}{1+\hat\xi}-(1+\hat\xi)^2\Big(\ln\frac{\hat\xi}{1+\hat\xi}+\frac{1}{1+\hat\xi}\Big)\Big]
=\begin{cases}-\frac13-\hat\xi\ln\hat\xi+O(\hat\xi),&\hat\xi\to0,\\[2pt]
\frac{\ln\hat\xi}{3\hat\xi}+\frac12+O(\hat\xi^{-1}),&\hat\xi\to\infty.\end{cases}
\end{equation}
The small-$\hat\xi$ form reproduces the term $-\frac12\omega^2u^2\ln u$ of section~\ref{sec:conductivity}, and $A^{(b)}_x=C_1(1-\frac13\hat\omega^2)+C_2$.

\paragraph{Outer region and matching.} For $u\gg1/\mu$ the outgoing solution is $b=uH^{(1)}_3(\omega u)$, up to corrections in powers of $1/(\mu u)^2$. It carries the flux $\mathcal{F}=2/\pi$, and for $\omega u\ll1$ it is $-16i/(\pi\omega^3u^2)-2i/(\pi\omega)+O(u^2)$. The $u^{-2}$ terms match for $C_1=-4i\mu^2/(\pi\omega^3)$, and the inner constant $\frac12\hat\omega^2C_1$ then reproduces $-2i/(\pi\omega)$; the $u^4$ terms give a negligible $C_2=O(\omega^3/\mu^4)$. The inner term $\frac13\hat\omega^2C_1\ln\hat\xi/\hat\xi$ contains $\ln\hat\xi=2\ln(\mu/\omega)+2\ln(\omega u/2)$, while the outer solution, whose logarithms are $\ln(\omega u)$, depends on $\mu$ only through powers of $1/(\mu u)^2=(\omega/\mu)^2/(\omega u)^2$ and so contains no $\ln\mu$. The $\ln(\mu/\omega)$ term is therefore absorbed by the normalization:
\begin{equation}
C_1=-\frac{4i\mu^2}{\pi\omega^3}\Big[1-\frac23\hat\omega^2\ln\frac1{\hat\omega}+O(\hat\omega^2)\Big],\qquad
{\rm Re}\,\sigma=\frac{L}{e^2}\,\frac{2/\pi}{\omega\,|A^{(b)}_x|^2}=\frac{\pi L}{8e^2}\,\frac{\omega^5}{\mu^4}\Big[1+\frac43\hat\omega^2\ln\frac1{\hat\omega}+O(\hat\omega^2)\Big],
\end{equation}
using \eqref{eq:flux} and $A^{(b)}_x=C_1[1+O(\hat\omega^2)]$. This is \eqref{eq:T0cond}.

\paragraph{Leading logarithms.} For $|\hat\xi|>1$, eq.~\eqref{eq:Heun} has no singular point, so it has a solution $\hat\xi^{-1+\nu}$ times a Laurent series in $\hat\xi$ with constant term $1$, where $e^{2\pi i\nu}$ is an eigenvalue of its monodromy around $\hat\xi=0$ in the region $|\hat\xi|>1$. The exponent $\nu$ and the coefficients of the series are power series in $\hat\omega^2$, and $y_1+\hat\omega^2y_1^{(1)}$ gives $\nu=\frac13\hat\omega^2+O(\hat\omega^4)$. Since the inner solutions are power series in $\hat\omega^2$ at fixed $\hat\xi$, and the outer ones at fixed $\omega u$, the matching produces logarithms of $\mu/\omega$ only through $\hat\xi^{\nu}=(\mu/2\omega)^{2\nu}(\omega u)^{2\nu}$, until the solution that grows as $u^4$ adds one of its own at relative order $\hat\omega^6$. The conductivity is therefore
\begin{equation}
{\rm Re}\,\sigma=\frac{\pi L}{8e^2}\,\frac{\omega^5}{\mu^4}\Big(\frac{\mu}{2\omega}\Big)^{4\nu}\big[1+O(\hat\omega^2)\big],
\label{eq:T0resum}
\end{equation}
with no logarithm in the bracket through order $\hat\omega^4$. The leading logarithms exponentiate, as the low-frequency logarithms of black-hole perturbations do in the renormalized angular momentum of \cite{MST1996}, and the coefficient of $\hat\omega^4\ln^2(1/\hat\omega)$ is $\frac12(\frac43)^2=\frac89$.

\paragraph{Amplitude channel.} At $\vk=0$ the amplitude couples to $b_t$, while $b_u$ follows from the $u$-component of \eqref{eq:b}. With $\tilde b\equiv b_t/\mu$ and $\tilde\tau\equiv eL\tau/(\sqrt2\,\hat\xi)$ the static equations read
\begin{equation}
\partial_{\hat\xi}^2\tilde b=\frac{2(\tilde b+\tilde\tau)}{(1+\hat\xi)^2},\qquad(\hat\xi\partial_{\hat\xi})^2\tilde\tau=\frac{\hat\xi(3\hat\xi-1)\tilde\tau-4\hat\xi\tilde b}{(1+\hat\xi)^2},
\label{eq:ampstatic}
\end{equation}
and for two static solutions the combination $2(\tilde b_1\partial_{\hat\xi}\tilde b_2-\tilde b_2\partial_{\hat\xi}\tilde b_1)-\hat\xi(\tilde\tau_1\partial_{\hat\xi}\tilde\tau_2-\tilde\tau_2\partial_{\hat\xi}\tilde\tau_1)$ is constant. Two solutions are rational: $\Upsilon_-=\mu\,\partial_\mu(\Bb_t,\rhob)$, with $(\tilde b,\tilde\tau)=(1-\hat\xi,4)/(1+\hat\xi)^2$, which decays as $\hat\xi^{-1}$, and $\Upsilon_+=(\hat\xi^4+2\hat\xi^3-3,\,2(1+\hat\xi)^3-8)/(1+\hat\xi)^2$, which grows as $\hat\xi^2$; with $\Upsilon_-$ and $\Upsilon_+$ as solutions $1$ and $2$ the combination is $-6$. The other two behave as $\hat\xi^{\pm\sqrt3}$ for large $\hat\xi$. The source of $O_2$ is the coefficient of $\ln\hat\xi$ in $\tilde\tau$. At $\vk=0$ the Ward identity ties the change of the charge density to the source of the phase, which is zero here, so $\partial_{\hat\xi}\tilde b=0$ at the boundary. The solution with $\tilde\tau\simeq\ln\hat\xi$ that decays in the interior, taken as solution $1$, gives $0$ with $\Upsilon_-$, which fixes $\tilde b=\frac23$ at the boundary, and then $2$ with $\Upsilon_+$. It is therefore $-\frac13\Upsilon_-$ plus a solution whose combinations with $\Upsilon_-$ and $\Upsilon_+$ vanish. These solutions satisfy $\tilde\tau=(1+\hat\xi)[\tilde b-(\hat\xi-1)\partial_{\hat\xi}\tilde b]/(2\hat\xi)$ and the equation
\begin{equation}
\hat\xi(1+\hat\xi)^2\partial_{\hat\xi}^2\tilde b+(\hat\xi^2-1)\partial_{\hat\xi}\tilde b-(3\hat\xi+1)\tilde b=0,
\end{equation}
whose solutions are $(1+\hat\xi)^{-2}$ times hypergeometric functions of $-\hat\xi$. The one that decays, $\tilde b=\hat\xi^{2-\sqrt3}(1+\hat\xi)^{-2}\,{}_2F_1(\sqrt3-2,\sqrt3;1+2\sqrt3;-1/\hat\xi)$, has $\tilde\tau\simeq\Gamma(1+2\sqrt3)\ln\hat\xi/[\Gamma(\sqrt3)\,\Gamma(3+\sqrt3)]$ at the boundary, by the connection formula, and $\tilde\tau\simeq\frac12(1+\sqrt3)\,\hat\xi^{-\sqrt3}$ in the interior. In the interior the solution with $\tilde\tau\simeq\ln\hat\xi$ is therefore
\begin{equation}
\tilde b\simeq\frac{1}{3\hat\xi},\qquad\tilde\tau\simeq\frac{(1+\sqrt3)\,\Gamma(\sqrt3)\,\Gamma(3+\sqrt3)}{2\,\Gamma(1+2\sqrt3)}\,\hat\xi^{-\sqrt3}=\frac{(2+\sqrt3)^2\,\Gamma(\sqrt3)^2}{2\,\Gamma(2\sqrt3)}\,\hat\xi^{-\sqrt3}.
\end{equation}
For $u\gg1/\mu$ the outgoing solutions are $\tau\propto u^2H^{(1)}_{2\sqrt3}(\omega u)$ and $b_u\propto u^2H^{(1)}_3(\omega u)$, up to corrections in powers of $1/(\mu u)$. Matching their decaying terms to these interior coefficients, with the fluxes given by Wronskians of Hankel functions, gives the coefficients of \eqref{eq:T0amp}: for the vector the square of $\frac13/\Gamma(3)$, and for the scalar the square of the ratio of the coefficient of $\hat\xi^{-\sqrt3}$ above to $\Gamma(2\sqrt3)$. The conductivity has the same form, ${\rm Re}\,\sigma/\sigma_\infty=[1/\Gamma(3)]^2\hat\omega^4$, from $y_1\simeq\hat\xi^{-1}$.

\section{Sound attenuation, pole skipping and axion branes}
\label{app:attenuation}

This appendix derives \eqref{eq:horizonGamma} and evaluates it on the planar brane, in units $e=L=u_h=1$, so that $\pi T=1$, $\xi=u^2$ and $f=1-\xi^2$, with $\Th\equiv(\Tc/T)^2$; its last two paragraphs give the pole-skipping points and, with units restored, the transport on the axion branes. The longitudinal perturbations are $\propto e^{-i\omega t+ikx}$ with $\omega=O(k)$, so the expansion is in $k$; the mode is normalized by the boundary value $\theta_0$ of the Goldstone field.

\paragraph{Radial equations.} With the radial electric fluxes $\Pi_t\equiv f_{ut}/u$ and $\Pi_x\equiv f\,f_{ux}/u$, which tend to $-\delta n$ and $\delta\langle J_x\rangle$ at the boundary (appendix~\ref{app:asymptotics}), the $t$-, $x$- and $u$-components of \eqref{eq:b} read
\begin{gather}
u\,\partial_u\Pi_t=\frac{\rhob^2b_t+2\rhob\Bb_t\tau}{u^2f}+\frac kf\,(\omega b_x+kb_t),\qquad
u\,\partial_u\Pi_x=\frac{\rhob^2b_x}{u^2}-\frac\omega f\,(\omega b_x+kb_t),\\
C\equiv i\omega\Pi_t+ik\Pi_x=-\frac{f\rhob^2}{u^3}\,b_u,\qquad
u\,\partial_uC=i\omega\,\frac{\rhob^2b_t+2\rhob\Bb_t\tau}{u^2f}+ik\,\frac{\rhob^2b_x}{u^2}.
\label{eq:Cflux}
\end{gather}
The last equation is the constraint \eqref{eq:constraint}. At the boundary $C=-i\omega\,\delta n+ik\,\delta\langle J_x\rangle$ vanishes for a normal mode, by charge conservation, and at the horizon \eqref{eq:ingoing} gives $fb_u\to-b_t$, and hence the horizon value $C(u_h)=\rhob_h^{\,2}\,b_t(u_h)$.

\paragraph{Leading order and energy.} At first order in $k$ the equations for $b_t$, $\tau$ and $b_x$ are static. With the conditions \eqref{eq:bc}, regularity and the boundary values $b_t\to i\omega\theta_0$ and $b_x\to-ik\theta_0$,
\begin{equation}
\tau=\delta\mu\,\partial\rhob/\partial\mu,\qquad b_t=\delta\mu\,\partial\Bb_t/\partial\mu,\qquad b_x=-ik\theta_0\,Y(u),\qquad\delta\mu=i\omega\theta_0,
\end{equation}
with derivatives at fixed $T$, and $Y$ the static London profile with $Y(0)=1$, eq.~\eqref{eq:London}, which on the exact solution is $Y=(1+\xi/\Th)/(1+\Th\xi)$. Then $\delta n=\chi\,\delta\mu$ and $\delta\langle J_x\rangle=-(n_s/\mu)\,b_x(0)$, so charge conservation is the dispersion relation $\chi\omega^2=(n_s/\mu)k^2$. The energy \eqref{eq:E2positive} is of order $k^2$, since the terms with $\dot\tau$, $f_{tx}$ and $b_u$ are of higher order. Integrating by parts with the static radial equations, the London profile contributes $-\frac12b_x\Pi_x|_{u\to0}=\frac12(n_s/\mu)\,b_x(0)^2$ and the $\delta\mu$ perturbation $-\frac12b_t\Pi_t|_{u\to0}=\frac12\chi\,\delta\mu^2$; the mixed terms $\mp\int du\,\rhob\Bb_t\tau b_t/(u^3f)$, from the electric and Proca terms and from the scalar terms with $\ee_\bdry$, cancel. Averaged over a period,
\begin{equation}
\langle\ee_2\rangle=\frac14\Big(\chi\omega^2+\frac{n_s}{\mu}k^2\Big)|\theta_0|^2=\frac12\,\frac{n_s}{\mu}\,k^2|\theta_0|^2 .
\label{eq:Eavg}
\end{equation}

\paragraph{Horizon flux.} For ingoing fields $\partial_u\simeq-f^{-1}\partial_t$, $f_{ux}\simeq-f^{-1}f_{tx}$ and $fb_u\simeq-b_t$ near the horizon, so the flux \eqref{eq:Su} there is $S^u=(\partial_t\tau)^2+f_{tx}^2+\rhob_h^{\,2}b_t^2$, each term of order $k^4$. The leading-order fields give $\partial_t\tau=\omega^2\theta_0\,\partial\rhob_h/\partial\mu$ and $f_{tx}=-\omega k\theta_0Y_h$, with $Y_h\equiv Y(u_h)$. Since $\partial\Bb_t/\partial\mu$ vanishes at the horizon, $b_t(u_h)$ is needed at order $k^2$. Integrating the equation for $C$ from the boundary, with the leading-order fields and the static equations for $\Pi_t$ and $\Pi_x$,
\begin{equation}
C(u_h)=i\omega\big[\Pi_t(u_h)-\Pi_t(0)\big]+ik\big[\Pi_x(u_h)-\Pi_x(0)\big]=i\omega\Pi_t(u_h)-\Big(\chi\omega^2-\frac{n_s}{\mu}k^2\Big)\theta_0,
\end{equation}
where the bracket vanishes by the dispersion relation, and $\Pi_x(u_h)=0$, $\Pi_t(u_h)=\delta\mu\,\partial E_h/\partial\mu$, so that $C(u_h)=i\omega\,\delta\mu\,\partial E_h/\partial\mu$. Hence $b_t(u_h)=-\omega^2\theta_0\,\rhob_h^{-2}\,\partial E_h/\partial\mu$, and
\begin{equation}
\langle S^u\rangle_{u_h}=\frac12|\theta_0|^2\Big[\omega^4(\partial\rhob_h/\partial\mu)^2+\omega^2k^2Y_h^2+\omega^4\,\frac{(\partial E_h/\partial\mu)^2}{\rhob_h^{\,2}}\Big].
\end{equation}

\paragraph{Result.} No energy flows through the boundary (appendix~\ref{app:energy}), so $2\Gamma k^2\langle\ee_2\rangle=\langle S^u\rangle_{u_h}$, which with $\omega^2=c_s^2k^2$, $c_s^2=(n_s/\mu)/\chi$ and $\sigma_\reg=Y_h^2$ (section~\ref{sec:conductivity}) is \eqref{eq:horizonGamma} in these units. The derivation uses $f$ only through the ingoing condition, the horizon area density and the static solutions, so the formula holds for any $f$ with flat spatial sections, such as the axion brane of section~\ref{sec:soundaxion}. The energy balance is used at leading order in $k$; the near-horizon divergence of the energy of a decaying mode first appears at order $k^4$, in the terms with $|g^{tt}|$ or $g^{uu}$ (for example $|g^{tt}|g^{xx}f_{tx}^2$), and does not affect \eqref{eq:Eavg}.

\paragraph{Exact solution.} With $\rhob=2\sqrt2\sqrt{\Th^2-1}\,\xi/(1+\Th\xi)$, $\Bb_t=2\sqrt\Th\,(1-\xi^2)/(1+\Th\xi)$, $\mu=2\sqrt\Th$ and the derivative $\partial/\partial\mu=\sqrt\Th\,\partial/\partial\Th$,
\begin{gather}
\rhob_h=2\sqrt2\,\sqrt{\frac{\Th-1}{\Th+1}},\qquad\frac{\partial\rhob_h}{\partial\mu}=\frac{\sqrt\Th\,\rhob_h}{\Th^2-1},\qquad
E_h=-\frac{8\sqrt\Th}{1+\Th},\qquad\frac{\partial E_h}{\partial\mu}=\frac{4(\Th-1)}{(1+\Th)^2},\nonumber\\
(\partial\rhob_h/\partial\mu)^2+\frac{(\partial E_h/\partial\mu)^2}{\rhob_h^{\,2}}=\frac{2}{\Th^2-1}.
\label{eq:horizondata}
\end{gather}
With $\chi=6\Th$, $n_s/\mu=2(\Th^2-1)/\Th$ and $Y_h=1/\Th$, \eqref{eq:horizonGamma} gives \eqref{eq:Gamma}, $\Gamma=5/(36\Th^3)$: the normal current gives $1/(12\Th^3)$, $3/5$ of it at every temperature, and the horizon term $1/(18\Th^3)$.

\paragraph{Near $\Tc$, general $\lambda$.} On the background of appendix~\ref{app:nearTc} with $\Sigma=1$, that is $\rhob=\epsilon\,\xi/(1+\xi)$, $\mu=2+\epsilon^2\mu_2^{\rm hom}$ and $\Bb_t=\mu(1-\xi)-\epsilon^2\xi(1-\xi)/[8(1+\xi)]$, one finds to leading order in $\epsilon$: $\rhob_h=\epsilon/2$, $E_h=-2\mu+\epsilon^2/8$, $n=2\mu+\epsilon^2/4$, $Y_h=1$ and $n_s/\mu=\epsilon^2\int_0^1d\xi\,[2(1+\xi)^2]^{-1}=\epsilon^2/4$. With $\partial\epsilon^2/\partial\mu=1/\mu_2^{\rm hom}$, this gives
\begin{gather}
\chi=\frac{2(7+4\lambda)}{1+4\lambda},\qquad(\partial\rhob_h/\partial\mu)^2=\frac{1}{16\epsilon^2(\mu_2^{\rm hom})^2},\qquad
\frac{(\partial E_h/\partial\mu)^2}{\rhob_h^{\,2}}=\frac{(1-16\mu_2^{\rm hom})^2}{16\epsilon^2(\mu_2^{\rm hom})^2},\nonumber\\
\Gamma=\frac{1+4\lambda}{4(7+4\lambda)}+\frac{9+(2-4\lambda)^2}{2(7+4\lambda)^2},
\end{gather}
which is \eqref{eq:GammaTc}; and $c_s^2=(n_s/\mu)/\chi$ reproduces the sound speed of \cite{Herzog2010} at $\CA=0$, $\CB=\lambda$. With units restored, at $\lambda=0$ eq.~\eqref{eq:zeta3} gives
\begin{equation}
\zeta_3\simeq\frac{e^2}{L}\,\frac{13}{588\,\pi^2T^2(\mu-2\pi T)},
\end{equation}
which reproduces the leading near-critical value found for this model in \cite{HLSY2011}, once the normalizations are matched (their $\zeta_3$ is ours over $q^2$, and $e^2=2\kappa_5^2q^2$ with their charge $q$).

\paragraph{Relation to the horizon formulas of Donos and Kailidis.} Eqs.~\eqref{eq:horizonGamma} and \eqref{eq:zeta3} are the probe-limit, AdS$_5$ form of the horizon formulas of \cite{DonosKailidis2022a} for the incoherent conductivity and the third bulk viscosity, whose AdS$_4$ derivation carries over to AdS$_5$. In their notation the horizon area density $s/4\pi$ becomes $(L/u_h)^3=1$, the derivatives with respect to their charge density at fixed $s$ become $\chi^{-1}\partial/\partial\mu$ at fixed $T$, the horizon scalar $\rho^{(0)}$, with the same kinetic normalization $\frac12(\partial\rho)^2$, is $\rhob_h$, and the horizon charge density $e^2g^{(0)}\tau^{(0)}a^{(0)}$ becomes $-E_h$, with $q_e=1$ and $\tau^{\rm DK}=1/e^2=1$; their neutral scalar is absent. Their zero-density limit has the coefficient $2\pi$ in the horizon-charge term, appropriate to the normalization of the scalar in \cite{DKP2021}, which differs by $\sqrt2$; in the normalization used here it is the $4\pi$ of their horizon formula at finite density, consistent with \eqref{eq:zeta3}, with the quasinormal modes and with the value $33/196$ of \cite{Herzog2010}.

\paragraph{Pole-skipping points.} Near the horizon the ingoing solution of the longitudinal equations fails to be unique at $\omega=0$ and at $\omega=-2i$, that is, $\omega=-2\pi iT$, for special values of $k$ \cite{NatsuumeOkamura2026}. The residues of their expansion, whose $\rho$ is $\hat\rho=\rhob/\sqrt2$, fix these values in terms of $\rhob_h$ and $E_h$ once the quartic term is added. By \eqref{eq:horizondata}, $\hat\rho_h^{\,2}=4(\Th-1)/(\Th+1)$ and $E_h^{\,2}=16-\hat\rho_h^{\,4}$, so the values are $k^2=-2\hat\rho_h^{\,2}$ at $\omega=0$ and, at $\omega=-2i$, the three roots of
\[
k^6+2(1+4\hat\rho_h^{\,2})\,k^4+4(3+2\hat\rho_h^{\,2}+4\hat\rho_h^{\,4})\,k^2+8(1+\hat\rho_h^{\,2})(\hat\rho_h^{\,4}-\hat\rho_h^{\,2}-5)=0.
\]
At $\Tc$ the roots are $k^2=2$ and $k^2=-2\pm4i$, the values of the normal phase. Near $\Tc$ the point at $\omega=0$ is $k^2\simeq-\epsilon^2/4$, as in \cite{NatsuumeOkamura2026,Natsuume2026b}, whose condensate is $\epsilon/\sqrt2$. As the temperature falls, the root that starts at $k^2=2$ reaches $k=0$ at $\hat\rho_h^{\,4}-\hat\rho_h^{\,2}-5=0$, that is, $7\Th^2-42\Th+15=0$, with the solution $\Th=3+4\sqrt{21}/7$ above one. A mode at $k=0$ then has $\omega=-2i$. Appendix~\ref{app:numerics} identifies it as the amplitude mode, so $\gamma=2\pi T$ at this temperature.

\paragraph{Axion branes.} On the axion brane of section~\ref{sec:axion}, whose metric is homogeneous and isotropic, the London profile is again rational, $Y(u)=1-(e^2n_s/2L\mu)\,u^2/(1+\mu^2u^2/4)$, and the derivations of \eqref{eq:sigmaDC} and of \eqref{eq:horizonGamma} use the blackening function only through the ingoing behavior and the static solutions. With $n=(L/e^2)\,\mu(\mu^2+\alpha^2)/2$ and $\chi=(L/e^2)(3\mu^2+\alpha^2)/2$,
\begin{gather}
\frac{n_s}{\mu}=\frac{4\langle O_2\rangle^2}{2\mu^2+\alpha^2},
\label{eq:nsaxion}\\
\frac{\sigma_\reg}{\sigma_\N(0)}=Y(u_h)^2=\Big(\frac{8\pi T}{u_h\,(2\mu^2+\alpha^2)}\Big)^2,
\label{eq:sigmaDCaxion}\\
c_s^2=\frac{4p^2}{(2\mu^2+\alpha^2)(3\mu^2+\alpha^2)},\qquad
\Gamma=\frac{64\pi^2T^2\,(5\mu^2+2\alpha^2)}{u_h^3\,(2\mu^2+\alpha^2)^2\,(3\mu^2+\alpha^2)^2},
\label{eq:Gammaaxion}
\end{gather}
with $\sigma_\N(0)=L/e^2u_h$; at $\alpha=0$ these are \eqref{eq:ns}, \eqref{eq:sigmaDC}, \eqref{eq:cs} and \eqref{eq:Gamma}. The relation $\mu(n-n_s)=Ts$ of section~\ref{sec:transverse} fails by terms of order $\alpha^2$: at $T=0$, $\mu(n-n_s)=L\mu^2\alpha^2/4e^2$, while $Ts$ vanishes. Longitudinal quasinormal modes reproduce $c_s^2$ and $\Gamma$ (appendix~\ref{app:numerics}).

\section{Perturbations on the sphere}
\label{app:perturbations}

This appendix uses the perturbations $\tau$ and $b_a$ and the quadratic Lagrangian \eqref{eq:L2}, with $L=R=1$ and $e$ absorbed into the scalar, $eL\rho\to\rho$, so that its factor $1/e^2$ is absent. In the coordinates $ds^2=\sec^2\zeta\,(-dt^2+d\zeta^2+\sin^2\zeta\,d\Omega_3^2)$ of global AdS$_5$, with $u=\cot\zeta$ and $r=1/u=\tan\zeta$, the background is
\begin{equation}
\rhob=\frac{p\cos^2\zeta}{\sin^2\zeta+K\cos^2\zeta},\qquad \Bb_t=\frac{\mu}{\sin^2\zeta+K\cos^2\zeta},\qquad K=\frac{\mu^2}{4},\quad p^2=\frac{\mu^2(\mu^2-4)}{2}.
\end{equation}
The perturbations are expanded in harmonics on the unit $S^3$ with time dependence $e^{-i\omega t}$.

\paragraph{Transverse vector harmonics.} Let $\mathcal{V}_i$ be a transverse vector harmonic on the unit $S^3$,
\[
\nabla^i\mathcal{V}_i=0,\qquad\nabla^j(\nabla_j\mathcal{V}_i-\nabla_i\mathcal{V}_j)=-(\ell+1)^2\mathcal{V}_i,\qquad\ell\ge1.
\]
With $b_i=b(\zeta)\,\mathcal{V}_i\,e^{-i\omega t}$ all other components vanish and
\begin{equation}
\frac{1}{\tan\zeta}\,\partial_\zeta\big(\tan\zeta\,\partial_\zeta b\big)+\Big[\omega^2-\frac{(\ell+1)^2}{\sin^2\zeta}-\frac{\rhob^2}{\cos^2\zeta}\Big]b=0 .
\label{eq:transverseS3}
\end{equation}
A normal mode behaves as $b\propto\zeta^{\ell+1}$ at the center and has no source at the boundary, where $b\propto\cos^2\zeta$. With $\rhob=0$ this is the Maxwell field in AdS$_5$, with $\omega=3+\ell+2n$, and with $\rhob^2\equiv8$ a Proca field of mass squared $8$, with $\omega=5+\ell+2n$; the normal modes are $b=\sin^{\ell+1}\zeta\cos^2\zeta\,P_n^{(\ell+1,1)}(\cos2\zeta)$ and $b=\sin^{\ell+1}\zeta\cos^4\zeta\,P_n^{(\ell+1,3)}(\cos2\zeta)$, respectively, with $P_n^{(a,b)}$ the Jacobi polynomials. With $\rhob^2\equiv8$, $\sqrt{\tan\zeta}\,b$ obeys the P\"oschl--Teller equation
\begin{equation*}
-\partial_\zeta^2\big(\sqrt{\tan\zeta}\,b\big)+\mathcal{P}\,\sqrt{\tan\zeta}\,b=\omega^2\sqrt{\tan\zeta}\,b,\qquad\mathcal{P}=\frac{(\ell+\frac12)(\ell+\frac32)}{\sin^2\zeta}+\frac{\frac72\cdot\frac52}{\cos^2\zeta},
\end{equation*}
whose eigenvalues $\omega=\ell+\frac32+\frac72+2n$ are linear in the indices $\ell+\frac32$ and $\frac72$ of its two terms. In these modes, with the weight $\tan\zeta$ for $b$, the Hellmann--Feynman theorem in the coefficient of $\sec^2\zeta$ gives $\langle\sec^2\zeta\rangle=2\omega/(2\cdot\frac72-1)=\omega/3$, and the hypervirial relation $\frac12\langle f'''\rangle+2\langle f'(\omega^2-\mathcal{P})\rangle=\langle f\mathcal{P}'\rangle$ with $f=\tan\zeta$, whose boundary terms vanish because both indices exceed $\frac32$, gives $32\langle\sec^4\zeta\rangle=2[\omega^2-(\ell+1)^2+8]\langle\sec^2\zeta\rangle$. The perturbation $-8\sec^2\zeta\,(1+2\tan^2\zeta)/K$ of $W$ in section~\ref{sec:transverseS3} therefore shifts $\omega^2$ by $-(8/K)\big(2\langle\sec^4\zeta\rangle-\langle\sec^2\zeta\rangle\big)=-\omega[\omega^2-(\ell+1)^2]/3K$, which with $K=\mu^2/4$ is \eqref{eq:transverseshift} for all $\ell$ and $n$. Near the onset $p^2=\mu^2(\mu^2-4)/2$ vanishes linearly, so that $\rhob^2/\cos^2\zeta=8(\mu-2)\cos^2\zeta+O\big((\mu-2)^2\big)$, and in the Maxwell modes, with the weight $\tan\zeta$, $\partial_\mu(\omega^2)=8\langle\cos^2\zeta\rangle$ at $\mu=2$. In $x=\cos2\zeta$, $\tan\zeta\,b^2\,d\zeta$ is proportional to $(1-x)^{\ell+1}(1+x)\,P_n^{(\ell+1,1)}(x)^2\,dx$, so $\langle x\rangle$ is the diagonal coefficient of the three-term recurrence of these Jacobi polynomials, $-\ell(\ell+2)/[(2n+\ell+2)(2n+\ell+4)]$, and $\langle\cos^2\zeta\rangle=(1+\langle x\rangle)/2=2(n+1)(n+\ell+2)/[(2n+\ell+2)(2n+\ell+4)]$; with $\omega=3+\ell+2n$, $d\omega/d\mu=4\langle\cos^2\zeta\rangle/\omega$ is \eqref{eq:onsetslope} for all $\ell$ and $n$.

\paragraph{Scalar harmonics.} With $\mathcal{Y}$ a harmonic of degree $\ell$ on the unit $S^3$, $\nabla^2\mathcal{Y}=-\ell(\ell+2)\mathcal{Y}$,
\begin{equation}
\tau=\hat\tau(\zeta)\,\mathcal{Y} e^{-i\omega t},\quad b_t=\hat b_t(\zeta)\,\mathcal{Y}e^{-i\omega t},\quad b_\zeta=\hat b_\zeta(\zeta)\,\mathcal{Y}e^{-i\omega t},\quad b_i=\hat b_s(\zeta)\,\partial_i\mathcal{Y} e^{-i\omega t},
\end{equation}
with $\hat b_s$ absent for $\ell=0$. The radial component of the Maxwell equation is algebraic in $\hat b_\zeta$, so that the four fields carry three degrees of freedom for $\ell\ge1$ and two for $\ell=0$; the constraint \eqref{eq:constraint} follows from the other equations. At the boundary, a normal mode has no source for $O_2$, no external field strength and no source for the phase,
\begin{equation}
\tau\simeq\tau_\vev\cos^2\zeta,\qquad \hat b_t+i\omega\,\hat b_s\to0,\qquad \hat b_\zeta\to0,
\label{eq:bcS3}
\end{equation}
while $\hat b_t$ and $\hat b_s$ tend to derivatives of the Goldstone field, $b_\mu\to-\partial_\mu\theta_0$, as in section~\ref{sec:bcs}. For $\ell=0$ and $\omega\neq0$, charge conservation forces the fluctuation of the charge density to vanish.

\paragraph{The lowest amplitude mode at large charge.} At $\ell=0$, with $'=d/d\zeta$, $\lambda/e^2=\frac12$ and the background expanded in $1/K$, eq.~\eqref{eq:tau} divided by $\cos^2\zeta$ becomes
\begin{equation*}
-\frac{(\tan^3\zeta\,\hat\tau')'}{\tan^3\zeta}+8\sec^2\zeta\,\hat\tau-\frac{28\sec^4\zeta-12\sec^2\zeta}{K}\,\hat\tau-\frac{8\sqrt2\sec^2\zeta}{K^{1/2}}\,\hat b_t=\omega^2\hat\tau+O\big(K^{-3/2}\big),
\end{equation*}
whose unperturbed lowest mode is $\tau_0=\cos^\Delta\zeta$, with $\omega_0=\Delta=2+2\sqrt3$ and the weight $\tan^3\zeta$. The $\zeta$ component of eq.~\eqref{eq:b} gives $\hat b_\zeta=-i\omega\cos^2\zeta\,\hat f/\rhob^2$, with $\hat f=\hat b_t'+i\omega\hat b_\zeta$, and with the radial flux $\Pi=\tan^3\zeta\cos^2\zeta\,\hat f$ the time component gives $\Pi'=\tan^3\zeta\,(\rhob^2\hat b_t+2\rhob\Bb_t\hat\tau)$. At $\rhob^2=8$ they combine into one equation for $\Pi$ and an expression for $\hat b_t$,
\begin{equation*}
\Big(\frac{\Pi'}{\tan^3\zeta}\Big)'-\frac{(8-\omega^2\cos^2\zeta)\,\Pi}{\tan^3\zeta\cos^2\zeta}=\frac{8\sqrt2}{K^{1/2}}\big(\sec^2\zeta\,\hat\tau\big)',\qquad\hat b_t=\frac{\Pi'}{8\tan^3\zeta}-\frac{\sqrt2\sec^2\zeta}{K^{1/2}}\,\hat\tau.
\end{equation*}
For $\hat\tau=\tau_0$ and $\omega=\omega_0$ the solution regular at the center and without a source at the boundary is $\Pi=2\sqrt2\,(3+\sqrt3)\,K^{-1/2}\sin^4\zeta\cos^{2\sqrt3-2}\zeta$; the homogeneous solutions behave there as $\cos^2\zeta$ and $\cos^{-4}\zeta$. First-order perturbation theory then gives
\begin{equation*}
K\,\delta(\omega^2)=12\langle\sec^2\zeta\rangle-12\langle\sec^4\zeta\rangle+\frac{\sqrt2\,K^{1/2}}{\|\tau_0\|^2}\int_0^{\pi/2}d\zeta\,\big(\sec^2\zeta\,\tau_0\big)'\,\Pi,
\end{equation*}
where $\|\tau_0\|^2=\int_0^{\pi/2}d\zeta\,\tan^3\zeta\,\tau_0^2$, the part of $\hat b_t$ proportional to $\hat\tau$ has turned $-28\langle\sec^4\zeta\rangle$ into $-12\langle\sec^4\zeta\rangle$, and the exchange term has been integrated by parts. The integrals are Beta functions: $\langle\sec^2\zeta\rangle=\Delta/(\Delta-2)$, $\langle\sec^4\zeta\rangle=\Delta(\Delta-1)/[(\Delta-2)(\Delta-3)]$ and the exchange term is $-(480+256\sqrt3)/11$. So $K\delta(\omega^2)=-(552+312\sqrt3)/11$, and with $K=\mu^2/4$ the shift of this mode is $\delta\omega=-\frac{24}{11}(8+5\sqrt3)/\mu^2$, at first order in $1/K$.

\paragraph{The descendant mode.} In the embedding coordinates of AdS$_5$, $X_0=\sqrt{1+r^2}\cos t$, $X_5=\sqrt{1+r^2}\sin t$ and $X_i=r\,\hat n_i$ with $\hat n\in S^3\subset\mathbb{R}^4$, the boost $J_{01}$ is the Killing vector
\begin{equation}
\mathcal{K}=\sqrt{1+r^2}\,\cos t\;\mathcal{Y}\,\partial_r-\frac{r\sin t}{\sqrt{1+r^2}}\;\mathcal{Y}\,\partial_t+\frac{\sqrt{1+r^2}\cos t}{r}\;\gamma^{ij}\partial_j\mathcal{Y}\,\partial_i,\qquad \mathcal{Y}=\hat n_1,
\end{equation}
with $\gamma_{ij}$ the metric of the unit $S^3$ and $\mathcal{Y}$ a harmonic with $\ell=1$. Since $\mathcal{K}$ preserves the metric, $\tau=\mathcal{K}^r\rhob'$ and $b=\mathcal{L}_{\mathcal{K}}\Bb$ solve the linearized equations. In the coordinate $r$, and in complex notation with time dependence $e^{-it}$,
\begin{equation}
\hat\tau=\sqrt{1+r^2}\,\rhob',\qquad \hat b_t=\sqrt{1+r^2}\,\Bb_t'-\frac{r\,\Bb_t}{\sqrt{1+r^2}},\qquad \hat b_r=-\frac{i\,\Bb_t}{(1+r^2)^{3/2}},\qquad \hat b_s=-\frac{i\,r\,\Bb_t}{\sqrt{1+r^2}},
\label{eq:descendant}
\end{equation}
which is regular at the center and satisfies \eqref{eq:bcS3}: $\hat\tau\simeq-2p/r^2$ has no logarithm, $\hat b_t\to-\mu$ and $\hat b_s\to-i\mu$, so that $\hat b_t+i\hat b_s\to0$, with the boundary Goldstone field $\theta_0=\mu\,\mathcal{Y}\sin t$. It is therefore a normal mode with $\omega=1$ and $\ell=1$ for every $\mu$: the descendant $P_\mu$ of the lowest state.

\section{Phonons at next-to-leading order}
\label{app:phonon}

This appendix derives eq.~\eqref{eq:phononS3} in two steps and ends with the coefficient $c_4$ at general couplings.

\paragraph{Structure on the sphere.} With the Goldstone field $\chi=\mu t+\pi$, the quadratic action of the phonon $\pi$ has two derivatives at leading order. At relative order $\mu^{-2}$ it has terms with four derivatives and, on $\mathbb{R}\times S^3$, curvature corrections to those with two. Since $\pi$ enters only through its derivatives, terms odd in $\partial_t$ are total derivatives and the Riemann tensor of $S^3$ is proportional to its metric, every term is $\omega^2$ or $k_\ell^2\equiv\ell(\ell+2)/R^2$, the eigenvalue of $-\nabla^2$, times a polynomial in both. With $\omega_\ell^2=k_\ell^2/3$ at leading order,
\begin{equation}
\omega_\ell^2=\frac{k_\ell^2}{3}\Big[1+\beta_1\frac{k_\ell^2}{\mu^2}+\frac{\beta_0}{\mu^2R^2}+O\big(\mu^{-4}\big)\Big],
\label{eq:phonontwoconst}
\end{equation}
with two constants. As $R\to\infty$ at fixed $k_\ell$ this becomes zero-temperature sound on the plane,
\begin{equation}
\omega=\frac{k}{\sqrt3}\Big(1+\Xi\,\frac{k^2}{\mu^2}+O(k^4)\Big),
\label{eq:alphadef}
\end{equation}
so $\beta_1=2\Xi$. The descendant at $\ell=1$ has $\omega R=1$ at every charge (section~\ref{sec:spectrum}) and $k_1^2R^2=3$, so $\beta_0=-3\beta_1$. Since $k_\ell^2R^2-3=(\ell-1)(\ell+3)$, for any conformal superfluid in four dimensions
\begin{equation}
\omega_\ell R=\sqrt{\frac{\ell(\ell+2)}{3}}\,\Big[1+\Xi\,\frac{(\ell-1)(\ell+3)}{\mu^2R^2}+O\big((\mu R)^{-4}\big)\Big].
\label{eq:phononalpha}
\end{equation}
This is the four-dimensional case of the dispersion relation obtained from the effective action in \cite{Cuomo2020}, where $\Xi$ is also given in terms of the Wilson coefficients (eq.~\eqref{eq:alphaWilson} below).

\paragraph{Effective action.} Through next-to-leading order, the effective Lagrangian in terms of the Weyl-invariant metric $\hat g_{ab}=Xg_{ab}$ \cite{Monin2017,Cuomo2020} is
\begin{equation}
\mathcal{L}=\sqrt{-\hat g}\,\Big(c_1+c_2\,\hat{\mathcal R}+c_4\,\hat{\mathcal R}^{ab}\partial_a\chi\,\partial_b\chi\Big)+\cdots,
\label{eq:Lnlo}
\end{equation}
where the curvatures are those of $\hat g_{ab}$, whose inverse raises the indices, and the omitted terms are of relative order $\mu^{-4}$. In the notation of \cite{Cuomo2020}, continued to Lorentzian signature, $c_2$ here is $-c_2$ there and $c_4$ is $c_3$. On the homogeneous state $\hat g_{ab}=\mu^2g_{ab}$ and the last term is proportional to $\mathcal{R}_{tt}=0$, so the first two terms give those of \eqref{eq:PEFT}, with the same $c_1$ and $c_2$, and $\Delta(Q)$ does not fix $c_4$. For a phonon on the plane the quadratic part of \eqref{eq:Lnlo} gives
\begin{equation}
c_1\mu^2\big(3\omega^2-k^2\big)=3c_2\,\omega^2\big(\omega^2-k^2\big)+c_4\,\omega^2\big(2k^2-3\omega^2\big),
\end{equation}
so that the coefficient in \eqref{eq:alphadef} is
\begin{equation}
\Xi=\frac{c_4-2c_2}{6c_1}.
\label{eq:alphaWilson}
\end{equation}
With \eqref{eq:EFTcoeffs} and $\Xi=1/9$ from \eqref{eq:soundk3} below, the coefficient that $\Delta(Q)$ leaves free is
\begin{equation}
c_4=-2c_1=-\frac{L}{4e^2}.
\label{eq:c4}
\end{equation}

\paragraph{Sound on the plane: setup.} At $T=0$, where $f=1$, take $e=L=1$, fields proportional to $e^{-i\omega t+ikx}$ and the gauge $a_u=0$, in which $b_t=a_t+i\omega\theta$, $b_x=a_x-ik\theta$ and $b_u=-\partial_u\theta$. Eqs.~\eqref{eq:tau} and \eqref{eq:b} then read
\begin{gather}
u^3\partial_u\Big(\frac{\partial_ua_t}{u}\Big)-u^2k\,(\omega a_x+ka_t)=\rhob^2\,b_t+2\rhob\Bb_t\,\tau,\qquad
u^3\partial_u\Big(\frac{\partial_ua_x}{u}\Big)+u^2\omega\,(\omega a_x+ka_t)=\rhob^2\,b_x,\nonumber\\
u^5\partial_u\Big(\frac{\partial_u\tau}{u^3}\Big)-u^2\bar k^2\tau=\Big(-4+\frac32\rhob^2-u^2\Bb_t^2\Big)\tau-2u^2\rhob\Bb_t\,b_t,
\label{eq:planarT0}
\end{gather}
with $\bar k^2=k^2-\omega^2$, and the $u$-component of the Maxwell equation is
\begin{equation}
\rhob^2\,\partial_u\theta=u^2\,\partial_u\big(i\omega a_t+ika_x\big).
\label{eq:radialphonon}
\end{equation}
The residual gauge freedom is fixed by $a^{(0)}_\mu=0$, the mode is normalized by $\theta\to1$ at the boundary, and the source of $O_2$ vanishes (section~\ref{sec:bcs}). Near the boundary $\rhob^2\simeq\mu^4u^4/2$ and $a_\mu\simeq a^{(2)}_\mu u^2$, so by \eqref{eq:radialphonon} $\theta$ has no term in $\ln u$, that is, no source, if and only if
\begin{equation}
\omega\,a^{(2)}_t+k\,a^{(2)}_x=0,
\label{eq:wardphonon}
\end{equation}
the conservation of the charge. Since $\omega<k$, $\bar k^2>0$, and $b_a$ and $\tau$ decay in the interior as $e^{-\bar ku}$.

\paragraph{Static problems and the interior.} Expanded in $\omega\sim k$, the problem becomes a sequence of static problems with sources from the lower orders. For $b_x$ the static homogeneous solutions are the London profile $\Bb_t/\mu$ \eqref{eq:London}, which decays as $u^{-2}$, and one that grows as $u^4$. For $(b_t,\tau)$ they are $(\partial\Bb_t/\partial\mu,\partial\rhob/\partial\mu)$, the response to a change of the chemical potential, which decays as $u^{-2}$, one that grows as $u^4$ and two that behave as $u^{2\pm2\sqrt3}$. Where $u\gg1/\mu$ the background is the infrared fixed point, at which both fields have mass squared $8$, and the solutions that decay at finite $\bar k$ are powers of $u$ times $K_3(\bar ku)$ and $K_{2\sqrt3}(\bar ku)$. Expanded in $\bar ku$, those for $b_t$ and $\tau$ begin as $u^{-2}$ and $u^{2-2\sqrt3}$, and the terms that grow as $u^4$ and $u^{2+2\sqrt3}$ appear only at relative order $(\bar ku)^6\ln\bar ku$ and $(\bar ku)^{4\sqrt3}$. Through relative order $k^4$ the decay in the interior therefore admits, at each order, only solutions that grow more slowly than $u^4$. The source-free conditions fix them, since a static solution without sources that decays in the interior would be a static normal mode, excluded in section~\ref{sec:energy}. The infrared region therefore cannot affect the dispersion relation before relative order $(k/\mu)^6$.

\paragraph{Sound on the plane: the expansion.}
\begin{itemize}
\item At order $k$ the solution is the static response to the boundary values $b_\mu=(i\omega,-ik)$: a change of the chemical potential by $i\omega$, $b_t=i\omega\,\partial\Bb_t/\partial\mu$ and $\tau=i\omega\,\partial\rhob/\partial\mu$, and the London profile of section~\ref{sec:transverse}, $b_x=-ik\,\Bb_t/\mu$.
\item At order $k^2$, eq.~\eqref{eq:radialphonon} gives
\begin{equation}
\partial_u\theta=-\frac{(\omega^2+k^2)\,\mu^2u^2+4\,(k^2-3\omega^2)}{\mu^2u\,(\mu^2u^2+4)} .
\end{equation}
Its term in $1/u$ vanishes if and only if $\omega^2=k^2/3$, and then
\begin{equation}
\theta=1-\frac{2k^2}{3\mu^2}\ln\Big(1+\frac{\mu^2u^2}{4}\Big).
\end{equation}
\item At order $k^3$ the terms with $\omega$ and $k$ in \eqref{eq:planarT0}, including those through $\theta$, act as sources for $(b_t,\tau)$ but, since $3\omega^2=k^2$ at this order, not for $b_x$. A constant $b_t=i\delta$ with $\tau=0$, where $\delta\equiv2k^3/(3\sqrt3\,\mu^2)$, solves the equations with these sources; for $\tau$ this rests on $\rhob\Bb_t=\frac12\mu^2\partial\rhob/\partial\mu$. Adjusting the solution of order $k$ so that $b_t\to i\omega$ at the boundary gives
\begin{equation}
b_t=i(\omega-\delta)\,\partial\Bb_t/\partial\mu+i\delta,\qquad \tau=i(\omega-\delta)\,\partial\rhob/\partial\mu,\qquad b_x=-ik\,\Bb_t/\mu ,
\end{equation}
which is bounded as $u\to\infty$, as the interior requires.
\item At order $k^4$, eq.~\eqref{eq:wardphonon} with $a_t=b_t-i\omega\theta$ and $a_x=b_x+ik\theta$, where $\partial\Bb_t/\partial\mu\simeq1-\frac34\mu^2u^2$, $\Bb_t/\mu\simeq1-\frac14\mu^2u^2$ and $\theta\simeq1-\frac16k^2u^2$ near the boundary, gives
\begin{equation}
3\omega^2-k^2-3\omega\delta-\frac{2k^2}{3\mu^2}\big(\omega^2-k^2\big)=0,
\end{equation}
that is, to this order, $3\omega^2=k^2+2k^4/9\mu^2+O(k^6)$.
\end{itemize}
Hence the dispersion relation and $\Xi$ are
\begin{equation}
\omega=\frac{k}{\sqrt3}\Big(1+\frac{k^2}{9\mu^2}+O(k^4)\Big),\qquad \Xi=\frac19 .
\label{eq:soundk3}
\end{equation}
In units of $k^4/\mu^2$, the constant $\delta$ contributes $\frac23$ to $3\omega^2-k^2$, and the logarithm in $\theta$ contributes $\frac29$ through $a_t$ and $-\frac23$ through $a_x$. The contributions of $\delta$ and $a_x$ cancel, so $\Xi$ comes from the logarithm through $a_t$; without the logarithm it would be $1/3$. With \eqref{eq:phononalpha}, $\Xi=1/9$ gives \eqref{eq:phononS3}.

\paragraph{The next order.} The expansion continues in the same way. At order $k^4$, eq.~\eqref{eq:radialphonon} gives $\theta$ in closed form, a term in $u^2$ and a multiple of $\ln(1+\mu^2u^2/4)$, and at order $k^5$ the static problems have rational solutions, which grow as $u^2$ in the interior. Eq.~\eqref{eq:wardphonon} at order $k^6$ needs only their coefficients of $u^2$ at the boundary. Green's identities give these, without solving, as integrals of the sources against the two static solutions that decay in the interior without a source of $O_2$, $(\partial\Bb_t/\partial\mu,\partial\rhob/\partial\mu)$ and $\Bb_t/\mu$. The result is
\begin{equation}
\omega=\frac{k}{\sqrt3}\Big(1+\frac{k^2}{9\mu^2}+\frac{19\,k^4}{162\,\mu^4}+O(k^6\ln k)\Big).
\label{eq:soundk5}
\end{equation}
At order $k^7$ the logarithm in $K_3$ brings in the solution that grows as $u^4$, so the first term that is not analytic in $k^2$ is of relative order $(k/\mu)^6\ln(k/\mu)$.

\paragraph{The coefficient $c_4$ at general couplings.} Eq.~\eqref{eq:c4} does not rely on the Bogomolny coupling. Take the zero-temperature planar flow for any $V(\rho)$, with the canonical Maxwell term of \eqref{eq:action} and a condensate that does not vanish in the interior. With $s=\ln(\mu u)$ and $Y=\Bb_t/\mu$, the Maxwell equation is
\begin{equation*}
\partial_s^2Y-2\partial_sY=e^2L^2\rhob^2\,Y,\qquad Y\to1\ (s\to-\infty),\quad Y\to0\ (s\to\infty).
\end{equation*}
A slowly varying phonon is locally a boost and a rescaling of this flow, $B_a=Y\partial_a\chi$ and $\rho=\rhob$, with $s=\ln u+\frac12\ln X$. Corrections to these fields enter the action only at fourth order in derivatives, since the action is stationary. The $u$-component of the Maxwell equation gives $B_u=-uY\,\partial^a\chi\,\partial_aX/2X$, in which $\rhob^2$ cancels by the equation above. The two-derivative part of the action is then
\begin{equation*}
\frac{L}{8e^2}\Big[(I_Y-I_\rho)\frac{(\partial X)^2}{X}+(I_Y+I_{\rho Y})\frac{(\partial^a\chi\,\partial_aX)^2}{X^2}\Big],
\end{equation*}
with $I_Y=\int ds\,(\partial_sY)^2$, $I_\rho=\int ds\,(eL\,\partial_s\rhob)^2e^{-2s}$ and $I_{\rho Y}=\int ds\,e^2L^2\rhob^2Y^2$. On the leading equation of motion, the curvature terms of \eqref{eq:Lnlo} equal $[\frac32c_2-\frac12c_4](\partial X)^2/X-\frac12c_4(\partial^a\chi\,\partial_aX)^2/X^2$, so $c_4=-(L/4e^2)(I_Y+I_{\rho Y})$ and $c_2=-(L/12e^2)(I_\rho+I_{\rho Y})$. Multiplying the equation for $Y$ by $Y$ and integrating gives $I_Y+I_{\rho Y}=1$, whatever $V$ is. Hence, given the derivative expansion,
\begin{equation}
c_4=-\frac{L}{4e^2},\qquad\Xi=\frac{L}{36e^2c_1}\Big(I_\rho-I_Y-\frac12\Big).
\label{eq:c4general}
\end{equation}
At the Bogomolny point $I_Y=\frac13$, $I_{\rho Y}=\frac23$ and $I_\rho=\frac43$, which give $c_2=-L/6e^2$ and $\Xi=\frac19$. On the exact flow at $\lambda=8e^2/3$ of section~\ref{sec:othercouplings}, $I_Y=\frac14$, $I_{\rho Y}=\frac34$ and $I_\rho=\frac12$, which give $c_2=-5L/48e^2$ and, with $c_1=L/16e^2$, $\Xi=-\frac19$. The value of $c_4$ needs the canonical Maxwell term. With $Z(\rho)F^2$ and $Z\to1$ at the boundary, so that the two-point coefficient of the current is unchanged, the same steps give
\begin{equation}
c_4=-\frac{L}{4e^2}\int_0^1d(Y^2)\,Z,
\label{eq:c4Z}
\end{equation}
an average of $Z$ over the wall, so the two-point coefficient alone does not fix $c_4$. In the effective theory $c_4$ is a free Wilson coefficient: it vanishes at leading order in the U(1) model of \cite{BCMR2019}, as in section~\ref{sec:phononNLO}, and is of order $N^0$ in the O($N$) model at large $N$ in three dimensions \cite{DS2025}.

\section{Numerical methods}
\label{app:numerics}

The programs for the numerical computations and the symbolic checks were written by Claude Opus 5.5 (Anthropic) at the author's direction. The units are $e=L=u_h=1$ on the branes at $T>0$, $e=L=1$ and $\mu=2$ at $T=0$, and $L=R=1$, with $e$ absorbed into the scalar and $r=1/u$, on the curved slicings.

\paragraph{Symbolic checks.} The residuals of \eqref{eq:radial} on \eqref{eq:family} vanish identically in $\kappa$, $u_h$, $R$ and $\mu$, and on the axion branes with $\kappa/R^2=-\alpha^2/4$. Gr\"obner bases give $q_{\rm RN}=0$ for the coefficient equations of section~\ref{sec:familysol} with a term $q_{\rm RN}^2u^6$ in $f$, and no real solution on the AdS$_5$ soliton (section~\ref{sec:beyondfamily}). Eq.~\eqref{eq:firstorderT} with \eqref{eq:ckappa} holds on the family for general $\kappa$, $R$ and $u_h$, preserves the constraint and has residuals proportional to it. Neither first-order system, as written, satisfies the integrability condition for a superpotential with the field-space metric of the radial action, $G_{\hat\rho\hat\rho}:G_{\Phi\Phi}=-2:f$; at $T=0$ the superpotential of section~\ref{sec:structure} and the completion of squares hold, with the exact solution as a control. Substitution verifies the order-$T^4$ boosted fields \eqref{eq:lowTfields}, their boundary data and \eqref{eq:PXT}, the order-by-order solutions of \eqref{eq:planarT0} for general $\mu$, and the solution at $\lambda=8e^2/3$ of section~\ref{sec:othercouplings} with its first-order equations; the solution fails at other $\lambda$. The identity of section~\ref{sec:closure} holds on rational test functions with a nonquadratic $f$ and general $m^2$. The slope in the first open question of section~\ref{sec:discussion} follows from the near-critical data of appendix~\ref{app:nearTc}, with the controls $0$ at $\lambda=e^2/2$ and $2$ as $\lambda\to\infty$, the normal phase, and its converse holds order by order through $u^{10}$. The identity and the converse fail at $\lambda/e^2=\frac35$.

\paragraph{Quasinormal modes.} The linearized equations, derived symbolically from \eqref{eq:action} in ingoing Eddington--Finkelstein coordinates for the scalar perturbation and $a=A-\bar A$, in the gauge $a_u=0$ imposed after the variation and with the $a_u$ equation kept as the constraint, are discretized by Chebyshev--Gauss--Lobatto collocation on $u\in[0,1]$. The fields are rescaled by $u^2$, so that the source-free conditions \eqref{eq:bc} and the ingoing condition are natural; the residual gauge is fixed by $a^{(0)}_\mu=0$, and at the horizon the row of the $a_v$ equation is replaced by the radial constraint, without which the problem is degenerate at every frequency. The solver reproduces the diffusion constant $D=1/2$ and the Maxwell frequencies $\pm2-2i$ of the normal phase. At eight temperatures between $0.4\,\Tc$ and $0.98\,\Tc$, fits of the sound mode at small $k$ agree with \eqref{eq:cs} and \eqref{eq:Gamma} within $10^{-7}$, relative. On the axion branes, at $\alpha=1$, $2$ and $2.5$ and two values of $\mu$ each, the fits reproduce $c_s^2$ and $\Gamma$ of \eqref{eq:Gammaaxion}, checked symbolically, to $10^{-8}$. At $\vk=0$, at eight temperatures between $0.3\,\Tc$ and $0.95\,\Tc$, the next modes after the amplitude mode are complex pairs with $|\omega|>2\pi T$. As the resolution is raised, the relaxation rate of the amplitude mode in figure~\ref{fig:qnm} changes by at most $3\times10^{-10}$, relative; at $(\Tc/T)^2=3+4\sqrt{21}/7$ it is $\gamma/\pi T=2$ to $10^{-8}$, so the mode at $\omega=-2\pi iT$ of appendix~\ref{app:attenuation} is the amplitude mode; the near-horizon expansion there, done symbolically for any $\lambda$, reproduces at $\lambda=0$ the residue determinant of \cite{NatsuumeOkamura2026}.

\paragraph{General quartic coupling.} At $\lambda/e^2=0$, $\frac14$, $1$ and $2$, with $m^2L^2=-4$, the background is constructed by spectral Newton iteration at fixed $\mu$. The attenuation from the quasinormal modes, extrapolated in $\epsilon^2$ from five backgrounds with $0.1\le\epsilon\le0.4$, agrees with \eqref{eq:GammaTc} to $10^{-5}$ at these couplings and at $\lambda=1/2$. The ratio $Y(u_h)^2/(1-n_s/n)$ of section~\ref{sec:conductivity} is obtained by shooting, and $\mu(n-n_s)=Ts$ is tested with $P=-\Om$ from \eqref{eq:Omega} and $Ts=4P-\mu n$. The failing relations of table~\ref{tab:generic}, from $n$ independent of $T$ to the split of $\Gamma$, were checked on these backgrounds at $0.9\,\Tc$ and $0.7\,\Tc$, with finite differences in $\mu$. At $\lambda=1/2$ these methods reproduce the exact results to $4\times10^{-4}$ or better.

\paragraph{Conductivity.} At $T>0$, eq.~\eqref{eq:transverse} at $\vk=0$ is integrated from ingoing horizon data with an adaptive Runge--Kutta method, which reproduces \eqref{eq:sigmaN} to $10^{-13}$. At $T=0$ it is integrated inward from $uH^{(1)}_3(\omega u)$, and ${\rm Re}\,\sigma$ is evaluated from the conserved flux \eqref{eq:flux}, since at small frequency ${\rm Re}\,\sigma/|{\rm Im}\,\sigma|$ falls below double precision. The Ferrell--Glover--Tinkham sum rule $\int_0^\infty d\omega\,[{\rm Re}\,\sigma_\N-{\rm Re}\,\sigma_\reg]=\pi n_s/2\mu$ \cite{HHH2008b,GHK2011} holds to $10^{-8}$ at $T=0$ and at $0.5$, $0.7$ and $0.9\,\Tc$.

\paragraph{Amplitude channel at $T=0$.} With $b_u$ eliminated through the constraint, the equations for $\tau$ and $b_t$ become a regular first-order system for $(\tau,\partial_u\tau,b_t,\Pi_t)$, with $\Pi_t=f_{ut}/u$ as in appendix~\ref{app:attenuation}. The system is integrated inward from the outgoing infrared solutions, $\tau\propto u^2H^{(1)}_{2\sqrt3}(\omega u)$ in the scalar channel and $b_u\propto u^2H^{(1)}_3(\omega u)$ in the vector channel, with their asymptotic corrections in $1/\mu u$, and ${\rm Im}\,G^R_{O_2O_2}$ is evaluated from the conserved radial flux, which separates the two infrared contributions, for $\omega\le3\mu$. Fits of each contribution in the powers of $\omega$ that the matching predicts reproduce its coefficient in \eqref{eq:T0amp} to $10^{-9}$, and the results change by less than $2\times10^{-9}$, relative, under changes of the infrared start, the tolerance and the cutoff.

\paragraph{Nodal branches.} The static equations \eqref{eq:radial} on global AdS$_5$ are integrated from the center with an adaptive Runge--Kutta method, in the variables $\ln r$ and $r^2\rho$, and the central data are tuned to the zeros of the coefficient of the logarithm. The integration reproduces the exact branch to $3\times10^{-11}$ for $\mu R\le20$, and the nodal branches are accurate to $2\times10^{-8}$.

\paragraph{Transverse spectrum on $S^3$.} A Jacobi--Galerkin method solves eq.~\eqref{eq:transverseS3} in $x=r^2/(r^2+r_0^2)$, with $r_0^2=\mu/2$, which resolves both $r\sim1$ and the scale $r\sim\mu/2$ of the condensate, and with the behaviors $x^{(\ell+1)/2}$ at the center and $1-x$ at the boundary factored out. It reproduces the Maxwell and Proca spectra to $10^{-10}$, and shooting from the center agrees to the six digits it resolves. Richardson extrapolation of $(\omega R-5-\ell-2n)\mu^2R^2$ over $64\le\mu R\le128$ gives the coefficient of \eqref{eq:transverseshift} to $6\times10^{-6}$ for all sixteen $(\ell,n)$ with $\ell\le4$ and $n\le3$.

\paragraph{Scalar spectrum on $S^3$.} In the linearized equations, derived symbolically from \eqref{eq:L2}, eliminating $b_r$ through the radial Maxwell equation introduces singular points at $\omega$-dependent radii. The collocation method therefore uses the radial electric flux $\Pi=r^3(\hat b_t'+i\omega\hat b_r)$, in terms of which $\hat b_r$ and $\hat b_t$ are algebraic and the system for $(\Pi,\hat b_s,\hat\tau)$ is regular for $r>0$ and quadratic in $\omega$. It is discretized in $x$, with $r_0^2=\max(1,\mu R/2)$ and the regular and source-free behaviors factored out, so that \eqref{eq:bcS3} is imposed by the equations at the end points; in extended precision, two resolutions filter spurious modes. It reproduces the frequency of the exact mode \eqref{eq:descendant} to $10^{-12}$ for all $\mu R\le80$. Shooting in unitary gauge, complete since $\rhob>0$ everywhere, agrees to $1.3\times10^{-10}$ on all 556 frequencies computed by both, up to $\mu R=40$, and a separate shooting program, compared only afterwards, to $2\times10^{-9}$ up to $\mu R=80$. Fits of the phonon gaps from $\mu R=16$ that include a term $(\mu R)^{-6}\ln\mu R$, the order at which the infrared region can first enter on the plane (appendix~\ref{app:phonon}), reproduce the coefficient $(\ell-1)(\ell+3)/9$ of \eqref{eq:phononS3} to $2\times10^{-5}$, relative. The coefficient $-\frac{24}{11}(8+5\sqrt3)$ of the lowest amplitude mode is reproduced to about $0.01$ by fits in powers of $(\mu R)^{-2}$.

\paragraph{R\'enyi onset.} Shooting from the horizon for the static mode with $-\nabla^2=1/R^2$ gives $n_*$ at $\mu=0$ and, with $B_t=\mu(1-u^2/u_h^2)$, $\mu_{\rm norm}(n)$ as the lowest root in $\mu$, both stable to $10^{-6}$ under changes of the cutoff; for the homogeneous mode it reproduces $\muc(n)$ to $10^{-7}$. At $\Delta_O=2$, $n_*$ agrees with the onset plotted against the dimension of the scalar in \cite{BMM2013}, within the accuracy of the plot. First-order perturbation about the zero mode at $\muc$ gives $\partial\mu^2/\partial(-\nabla^2)=2$, the limit of $(\mu_{\rm norm}^2-\muc^2)R^2$ as $n\to0$; the difference decreases to $1.77$ at $n_*$.

\paragraph{Other dimensions.} The zero mode at $\Tc$, a generalized eigenvalue problem in $\mu^2$, gives $\mu_cu_h=2.47716$, $1.89647$ and $1.87662$ for $d=3$, $5$ and $6$, respectively, and the exact value $2$ at $d=4$. On the nonlinear backgrounds at $\lambda=e^2/2$, constructed by Newton iteration with the charge read off from the radial Gauss law, $n/\mu^{d-1}$ changes between $\Tc$ and $0.2\,\Tc$ by $-3.0\%$, $+7.9\%$ and $+19.7\%$ for $d=3$, $5$ and $6$, respectively; the values are stable in the resolution.

\paragraph{Other couplings and masses.} Table~\ref{tab:generic} rests on twenty pairs $(\Delta_O,\lambda/e^2)$: $\Delta_O=2$ with $\frac18$, $\frac14$, $\frac12$, $1$, $\frac54$, $\frac32$, $2$ and $4$; $\Delta_O=\frac52$ with $\frac14$, $\frac12$, $1$ and $2$; $\Delta_O=3$ with $\frac14$, $\frac12$, $\frac34$ and $\frac32$; $\Delta_O=\frac72$ with $\frac14$, $\frac12$, $1$ and $4$. Planar flows at $T=0$ are solved by multidomain collocation in $\ln u$; two integrals for $c_1$ and $c_2$ agree to $10^{-12}$. The planar branch is followed from $\Tc$ to $10^{-3}\Tc$ by Newton continuation in $\mu/T$, with exact horizon conditions and no infrared condition, and at $(2,\frac12)$ it reproduces the closed form to $4\times10^{-12}$ in $n$ and $P$. Only at $\Delta_O=2$ is the infrared bound crossed on a fine grid. Solitons are computed up to $\mu R=2000$ at $\Delta_O=2$ and $1000$ otherwise, and obey the first law to $7\times10^{-6}$. The value of $\Xi$ from the phonons on $S^3$ agrees with \eqref{eq:c4general} to $2\times10^{-5}$ at all twenty pairs, apart from $1.2\times10^{-4}$ at $(2,\frac54)$. Separate computations, mainly by Taylor integration in double-double arithmetic with shooting from the infrared and from the center and by a second spectral solver, reproduce the exponent of ${\rm Re}\,\sigma$ at eight of these pairs to $1\%$ and test the flows, solitons, normal modes and spectral bounds at further pairs; their phonons give $\Xi=-\frac19$ at $(2,\frac83)$ to $10^{-5}$. At $m^2L^2=-4$ they reach the fixed point directly only for $\lambda/e^2\le1.1$ or $\lambda/e^2\ge\frac32$; with the infrared condition imposed on the stable manifold of each fixed point, including its slow direction, the flow ends at the minimum for every $\lambda/e^2\le1.33$ computed and at the other fixed point for every $\lambda/e^2\ge1.337$. Very near the bound, quantities converge as small powers of $\mu$ or $T$, so the tower limits and some exponents are not resolved. Above the bound they give the coefficient of $1/\mu$ in the charge, solved without fits from three values of $\mu$ up to $\mu R=32768$, which agrees with \eqref{eq:lnmuR} to $6\times10^{-3}$ or better at $(2,\frac83)$, $(\frac72,4)$ and $(\frac52,2)$; at $(2,\frac7{10})$, below the bound, it is $2\times10^{-5}$.

\paragraph{D7 branes.} The scalars of dimension two are the $\ell=1$ components along $S^3$ of the gauge field on a D7 brane that wraps AdS$_5\times S^3$, a triplet of SU(2)$_R$ invariant under SU(2)$_L$ \cite{KMMW2003}. The Yang--Mills and Wess--Zumino terms of the D7-brane action are reduced on $S^3$ to the modes invariant under SU(2)$_L$, and the pair energies follow from the solvability condition at third order in the amplitude, on global AdS$_5$ with the metric frozen. This gives $(1+\lambda/e^2)/12\Nn$ for the Abelian Higgs model and, along a null $\vec z$, $\vec z\cdot\vec z=0$, the three contributions to $\Nn U$ of section~\ref{sec:topdown}, which cancel, as protection requires.

\bibliographystyle{unsrturl-compact}
\bibliography{refs}

\end{document}